\documentclass[fleqn,10pt]{wlscirep}
\RequirePackage[numbers,comma,sort&compress]{natbib}
\usepackage{microtype}
\usepackage[T1]{fontenc}
\usepackage{tikz}
\usepackage{amsmath}
\usepackage[utf8]{inputenc}
\usepackage{graphicx}   %
\usepackage{subcaption} %
\usepackage{hyperref}  %
\usepackage{etoolbox}
\AtBeginEnvironment{thebibliography}{%
  \setlength{\itemsep}{0pt}%
  \setlength{\parsep}{0pt}%
  \setlength{\baselineskip}{11pt}%
}
\title{Human mobility shapes how urban policy effects spread across cities}
\author[1,2]{Bijin Joseph}
\author[2]{Hamish Gibbs}
\author[3]{Takahiro Yabe}
\author[1,2, *]{Esteban Moro}
\affil[1]{Department of Physics, Northeastern University, Boston, MA, USA}
\affil[2]{Network Science Institute, Northeastern University, Boston, MA, USA}
\affil[3]{Tandon School of Engineering, New York University, Brooklyn, NY, USA}
\affil[*]{e.moroegido@northeastern.edu}

\usepackage{mathtools}
\usepackage{adjustbox}
\usepackage{multirow}
\usepackage{url}

\begin{abstract}
Urban policies are often evaluated assuming their effects decay with geographic distance. However, cities are systems of human movement, and policy effects propagate through individual mobility rather than spatial proximity alone. This study demonstrates that neglecting mobility can systematically distort estimates of how urban policies affect activity. Using large-scale mobility data, we estimate the causal effects of three transportation interventions in Boston and New York City using both conventional geographic and mobility-based definitions of exposure. Geographic designs detect no effects, or even negative ones. When exposure is defined through mobility networks, the same interventions reveal substantial behavioral change. Boston’s free-fare program increases overall activity by approximately 16\%, with more pronounced effects among lower-income communities. Open Streets programs significantly increase visits to nearby businesses, particularly food establishments. For congestion pricing, geographic approaches suggest a decline in activity in Manhattan, while a mobility-based perspective indicates no net reduction but a shift away from shopping and grocery destinations. Analysis across 138 metropolitan areas in eight countries further reveals that geographic and mobility-network distances diverge locally in every city, indicating that this discrepancy is a general characteristic of urban mobility. Collectively, these findings show that urban policies operate not only where they are implemented but where and how people move, so their impact must be measured through mobility rather than through space.
\end{abstract}
\begin{document}

\flushbottom
\maketitle
\thispagestyle{empty}

\section*{Introduction}

Cities are not collections of places arranged in space but complex systems of human movement \cite{bettencourt_introduction_2021,barbosa_human_2018,batty_cities_2009}. Daily mobility connects neighborhoods into dense behavioral networks that determine access to jobs, services, and opportunities \cite{batty_cities_2009,yabe2025behaviour,bassolas_hierarchical_2019}. Transportation policies act directly on this structure: they set the cost and reach of movement, so they determine who can get to what. Implemented locally, they propagate through these networks and produce effects that extend well beyond their immediate area.

Yet the methods used to evaluate these policies either assume that effects decay with physical distance and remain localized around the treated area \cite{zheng_effects_2024,butts_jue_2023,pollmann_causal_2023,volker_economic_2021}, or ignore spatial structure altogether, comparing treated units to a system-wide or city-wide aggregate \cite{boston_fare_free_mid_program_report,cook_short-run_2025}. The first is the more common choice. It estimates the policy effect by comparing treated units with controls that stand in for the counterfactual, and it assigns the two by distance, taking places far enough apart to be independent. That assumption fails in cities. Urban residents routinely commute tens of kilometers each day and move across neighborhoods for shopping, work, and social activity \cite{gonzalez_understanding_2008,batty_cities_2009,garcia2024effect}, linking distant areas while leaving some adjacent ones only weakly connected \cite{weng_beyond_2025}. Cities partitioned by these flows rather than by administrative boundaries break into functional regions that organize commuting and economic activity \cite{ratti_redrawing_2010,nelson_economic_2016}, a structure that bears no fixed relation to distance. Controls chosen by physical distance are therefore silently exposed through the mobility flows that policies act upon, violating standard assumptions in causal inference under interference \cite{hudgens_toward_2008,aronow_estimating_2017,ugander_graph_2013}. Policies that substantially reshape urban opportunity can then appear weak or ineffective when evaluated through a purely spatial lens \cite{zheng_effects_2024,cook_short-run_2025,volker_economic_2021,yoshimura_street_2022}.

To address this, we redefine exposure through the structure of human mobility itself. Using large-scale, anonymized mobile phone data, we construct city-wide mobility networks in which neighborhoods are connected by flows of people moving between them \cite{yabe2025behaviour,gonzalez_understanding_2008}. We then introduce a Network Distance Method (NDM) that assigns treatment and control status by position in this network rather than by physical distance, adapting the framework of causal inference under network interference \cite{hudgens_toward_2008,aronow_estimating_2017,ugander_graph_2013,eckles_design_2017} to the urban setting. Combined with synthetic control methods \cite{abadie2010synthetic, abadie2021using}, NDM recovers policy effects in the presence of interference that standard geographical methods attenuate, and in some cases invert. In simulated data where the true effect is known, NDM recovers it, while geographic designs return zero or the wrong sign at the short distances where buffers are drawn.

We apply the method to three transportation interventions in Boston and New York: a fare-free bus program, street pedestrianization, and congestion pricing. All three were prominent policy experiments, watched closely as models for other cities. Two of the three lower the cost of reaching a place and the third raises it, so together they test whether the same exposure problem arises no matter which way a policy pushes. We ask what each did to urban activity rather than to the service it altered: where residents went, what they did there, and whom they encountered. In all three cases, the conclusion changes with the definition of exposure. Geographic designs return null or inconsistent estimates, while the same data under a mobility-based definition indicate considerable reallocation of activity, concentrated in the populations each policy was designed to reach. The divergence between geographical and network distance is not particular to these cases. Across 138 metropolitan areas in eight countries and five independent data sources, the two diverge locally in every city, and most sharply at the short range where geographical buffer designs operate. Mobility, not geography, is the substrate through which urban policies propagate.

\begin{figure}[t!]
    \centering
    \includegraphics[width=0.9\linewidth]{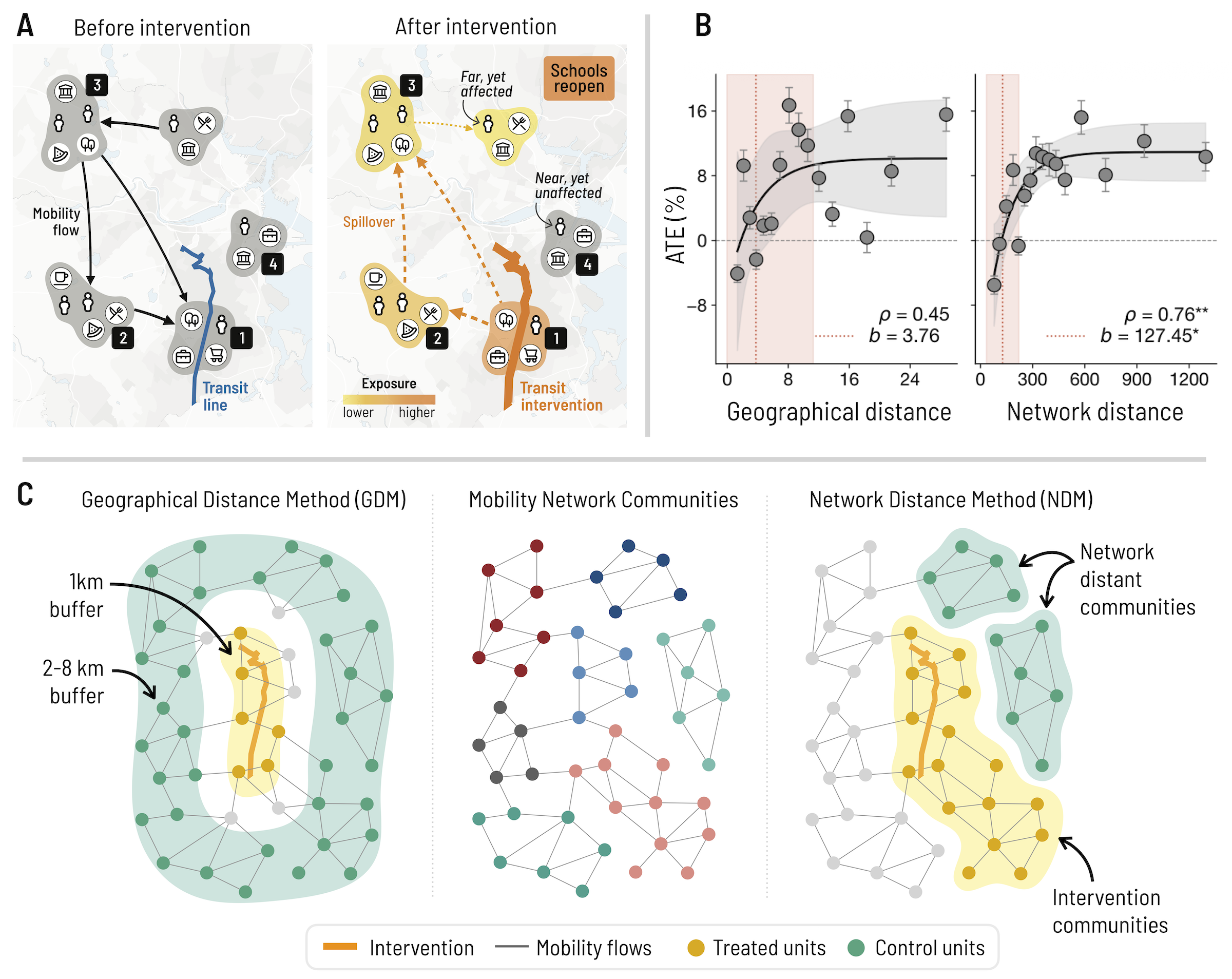}
    \caption{\textbf{A mobility-network approach to defining policy exposure.} \textbf{A,} Exposure under a mobility network. Individuals can be affected directly (1) or indirectly (2, 3) by an intervention through their daily movement, even when geographically distant from it, while geographically adjacent areas (4) may remain unexposed. Concurrent shocks such as school reopening act on the same units. \textbf{B,} Estimated average treatment effect (ATE) for the Route~28 fare-free program as a function of the distance between treated and control communities, measured geographically (left) and in the mobility network (right). Each point is the ATE recovered using controls in one distance bin, with whiskers showing the 95\% confidence interval; curves are saturating fits $c+a(1-e^{-d/b})$ with the shaded band indicating fit uncertainty, and the dotted vertical line marks the fitted characteristic distance $b$ and its 95\% confidence interval. Within each panel, $\rho$ is the Spearman correlation between the ATE and the distance across bins, with a two-sided permutation $p$-value. The ATE increases with network distance ($\rho=0.76$, $p=0.001$) but shows no significant relationship with geographical distance ($\rho=0.45$, $p=0.080$). \textbf{C,} Comparison of exposure definitions. The Geographical Distance Method (GDM) assigns treated and control units using geographic buffers, whereas the Network Distance Method (NDM) assigns them using mobility communities detected with the Leiden algorithm~\cite{traag_louvain_2019}, selecting control communities that are network-distant from the intervention but demographically comparable.}
    \label{fig1}
\end{figure}

\section*{Results}
We evaluate the effect of each policy under both geographic and mobility-based definitions of exposure, so the two designs are compared on identical data. We also go beyond the service-use measures that evaluations of transportation policy typically rely on, such as bus boardings \cite{boston_fare_free_mid_program_report}, toll payments \cite{cook_short-run_2025}, or visits to affected businesses \cite{zheng_effects_2024,de2025causal,volker_economic_2021}. We ask instead how activity is reallocated across the city \cite{liao_uncovering_2025}, measuring three dimensions of that reallocation for each intervention (Methods). Activity is the volume of visits to points of interest (POIs), reported in total and by POI category. Spend is the dollar volume those visits generate at businesses. The social dimension covers who benefits and what they reach: exploration counts the distinct venues residents visit, registering whether a policy widens their destinations rather than only the frequency of familiar trips, and experienced segregation measures the income mix encountered at those venues \cite{moro2021mobility}. We also break every outcome down by the income quartile of the resident's home neighborhood, since a policy can raise activity in aggregate while reaching only part of the population it was meant to serve.

\subsection*{Defining exposure through mobility networks}
The conventional answer is the Geographical Distance Method (GDM), which assigns treatment and control groups by spatial proximity to the intervention and treats units outside a buffer as unexposed (Fig.~\ref{fig1}A,C; SI Note~\ref{supp:sec:GDM}). Spatial distance is a poor proxy for exposure in cities. Individuals in large US cities routinely travel substantial distances within a single metropolitan area, around 4.4 km for grocery shopping and up to 7 km for fast food visits \cite{garcia2024effect}. Areas far from an intervention may therefore be exposed through mobility flows, while nearby areas may remain only weakly connected (Fig.~\ref{fig1}A). Some evaluations push the buffer further and compare outcomes across cities altogether \cite{cook_short-run_2025}, but cities differ in commuting patterns, daily rhythms, and the temporal structure of activity, so policies that operate at specific times of day may affect behavior differently across them even when pre-treatment trends look similar.

We define exposure through mobility networks instead. Using large-scale anonymized mobility data, we construct a network in which nodes are spatial units and edges are flows of people between them (Fig.~\ref{fig1}C), and assign treatment and control by position in this network rather than by physical distance. Ordering control communities by their network distance from the Route~28 intervention reveals a saturation pattern (Fig.~\ref{fig1}B). The recovered effect rises sharply, stabilizes at an asymptotic value, and is consistently significant beyond a characteristic distance. Geographic distance produces no such pattern. Controls at moderate geographic distance still register attenuated or negative effects, and even at large distances the recovered ATE often fails to reach significance. Controls that are behaviorally close to the intervention bias the recovered effect toward zero, and geographic separation does not prevent it.

Individual spatial units are the wrong scale for this. Nearby areas and businesses are densely connected by mobility flows, so effects spill freely across the boundary of any single unit. We instead work with mobility communities: groups of units within which movement is dense and across whose boundaries it is sparse. Most movement stays inside a community, so a community can be labeled treated or control as a near-closed unit, in direct analogy to cluster randomization in network experiments \cite{ugander_graph_2013,eckles_design_2017}. We detect communities with the Leiden algorithm \cite{traag_louvain_2019} and confirm the partition under Infomap \cite{rosvall2007maps,mirzaee2020urban}, and classify as treated any community the intervention passes through.

Selecting the control communities confronts two requirements that pull in opposite directions. Controls must be insulated from the intervention, which pushes them outward in the network, and comparable to the treated units on observables, which pulls them inward, since urban segregation makes demographically similar communities the ones most strongly connected to the treated region. No community satisfies both. We therefore select along the Pareto frontier of network distance and demographic similarity, retaining the communities that no alternative dominates on both criteria at once. We refer to this approach as the Network Distance Method (NDM; Methods and SI Note~\ref{supp:sec:NDM}).

A second problem remains. Cities experience many concurrent policies and shocks, so a before-and-after comparison cannot isolate a causal effect, however well exposure is defined. Recovering the effect requires a counterfactual for what would have happened without the intervention. We build that counterfactual with Synthetic Control Methods (SCM) \cite{abadie2010synthetic,abadie2021using}, which construct for each treated unit a weighted combination of unexposed control units fitted to its pre-treatment trajectory. We apply SCM under both exposure definitions. Both designs filter donors on demographic comparability, so they differ only in how separation from the intervention is measured: a geographic ring under GDM, network distance under NDM. Holding the estimator and the demographic filter fixed and varying only the separation criterion isolates the contribution of network-based assignment. Crossing each method's treated group with each method's donor pool shows that the difference between the two designs comes from the donor pool rather than from redefining the treated group (SI Note~\ref{supp:sec:decomposition}).

\subsection*{Boston's Free Fare program}
\begin{figure}[t]
    \centering
    \includegraphics[width=0.9\linewidth]{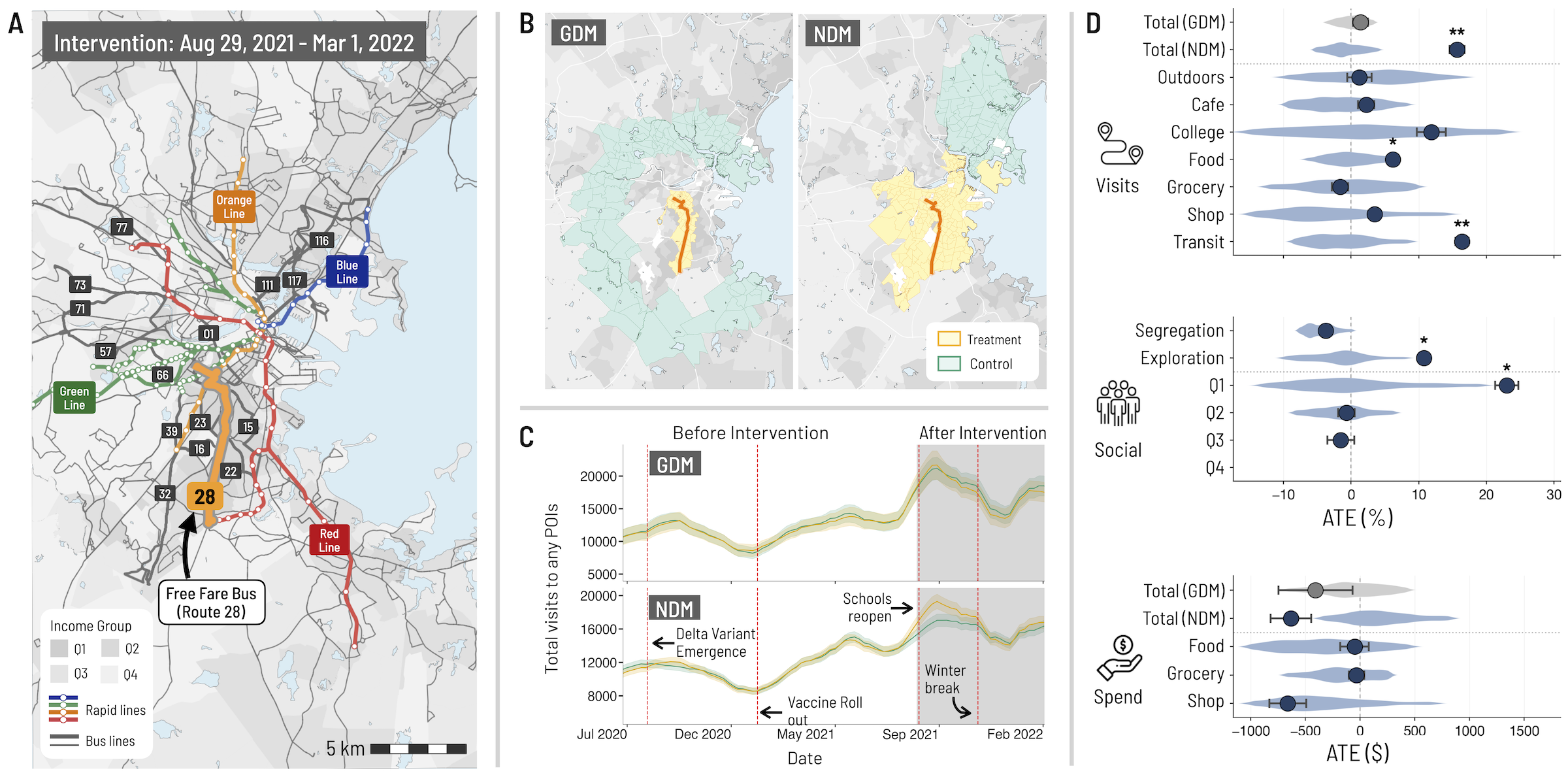}
    \caption{\textbf{Boston's fare-free bus program.} \textbf{A,} Map of the intervention. The bold line marks Route~28, on which fares were eliminated between 29 August 2021 and 1 March 2022, over MBTA rapid transit and bus routes labeled by route number, with census tracts shaded by income quartile. \textbf{B,} Treated and control  census tracts under the Geographical Distance Method (GDM) and the Network Distance Method (NDM). GDM treats tracts within 1\,km of Route~28 and draws controls from a 2--8\,km ring. NDM treats the mobility communities that intersect the route, which cover most of Boston and part of Brookline, and draws controls from communities that are network-distant but demographically comparable. \textbf{C,} Weekly visits to points of interest (POIs) for the treated group and its synthetic control under each design. Dashed lines mark the intervention window; annotations mark concurrent events. \textbf{D,} SCM--DiD average treatment effects (ATEs) for the intervention. Totals are shown under both designs; POI categories, income quartiles of the visitor's home tract, experienced segregation, exploration and spend are shown under NDM. Points give the estimate with whiskers showing the 95\% confidence interval, and violins give the placebo null distribution, from which significance is obtained by a two-sided empirical test ($^{*}p<0.05$, $^{**}p<0.01$).}
    \label{fig2}
\end{figure}

In August 2021, the City of Boston eliminated fares on Route 28, a corridor serving several socially vulnerable neighborhoods (Fig.~\ref{fig2}A). Simple before-and-after comparisons suggest that ridership on the fare-free routes grew up to three times as fast as the rest of the MBTA bus system \cite{boston_fare_free_mid_program_report}. The program also coincided with the lifting of COVID-19 restrictions and the reopening of schools, so much of that increase reflects the resumption of school commuting rather than the policy.

We first evaluate the impact using a geographic design (GDM), classifying census tracts within a 1 km buffer of Route 28 as treated and tracts within a 2–8 km buffer as potential controls, and constructing a synthetic counterfactual via SCM. We find no statistically significant increase in activity relative to the synthetic control (Fig. \ref{fig2}C,D). This null result matches the spillover-attenuation pattern in Fig. \ref{fig1}B. Control tracts within a few kilometers of Route 28 remain behaviorally close to the intervention, so the synthetic counterfactual is contaminated, and the recovered effect is biased toward zero.

The Network Distance Method resolves this ambiguity. Working at the level of mobility communities (Fig.~\ref{fig2}B), we classify communities intersecting the 1 km buffer as treated and select control communities that are distant in the mobility network but demographically comparable. The treated set covers most of the City of Boston and part of Brookline, a footprint the route geometry does not reveal. Applying SCM within this network-based design, we recover a significant increase in total visits among residents of treated communities ($15.65\%$, SE~$0.55$, placebo $p<0.01$). The gain extends beyond transit itself ($16.39\%$, SE~$0.51$, placebo $p<0.01$) to food establishments ($8.02\%$, SE~$2.09$, placebo $p<0.05$; Fig.~\ref{fig2}D), and residents also reach a wider set of destinations, with exploration rising by $10.77\%$ (SE~$0.33$, placebo $p<0.05$). The effect is sharply concentrated by income. Residents of the lowest-income quartile show significant gains (Q1: $23.00\%$, SE~$0.89$, placebo $p<0.01$), while higher-income residents show no detectable change. The result does not depend on how broadly we measure activity. Narrowing the outcome from all visits made by treated residents, to visits received by POIs in the treated tracts, to visits at venues along the bus network, to treated residents' visits along Route~28 itself leaves the NDM estimate positive at every step, while GDM recovers no coherent effect at any of them (SI Notes~\ref{supp:sec:ff_subsamples} and \ref{supp:sec:decomposition}).

The free-fare program reached the populations it was designed to serve, and did so by widening where they could go rather than only how often they traveled, but this is visible only when exposure is defined behaviorally rather than spatially.

\subsection*{New York City Open Streets}
In 2021, New York City's Open Streets program temporarily converted selected streets into pedestrian spaces \cite{nyc_dot_open_streets}. We focus on three sites where streets were closed at least four days per week between June and December 2021: Community League of the Heights, Washington Irving High School, and Downtown Brooklyn Partnership (Fig.~\ref{fig3}A-B). The units of analysis are individual businesses, and the outcome is visitation to them. The mobility network is correspondingly defined at the business level. Nodes are businesses and edges are the number of customers shared between them, so the network captures the co-visitation structure that links establishments through their common visitors \cite{yabe2025behaviour}.

Under a geographic design, we treat the businesses within 50\,m of the open street and draw controls from a 500--800\,m ring around it. GDM detects no significant change in total visitation at any of the three sites (Fig.~\ref{fig3}C), similar to analyses of other pedestrianization policies \cite{yoshimura_street_2022}. NDM recovers a different picture, with significant increases of $23.50\%$ (SE~$2.25$, placebo $p<0.01$) for Downtown Brooklyn Partnership, $27.27\%$ (SE~$5.25$, placebo $p<0.05$) for Washington Irving High School, and $32.60\%$ (SE~$4.86$, placebo $p<0.05$) for Community League of the Heights. Food establishments drive the pattern and are the only category with consistent coverage across all three sites. Visits to food POIs rise by $24.74\%$ (SE~$3.11$, placebo $p<0.01$), $31.92\%$ (SE~$5.11$, placebo $p<0.01$) and $41.94\%$ (SE~$7.75$, placebo $p<0.01$) at the three sites respectively. The total and food effects agree in direction and magnitude across all three locations, and the estimates hold across community-detection resolutions and algorithms (SI Note~\ref{supp:sec:sensitivity_community}).

Businesses are linked by shared customers, and those links do not track physical proximity: establishments hundreds of meters apart can draw on the same visitors while adjacent ones share almost none \cite{yabe2025behaviour}. When one location gains foot traffic, the effect propagates to whichever businesses its visitors also frequent. A donor ring therefore selects controls on a criterion unrelated to the contaminating channel, and the businesses it retains are indirectly treated, which biases the comparison toward zero. NDM selects controls from communities that share few customers with the treated businesses, recovering an effect that the geographic design attenuates below significance.

\subsection*{New York City Congestion Pricing}
Our third case study is structurally different from the first two. New York City's congestion pricing policy, which took effect in January 2025 \cite{mta_congestion_relief_zone}, imposes a fee on vehicles entering Manhattan south of 60th Street, an area designated the Congestion Relief Zone (CRZ); in its first year, 27 million fewer vehicles entered the zone, an 11\% reduction in traffic \cite{mta_anniversary_report}. The harder question is what happens to activity beyond the cordon, to visits, spending, and movement in the city center, and to the boroughs absorbing displaced trips. Public reports claim broad gains in Manhattan in foot traffic, office leasing, and sales tax receipts \cite{mta_anniversary_report}, but these rely on before-and-after comparisons or contrasts between Manhattan and other boroughs that ignore both spillovers and concurrent shocks. A generalized synthetic control comparing Manhattan with five other U.S. metropolitan areas recovers near-null effects on foot traffic and transactions \cite{cook_short-run_2025}. Those cities avoid the contamination, but they differ from Manhattan in density, transit access, the composition of their central business districts, and, above all, when people travel \cite{pandit_behavioural_2026}. The toll varies by time of day, at \$9 during peak hours and \$2.25 overnight, so it affects the morning and evening commute. A control city with a different peak structure need not track Manhattan through precisely those hours, and none registers the substitutions the policy triggers inside the zone.

Staying within New York removes that objection but introduces the opposite one. Under a geographic design, visits to Manhattan appear to decline by $6.46\%$ (SE~$0.39$, placebo $p<0.05$), visits originating from outside the congestion zone by $8.59\%$ (SE~$0.41$, placebo $p<0.01$), and total weekly spend per treated unit by $\$466{,}241$ (SE~$34{,}560$, placebo $p<0.05$; Table~\ref{supp:tab:gdm_ate_crz}). These estimates assume the control set absorbs none of the policy's effect. The median home-to-work distance for commutes terminating inside the zone is $13.1$\,km, and a tenth of these workers travel more than $40$\,km, so the 5 to 15\,km control ring sits well inside the catchment the policy acts on. Trips that previously went to Manhattan are partly redirected to the boroughs the design uses as controls \cite{pandit_behavioural_2026,ren_distributional_2026,li_public_2026}, so the controls gain activity the zone loses and the estimated decline combines both.

Defining controls through the mobility network addresses both problems. The treated communities extend north to 80th Street, so the population whose activity the toll reorganizes is not the population inside the cordon. At the mobility-community level, we find no significant change in overall visits to Manhattan or in visits originating from outside the congestion zone (Fig.~\ref{fig4}D). What changes is the composition of activity. Visits to grocery destinations decline by $7.87\%$ (SE~$0.84$, placebo $p<0.01$) and visits to shopping destinations by $8.54\%$ (SE~$0.76$, placebo $p<0.05$), while visits to transit hubs hold steady ($-0.25\%$, SE~$1.38$, not significant). This is consistent with independent evidence that commuters substituted public transportation for private vehicles \cite{zhang_congestion_2026} while curtailing discretionary trips, shopping, and grocery runs that are less compatible with transit-based access. Aggregate spending declines significantly under GDM but shows no significant effect under NDM. Failing to control for spillovers can therefore flip the apparent direction of a policy's economic impact, supporting opposite interpretations of whether congestion pricing harmed Manhattan's economy. The null overall effect holds across community-detection resolutions and algorithms and under random removal of treated tracts (SI Notes~\ref{supp:sec:sensitivity_community}, \ref{supp:sec:sensitivity_remove_tracts}).

The composition shift also has a distributional dimension. Visits from lower-income neighborhoods (Q1) decline by $4.31\%$ (SE~$0.46$, placebo $p<0.05$), while visits from higher-income neighborhoods (Q3, Q4) are unchanged. Experienced segregation in the visitor population rises by $4.20\%$ (SE~$1.11$, placebo $p<0.01$), so Manhattan now draws from a less diverse set of source neighborhoods than before. The policy did not reduce activity in the congestion zone but rather reorganized it, and the reorganization fell unevenly across the income distribution.

\subsection*{Geographic and mobility-network distance diverge across cities}
The three case studies establish that redefining exposure through the mobility network changes what we conclude about a policy. This is because mobility-network distances differ systematically from geographic ones, so units that are far in space may still be close in mobility-network distance. However, the observed mismatch between geography and mobility could be a particularity of our transit-dense examples or of our particular mobility dataset. To test this, we compare the two distance measures at the local scale across five data sources in eight countries and 138 metropolitan areas.

Geographic and network distance diverge across all ranges, and most sharply at short range. Tracts that are neighbors in space are often distant in the mobility network, and vice versa (Fig.~\ref{fig5}A-B). Short range is the regime in which buffer-based designs operate \cite{zheng_effects_2024,pollmann_causal_2023,volker_economic_2021}, so it is where geographic proximity is least informative about behavioral exposure. To quantify the divergence we compute, for each city, the average over tracts of the local rank correlation $\rho$ between geographic distance and network distance within a 5\,km neighborhood. A $\rho$ near one would mean that geographic proximity recovers the mobility-network ordering. This coefficient lies well below one in every city and data source examined (Fig.~\ref{fig5}C), so the mismatch is not a peculiarity of Boston, New York, or a single mobility dataset. Replacing Euclidean distance with road travel distance leaves the result unchanged, so the divergence is behavioral rather than an artifact of physical routing (SI Note~\ref{supp:sec:travel_distance}).

The severity of the divergence varies systematically across cities (Fig.~\ref{fig5}C). Mean local $\rho$ rises with the entropy of the visit distribution. Geographic and network distance align in cities where activity is spread across many destinations, and diverge where it concentrates on a few hubs. The same positive association appears in US, UK, Spanish, and Latin-American data, though its level varies across datasets (SI Note~\ref{supp:sec:rho_correlates}). Where flow is predominantly short-range, nearby places are strongly connected and geography tracks the network. Where it is long-range and directed toward distant centers, that local signal disappears and $\rho$ falls. US commuting flows (LODES) sit lowest for this reason, since commuting in US metros is long-range and concentrated on few job centers (mean $\sim$18~km), so little movement is local and proximity no longer predicts network closeness. Boston and New York are among the cities where geography best approximates the network, so our case studies were evaluated in relatively favorable conditions for geographic designs and the discrepancy we recover may be a lower bound on the error geographic evaluation incurs elsewhere.

\begin{figure}[t]
    \centering
    \includegraphics[width=\linewidth]{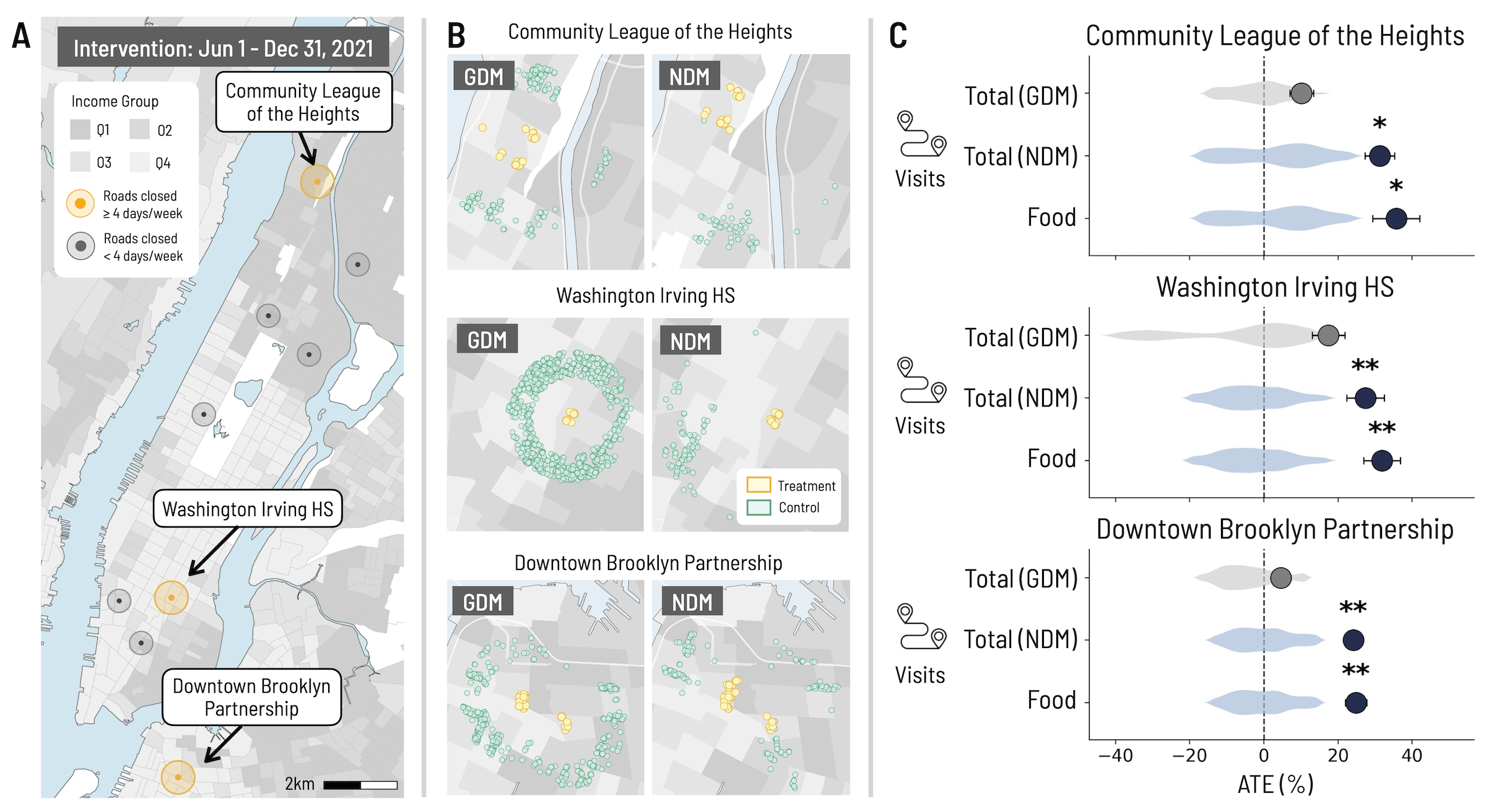}
    \caption{\textbf{New York City's Open Streets program.} \textbf{A}, Map of the intervention sites. Circles mark streets that participated in the program between 1 June and 31 December 2021; yellow circles are the three sites closed to traffic at least four days per week, which we analyze. Census tracts are shaded by income quartile. \textbf{B}, Treated and control businesses at each site under the Geographical Distance Method (GDM) and the Network Distance Method (NDM). GDM treats businesses within 50\,m of the open street and draws controls from a 500--800\,m ring. NDM draws controls from co-visitation communities that share few customers with the treated businesses, and its treated set can differ slightly. \textbf{C}, SCM--DiD average treatment effects (ATEs) on visits at each site. Totals are shown under both designs and food POIs under NDM. Points give the estimate with whiskers showing the 95\% confidence interval, and violins give the placebo null distribution, from which significance is obtained by a two-sided empirical test ($^{*}p<0.05$, $^{**}p<0.01$).}
    \label{fig3}
\end{figure}

\section*{Discussion}
Whether a policy appears to work depends on how exposure is defined, and that definition is a modeling choice rather than a property of the data. Across the interventions we study, geographic designs either fail to resolve an effect (Boston, Open Streets) or suggest a decline in activity (congestion pricing), whereas a mobility-based definition of exposure reveals substantial increases or no net change. The geographic estimates are also unstable: the recovered effect shifts in sign with the distance at which controls are drawn (Fig.~\ref{fig1}B), and its magnitude and significance shift with buffer width, so the congestion-pricing loss varies threefold across rings and the Open Streets effect at one site is significant at two widths and absent at a third (SI Note~\ref{supp:sec:sensitivity_buffer}). These discrepancies indicate that behavioral spillovers are the rule rather than the exception in complex urban systems, and we would expect them to be strongest where interventions sit on densely connected mobility corridors, near multimodal transfer hubs, or within central neighborhoods. This generality does not rest on our three cases alone. The decoupling between geographic and network distance that produces these spillovers holds across cities, countries, and data sources (Fig.~\ref{fig5}), and Boston and New York are among the cities where geography approximates the network best, so the discrepancies we recover are likely a lower bound on the error geographic evaluation incurs elsewhere. Policymakers relying on standard before-and-after comparisons or purely spatial difference-in-differences designs may therefore undervalue successful programs, misallocate resources, or fail to anticipate unintended consequences.

The community that Route 28 runs through spans 159 census tracts, four times the 40 the route physically passes. It covers most of the City of Boston and extends west into Brookline, so the population the free-fare reaches is not identifiable from a geographical map. The same holds for congestion pricing, where the treated communities extend twenty blocks north of the tolled boundary. Communities outside that set are connected to it by flows of people, and the strength of that connection, not the distance, determines how much of the policy's effect they absorb. Policy effects also spread across types of activity. An intervention that changes the friction of movement changes which activities are feasible, so it affects how people allocate time, where they socialize, and what they consume. Overall visitation to Manhattan is unchanged under a mobility-based framework, but the trips that disappeared were shopping and grocery runs while transit activity held steady, so the toll reallocated discretionary time rather than suppressing travel. Fare removal widened the set of distinct destinations residents reached, and congestion pricing narrowed the mix of neighborhoods from which Manhattan draws its visitors. The totals alone do not show this, and a geographic design inverts it.

The network-based approach also identifies areas that are behaviorally close to an intervention and which are distant from it. Networks built from mobility flows rather than administrative boundaries divide cities into functional regions that organize commuting and economic activity \cite{ratti_redrawing_2010,nelson_economic_2016}, and position in that structure predicts what a policy reaches better than distance from it does. Areas that are peripheral in the mobility network, weakly connected, or socially segregated may receive little benefit from a policy implemented nearby. A geographic design cannot detect this, since it treats proximity as the channel through which effects spread. These are areas where residents are cut off from a policy by mobility barriers in the network rather than by distance \cite{weng_beyond_2025}.

Our study has several limitations. First, the mobility networks aggregate over all trips, and we cannot distinguish visits of different natures. A visit-to-visit edge carries the same weight whether it reflects a routine commute or an exploratory trip. Decomposing the network by trip purpose or by traveler income \cite{moro2021mobility,yabe2023behavioral} would sharpen the picture of how a given policy spreads, since a fare change and a toll act on different kinds of trips, and is a natural extension of this work. We opted for the aggregate definition because it is portable: it requires only the same definition of origin-destination flows, which exist for hundreds of cities, so one network serves any intervention in a given city. Second, we treat the mobility network as fixed, estimate the partition on pre-intervention flows, and hold it constant. This holds here: the post-intervention partition agrees with the pre-intervention one at an Adjusted Rand Index of 0.96 (SI Note~\ref{supp:sec:sensitivity_flows_visits}), so these policies changed the volume of travel within existing communities rather than the communities themselves. Interventions that reshape the network, such as a new rapid transit line, would require dynamic modeling of the partition. Third, community membership is binary while exposure is continuous, so a tract at the edge of a community carries the same label as one at its center. We binarized for simplicity and for comparability with GDM, which also splits treated from control at a chosen distance. A dose-response formulation would be closer to the mechanism \cite{aronow_estimating_2017, pollmann_causal_2023}, and the diffusion kernel we already compute supplies the exposure measure at the unit level, so we leave this to future work.

How exposure is defined has implications for how urban policy is evaluated and, in turn, for which policies survive evaluation. When exposure is mis-specified, policies that expand access can appear ineffective, as in Boston and Open Streets, or harmful, as the geographic estimates of congestion pricing suggest. Boston's fare removal met its stated aim, raising activity among corridor residents by $16\%$ and concentrating the gain in the lowest-income quartile. Open Streets raised visits to the businesses on the closed streets by 23--33\%. Congestion pricing cut traffic without reducing activity inside the zone. A geographic evaluation records the first two as null and the third as an economic loss, and on that evidence a city would be justified in ending all three.

Mis-specified exposure is not particular to transportation. The framework should apply wherever a policy's effects travel with people rather than with distance: natural hazards, closures or openings of essential services, zoning and land-use reform, and changes in the retail or food environment. In cities, proximity in space is a poor proxy for proximity in experience. The question is not whether a policy works, but where, and for whom, it is felt. Policies do not only act where they are implemented. They act where people go.

\begin{figure}[t]
    \centering
    \includegraphics[width=\linewidth]{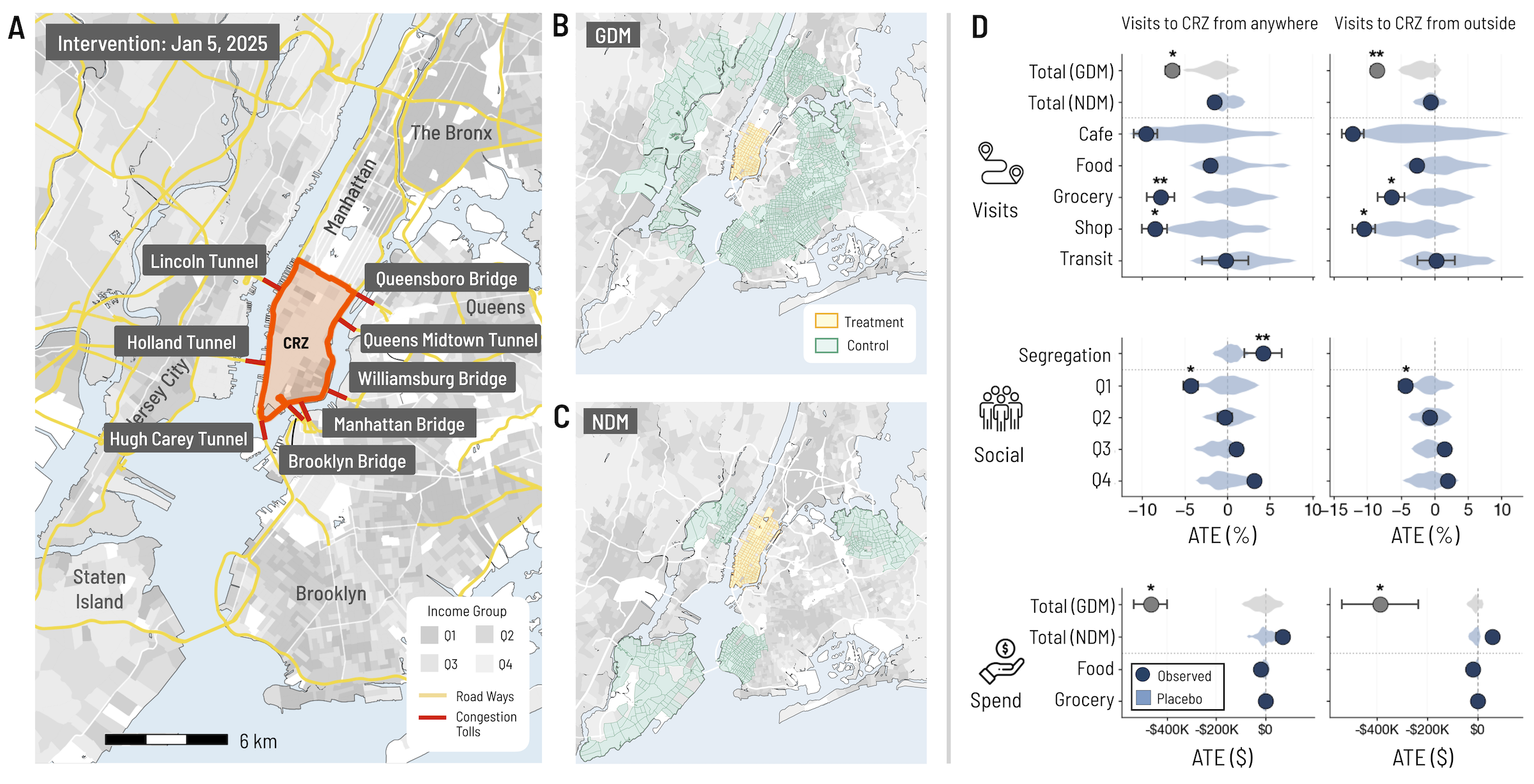}
    \caption{\textbf{New York Congestion Relief Zone}.
    \textbf{A}, Map of the intervention. The bold line marks the boundary of the tolled zone, in force from 5 January 2025, and its tolled entry points, over the road network, with census tracts shaded by income quartile.
    \textbf{B}, Treated and control census tracts under the Geographical Distance Method (GDM). GDM treats the tracts inside the zone boundary and draws controls from a 5--15\,km ring outside it.
    \textbf{C}, The same under the Network Distance Method (NDM). Tracts are grouped into mobility communities detected with the Leiden algorithm; treated communities are those intersecting the zone and extending north to 80th Street, and controls are communities that are network-distant but demographically comparable.
    \textbf{D}, SCM--DiD average treatment effects (ATEs) on visits to POIs inside the zone, from devices with home location anywhere in the New York CBSA (left) and outside the zone (right). Totals are shown under both designs; POI categories, income quartiles of the visitor's home tract, experienced segregation and spend are shown under NDM. Points show the estimate with whiskers indicating the 95\% confidence interval, and violins show the placebo null distribution; significance is obtained via a two-sided empirical test ($^{*}p<0.05$, $^{**}p<0.01$).}
    \label{fig4}
\end{figure}

\begin{figure}[t]
    \centering
    \includegraphics[width=\linewidth]{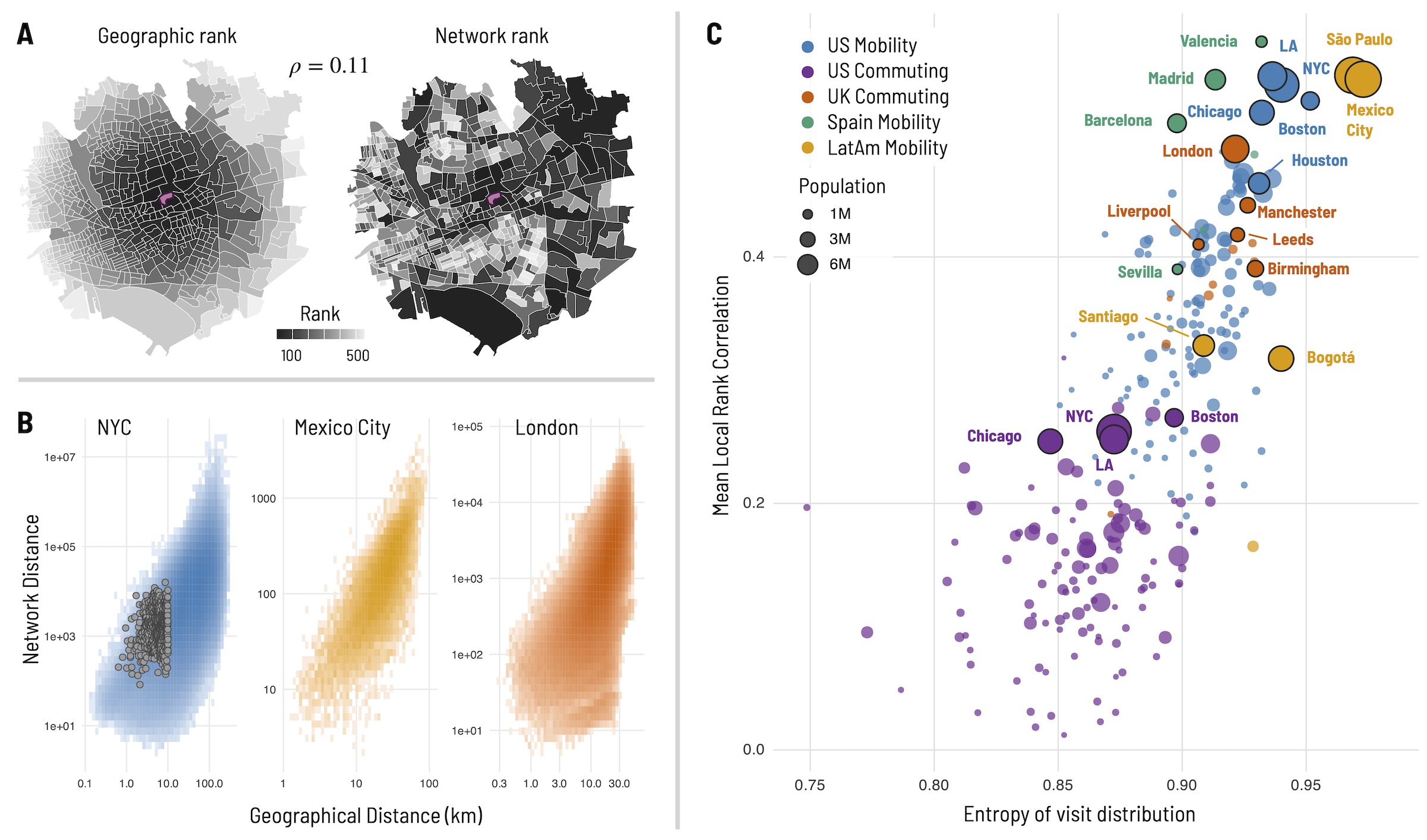}
    \caption{\textbf{Geographic and network distance diverge locally in every city and data source.}
    \textbf{A}, Census tracts surrounding a focal tract in New York City, shaded by rank in geographic distance (left) and network distance (right) from that tract. The nearest tracts by geography are not the nearest in the mobility network, and the rank correlation between the two orderings is $\rho=0.11 \pm 0.04$.
    \textbf{B}, Network distance against geographical distance for all tract pairs in New York, Mexico City and London. The two measures diverge across all ranges and most sharply at short range. Circled points in the New York panel are the pairs shown in \textbf{A}.
    \textbf{C}, City-level mean local rank correlation $\rho$ against the entropy of the visit distribution, for 138 metropolitan areas across eight countries and five data sources (US general mobility, Cuebiq; US commuting, LODES; UK commuting, ONS; Spain general mobility, MITMA; Latin-American general mobility, CEPAL). Point size scales with metro population and selected cities are labeled. Mean $\rho$ lies well below one in every city and rises with visit entropy. Cities measured with both general-mobility and commuting data, such as New York, Boston, Chicago and Los Angeles, appear twice; commuting flows give lower $\rho$ than general mobility in every case, because commuting in US metros is long-range and concentrated on few job centers.}
    \label{fig5}
\end{figure}

\section*{Methods}

\subsection*{Data}
We use anonymized, privacy-enhanced individual-level mobility data from Cuebiq (\url{https://cuebiq.com/}), covering 26.1 million opted-in users under GDPR- and CCPA-compliant frameworks. The analysis covers visits~\cite{moro2021mobility} across the Boston and New York Core-Based Statistical Areas (CBSAs)~\cite{census_cbsa_housing_patterns}, which together include approximately 36.44 million residents, over 2020--2023, totaling around 3 billion visits aggregated to the Census Tract level (SI Notes~\ref{supp:section:locationdata}, \ref{supp:subsection:representation}). We correct heterogeneous device-to-population sampling via the post-stratification scheme described in SI Note~\ref{supp:sec:post}. We validated this method in previous work~\cite{moro2021mobility,yabe2023behavioral} and validate it further here against MBTA route-level ridership (SI Note~\ref{supp:sec:mbta_validation}). We use weekly POI visitation counts from Placer.ai~\cite{placer_ai} in the Ne analysis. POI inventories combine Foursquare Open Source Places~\cite{foursquare} and SafeGraph~\cite{safegraph}, totaling approximately 607\,000 POIs across the two CBSAs, each mapped to one of 20 activity categories via a custom taxonomy (SI Note~\ref{supp:sec:poidata}). Demographic characteristics at the CBG level come from the ACS 5-year tables (2019--2021)~\cite{acs} (SI Note~\ref{supp:subsec:demographic}), weather data from~\cite{NRCC_CLIMOD2_2024} (SI Note~\ref{supp:subsec:weather}), and intervention metadata from MassDOT, the City of Boston, and the City of New York (SI Note~\ref{supp:section:interventiondata}). We draw monthly per-POI transaction records for the spend analysis from SafeGraph Spend Patterns through Dewey Data (SI Notes~\ref{supp:sec:spenddata} and~\ref{supp:sec:spend_methods}).

\subsection*{Interventions}
We study three transportation interventions implemented in Boston and New York City between 2021 and 2025: fare elimination, street reallocation, and congestion pricing. The Boston Free Fare Bus Program eliminated fares on MBTA bus route 28 from 2021-08-29 to 2022-03-01; the route serves Mattapan, Dorchester, Roxbury, and Nubian Square (Fig.~\ref{fig2}A). The New York City Open Streets Program temporarily converts selected streets into pedestrian and cyclist spaces from 2021-06-01 to 2021-12-31. We focus on streets closed at least four days per week, yielding three sites: Community League of the Heights (July--August), Washington Irving HS (September--December), and Downtown Brooklyn Partnership (June--November) (Fig.~\ref{fig3}A). New York City Congestion Pricing took effect on 2025-01-05 and charges tolls for vehicles entering Manhattan south of 60th Street (the Congestion Relief Zone, CRZ); the CRZ boundary follows the Lincoln, Holland, Queensboro, Williamsburg, Manhattan, and Brooklyn entry points (Fig.~\ref{fig4}A). See SI Note~\ref{supp:section:interventiondata} for more details about the treated and control windows for each intervention.

\subsection*{Outcomes}
Our primary outcome is the weekly count of post-stratified visits per unit, split by POI taxonomy and reported as a total. We construct three further outcomes from the same panels. Spend is a panel of estimated weekly dollar spend per tract. For each linked POI $i$ and calendar month $m$ we form a per-customer spend $s_{i,m}$ from the SafeGraph Spend Patterns ratio of monthly transaction dollars to unique cardholders, assign it to every week of $m$, and take $\widehat{D}_{i,w}=v_{i,w}\,s_{i,m(w)}$ as the weekly dollar volume at that POI, where $v_{i,w}$ counts weekly visits. Tract outcomes are the sum over POIs in the tract. Visits and spend are linked at the POI level, directly in Free Fare and via a one-time spatial and name match in the CRZ~\cite{rapidfuzz}. We report the dollar effect per treated unit-period as $\hat\beta_\$=\hat\beta_\%\,\overline{Y^{\text{cf}}_{\text{post}}}/100$ (SI Note~\ref{supp:sec:spend_methods}). The remaining two outcomes measure access rather than volume (SI Note~\ref{supp:sec:seg_exp_methods}). Exploration is the number of distinct POIs visited from a tract in a given week \cite{pappalardo_returners_2015}, post-stratified as the visit volumes are. Experienced income segregation~\cite{moro2021mobility} measures the imbalance of the income mix a tract's residents encounter through the venues they visit, running from zero when they meet a representative sample of the city's income groups to one when all co-visitors fall in a single quartile. Both are built at the tract-week level and enter the same SCM and gap regression as the visit panels. Segregation is reported for Free Fare and the CRZ, exploration for Free Fare only, since it requires individual-level data. For the CRZ we compute every outcome on all visits and separately on those originating inside and outside the zone.

\subsection*{Geographical Distance Method}
The Geographical Distance Method (GDM) defines treated and control units by physical distance from the intervention: tracts whose centroids fall within a small radius of the intervention are treated, and tracts in an outer ring around that radius form the control pool. We then filter that pool on the same propensity score used by NDM, retaining donors whose logit propensity lies within a fixed threshold of the treated-pool mean, so demographic comparability is enforced under both designs. Per-intervention buffer radii and propensity thresholds are listed in SI Note~\ref{supp:sec:GDM}. Sweeping alternative ring widths and a road travel-time variant leaves the GDM null in place where it is null, while the magnitude and significance of the non-null estimates depend on the ring (SI Notes~\ref{supp:sec:sensitivity_buffer} and~\ref{supp:sec:sensitivity_traveltime}).

\subsection*{Network distance}
We measure how close any unit (a tract, a business, or a community of such units) is to the treated region through the mobility network rather than physical space. Let $W$ be the row-normalized directed flow matrix on units, where $W[u,v]$ is the probability that a visit to unit $u$ is followed by a visit to unit $v$; the construction of the underlying flow network is described in SI Note~\ref{supp:sec:mobility_network}. To capture not only direct flows but also chains of flows, we form the truncated diffusion kernel \cite{katz_new_1953} adapted to mobility flows \cite{yabe2025behaviour} $W_{\text{diff}} \;=\; \sum_{k=1}^{K} \beta^{k}\, W^{k}$ with attenuation $\beta=0.8$ and depth $K=5$, so that $W_{\text{diff}}[u,v]$ measures the total mobility connectivity between $u$ and $v$ across direct and indirect paths. We define then network distance from a candidate donor $c$ to a treated set $T$ as the inverse of the mean connectivity over the $c \times T$ pairs  $d_{\text{net}}(c, T) = 1/\overline{W_{\text{diff}}[c, T]}$, so that units $c$ with strong flow to units in $T$ are close and mobility-isolated units are far. The ranking in network distance is insensitive to both kernel parameters. The Pareto front over donor communities contains the same members in the same order at every $\beta$ we test, and the ordering of control communities by $d_{\text{net}}$ is identical at every depth $K$ we tested, so donor selection is unchanged (SI Note~\ref{supp:sec:sensitivity_beta}).

\subsection*{Cross-metropolitan comparison}
The three case studies rest on two cities, so we test whether the divergence between geographic and network distance is general. We assemble five independent origin--destination datasets covering 138 metropolitan areas in eight countries (SI Note~\ref{supp:sec:metro_data}). For each metro we compute the network distance $d_{\text{net}}(u,v)$ for every pair of areas and compare it to the Euclidean distance, reporting the local Spearman correlation $\rho$ over the neighbors of each origin within 5\,km, averaged across origins, since the local scale is where buffer designs operate (SI Note~\ref{supp:sec:divergence_local}). Replacing Euclidean distance with road travel distance leaves the result unchanged, so the divergence is behavioral rather than an artifact of physical routing (SI Note~\ref{supp:sec:travel_distance}). We also regress each metro's mean local $\rho$ on the entropy of its visit distribution, $-\frac{1}{\log N}\sum_{u=1}^{N} p_u \log p_u$ with $p_u$ the share of the city's visits received by area $u$, together with population and ACS covariates (SI Note~\ref{supp:sec:divergence_local}).

\subsection*{Network Distance Method}
The Network Distance Method (NDM) selects control units that are weakly connected to the treated region in the visit-to-visit mobility network while remaining demographically comparable. We build a directed flow network on tracts, in which the edge weight $W_{uv}$ is the probability that a visit to tract $u$ is followed by a visit to tract $v$, and partition it via Leiden community detection \cite{traag_louvain_2019} at a fixed resolution $\gamma$ run for $N=100$ random seeds. The ensemble is collapsed into a consensus partition by hierarchical clustering on the co-assignment frequency, cut at threshold $0.5$ (SI Note~\ref{supp:subsec:consensus}). Communities in which at least $10\%$ of member tracts are treated are themselves treated; the remaining communities form the donor pool (SI Note~\ref{supp:sec:treated_donor}). Grouping donors into communities rather than binning them by network distance is what keeps boundary-crossing flow low and the donor pool internally coherent (SI Note~\ref{supp:sec:exposure}). For each candidate donor community $c$, we compute $d_{\text{net}}(c,T)$ and a propensity distance in standardized demographic feature space~\cite{rosenbaum1985constructing,austin2011introduction} (median household income, racial composition, public-transit commute share, and educational attainment; the exact per-intervention features are listed in SI Note~\ref{supp:sec:propensity}), and select the donor set from the Pareto front of these two scalars (SI Note~\ref{supp:secc:pareto_selection}). Selecting on either criterion alone fails in a different way. Propensity similarity alone fills the donor pool with indirectly-treated units and attenuates the estimate, while network distance alone yields demographically incomparable donors and biased SCM weights. Intervention-specific resolutions $\gamma$ and donor sets are listed in SI Note~\ref{supp:subsec:ndm_choices}.

The NDM estimates are stable across the construction choices of the design: the attenuation $\beta$ and kernel depth $K$, the detection algorithm and resolution, the definition of the mobility flows, administrative LODES commute flows in place of device-derived ones, and random removal of treated units (SI Notes~\ref{supp:sec:sensitivity_beta}--\ref{supp:sec:sensitivity_remove_tracts}). Because the true effect is unknown in all three interventions, we also validate both designs on synthetic data with a planted ground truth (SI Note~\ref{supp:sec:simulation}), and we cross each method's treated group with each method's donor pool to show that the difference between the designs comes from the donor pool (SI Note~\ref{supp:sec:decomposition}). SI Note~\ref{supp:sec:gdm_ndm_results} reports per-intervention estimates and donor-versus-treated demographic balance for every pool.

\subsection*{Synthetic control and Difference-in-Differences}
For each treated unit outcome $Y^\text{tr}_t$ we construct a synthetic counterfactual $Y^{\text{cf}}_{t}$ as a non-negative weighted combination of donor units, with weights fit by Lasso~\cite{tibshirani1996regression} on pre-treatment outcomes (SI Note~\ref{supp:subssection:scm}). We then estimate the treatment effect with a Difference-in-Differences (DiD) regression~\cite{roth_trending_2023} (SI Note~\ref{supp:section:did}) on the gap $G_{t}=Y^{\text{tr}}_{t}-Y^{\text{cf}}_{t}$:
\begin{equation*}
G_{t} \;=\; \beta_{0} \;+\; \beta\, \mathbf{1}\{t \ge t^{\ast}\} \;+\; \mathbf{x}_{t}^{\top}\boldsymbol{\gamma} \;+\; \varepsilon_{t},
\end{equation*}
where $t^{\ast}$ is the intervention date and $\mathbf{x}_{t}$ collects optional period-level controls (weekly weather variables for visits panels). For comparability across panels with different scales, we report the percentage gap $\hat\beta_{\%}=100\,\hat\beta/\overline{Y^{\text{cf}}_{\text{post}}}$, where $\overline{Y^{\text{cf}}_{\text{post}}}$ is the post-treatment mean of the synthetic counterfactual.

\subsection*{Placebo inference}
We do not read significance from the standard error of $\hat\beta$. With a single treated set and donor weights selected from the data, that standard error has no valid sampling interpretation~\cite{abadie2010synthetic,abadie2021using}. Panels built from millions of visits or transactions also make regression standard errors very small~\cite{zheng_effects_2024}, so they cannot separate a policy effect from a gap the design itself would produce by chance. We instead assess significance against an in-space placebo distribution built from the design~\cite{abadie2010synthetic,abadie2021using}. In each draw we relabel a random subset of the donor units as a pseudo-treated set, use the remaining units as the placebo donor pool, refit the SCM and the gap regression on this pseudo design, and record the pseudo-coefficient $\hat\beta_{(b)}$. Repeating for $B=100$ draws gives the two-sided empirical $p$-value, $\hat p = \bigl(1 + \#\{b : |\hat\beta_{(b)}| \ge |\hat\beta|\}\bigr)/(B+1)$, so an effect is flagged only when it exceeds what the same procedure produces on units that were never treated. The standard errors reported alongside each estimate describe the precision of the gap regression and are not used for inference (SI Note~\ref{supp:sec:placebo_test}).

\section*{Data and code availability}
Mobility data for the Boston Free Fare and New York Open Streets analyses were provided through Cuebiq's Social Impact program, and the Placer.ai data for the Congestion Relief Zone analysis under an academic research license. Both agreements preclude redistribution of the underlying individual-level data, so we cannot share it, but we provide the aggregated data required to reproduce the main results of the paper. Researchers may request access to the individual-level data directly from Cuebiq (\url{https://cuebiq.com/social-impact}) and from Placer Labs Inc.\ (\url{https://www.placer.ai/}), subject to their eligibility and data-use conditions. Point-of-interest data come from Foursquare Open Source Places (\url{https://docs.foursquare.com/data-products/docs/access-fsq-os-places}) and SafeGraph Global Places~\cite{safegraph}, obtained through Dewey Data (\url{https://app.deweydata.io}).

All remaining datasets are publicly available: American Community Survey tables and TIGER/Line shapefiles; LEHD LODES commute flows; U.K. Census 2021 workplace data from NOMIS; Spanish district-level trip matrices from the Ministerio de Transportes y Movilidad Sostenible; Latin-American origin--destination surveys from the ECLAC CISI repository; MBTA route geometries and ridership from MassGIS and the MassDOT Mobility Dashboard; the NYC DOT Open Streets inventory; the Congestion Relief Zone boundary and toll schedule from the MTA; and weather data from the Northeast Regional Climate Center. SI Notes~\ref{supp:section:locationdata}, \ref{supp:subsec:demographic}, \ref{supp:sec:poidata}, \ref{supp:sec:spenddata}, \ref{supp:section:interventiondata}, and~\ref{supp:sec:metro_data} give the provenance, extracts, and inclusion criteria for each source.

The aggregated data and code required to reproduce the main analyses are deposited at \url{https://github.com/SUNLab-NetSI/causal_inference_in_the_city}. The deposit contains the tract-to-tract and business-to-business mobility flow matrices used to build the networks and detect communities, the weekly visit counts by tract (Boston Free Fare and the CRZ) and by POI (Open Streets) over the study periods, and the origin--destination matrices of the cross-metropolitan comparison. It also contains the notebooks and scripts that build the designs and produce Figs.~\ref{fig2}--\ref{fig5}. None of the shared data contain individual-level information.

\section*{Acknowledgments}
We thank E.\ Then for her participation in the early stages of this project, D.\ T.\ O'Brien for discussions, and the City of Boston for feedback on the fare-free bus program analysis. We also thank Nicole Samay for assistance with figure design. E.M. acknowledges the support of NSF grants 2420945 and 2425023. T.Y. acknowledges the support of NSF grant 2425021.

\section*{Author contributions}
B.J. and E.M. conceived and designed the study, with input from T.Y. and H.G. B.J. curated and processed the mobility data, developed the software, and performed the analyses, with contributions from E.M. B.J. and E.M. wrote the original draft. E.M. supervised the project and acquired funding. All authors interpreted the results, reviewed and edited the manuscript, and approved the final version.

\section*{Competing interests}

The authors declare no competing interests.

\newpage

\clearpage
\setcounter{section}{0}
\setcounter{figure}{0}
\setcounter{table}{0}
\setcounter{equation}{0}
\renewcommand{\thefigure}{S\arabic{figure}}
\renewcommand{\thetable}{S\arabic{table}}
\renewcommand{\thesection}{\arabic{section}}
\renewcommand{\thesubsection}{\arabic{section}.\arabic{subsection}}
\begin{center}
{\LARGE\sffamily\bfseries Supplementary Information}
\end{center}
\vspace{1em}
\section{Data}
\subsection{Mobility data} \label{supp:section:locationdata}
Mobility data used in the project comes from two different providers. The Boston Free Fare and New York Open Streets analyses use privacy-enhanced individual-level stays at POIs from Cuebiq \cite{moro2021mobility}. The New York Congestion Relief Zone (CRZ) analysis uses total weekly POI visits from Placer.ai \cite{placer_ai}, disaggregated by the origin CBG of the visitors. We use Placer.ai for the CRZ analysis because Cuebiq panel coverage in 2025 declined substantially relative to earlier years, reducing representativeness and increasing sampling noise for weekly zone-level estimates. 

Cuebiq provides anonymized GPS stays under its Social Impact Program, which grants de-identified mobility data to academic research and humanitarian projects under a GDPR- and CCPA-compliant framework. All users opted in to share their locations anonymously, and researchers are contractually prohibited from redistributing the data or attempting re-identification. Placer.ai data is already anonymized at the POI-CBG level. A stay is an event $\mathbf{x}_{i,t}$ in which user $i$ remained stationary at location $\mathbf{x}$ at time $t$; we detect stays and infer home Census Block Group (CBG) labels following the methodology in~\cite{moro2021mobility}. We retain stays for users whose home CBG lies in the relevant Core Based Statistical Area (CBSA). Following the same stay-processing pipeline, we restrict to stays with dwell times between 5 and 720 minutes, exclude recurring stays at each user's inferred home and work locations, and drop stays at sensitive POIs: places of religious worship, locations primarily occupied by children, credit-repair and debt-services venues, and the other categories listed in the Cuebiq sensitive-POI policy~\cite{cuebiq_spoi_policy}. These filters are applied in preprocessing, before the post-stratification and aggregation steps below.
\begin{figure}[h!]
    \centering
    \includegraphics[width=\linewidth]{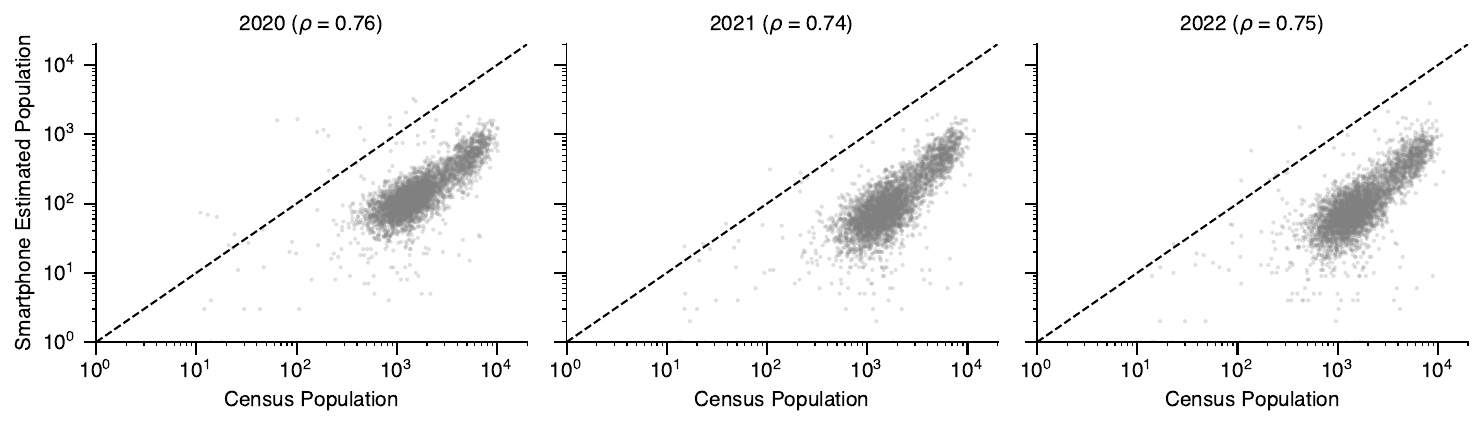}
    \caption[\bf Smartphone population vs Census population.]{\textbf{Smartphone population vs Census population.} Correlation between the smartphone population detected in our data and Census population. The Spearman correlation between them is shown at the top.}
    \label{supp:fig:representation}
\end{figure}
The two Cuebiq-based analyses span two metropolitan areas over 2020--2023: the Boston CBSA (Free Fare) and the New York CBSA (Open Streets). Although Cuebiq provides nationwide coverage, restricting to stays whose inferred home CBG lies in these two CBSAs yields about 3 billion tract-level visits across the study period. The Boston visit panel spans 2020-01 to 2023-03 and contains roughly 449 million POI-attributed weekly visits to 150{,}781 distinct POIs, while the New York panel contains roughly 2.8 billion visits to 456{,}108 distinct POIs. 

Placer.ai provides weekly visit counts for each POI, disaggregated by the visitor's origin home CBG. Counts are derived from an opted-in mobile-device panel and post-stratified to the U.S. population; no device-level trajectories are shared~\cite{placer_ai}. We accessed the data under an academic research license, and counts are subject to vendor-side small-cell suppression for privacy reasons.

For the CRZ analysis we pull Placer records for two disjoint regions: (i)~the CRZ treatment area, defined as the census tracts whose centroids lie inside the Congestion Relief Zone boundary (Manhattan south of 60th Street), and (ii)~a 20\,km donor buffer formed by the 20\,km ring around that boundary, from which the synthetic-control donor pool is drawn. These records span May 2023 to July 2025 and comprise 20{,}080 POIs across 1{,}805 census tracts, carrying approximately 7.5 billion post-stratified visits; because Placer reports vendor-aggregated counts already scaled to the U.S.\ population rather than raw device stays, these totals are not directly comparable to the observed Cuebiq stay counts.

The use of the data in this research was
reviewed by the Northeastern University Institutional Review Board (IRB \#24-03-43).

\begin{figure}[t!]
    \centering
    \includegraphics[width=\linewidth]{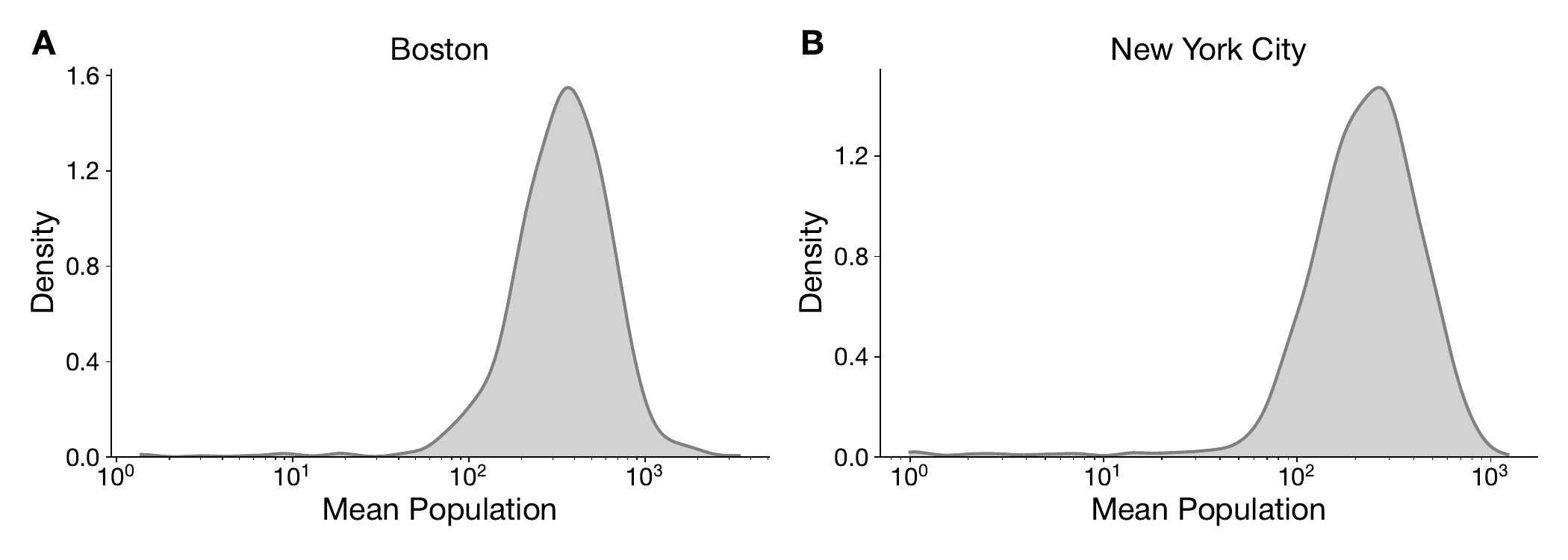}
    \caption[\bf Distribution of mean detected population per census tract.]{\textbf{Distribution of mean detected population per census tract.} Kernel-density estimates of the time-averaged detected population per tract, computed from the rolling monthly population panel and aggregated from block-group to tract level. (A)~Boston CBSA (916 tracts). (B)~New York City CBSA (4{,}912 tracts). The x-axis is on a log scale.}
    \label{supp:fig:population_distribution}
\end{figure}
\subsection{Representativeness of the location data} \label{supp:subsection:representation}
Both mobility datasets are derived from opted-in panels of mobile-device users rather than complete population counts, so we check how well the panels represent the underlying population. Smartphone penetration is high in urban populations but varies across neighborhoods. For Cuebiq, we compare the spatial distribution of devices in our dataset against population estimates from the American Community Survey (ACS) 5-year estimates spanning 2019-2021 at the census-tract level across the Boston and New York MSAs. The two are strongly correlated ($\rho = 0.75 \pm 0.01$ on average across the three years; Fig.~\ref{supp:fig:representation}). Even so, we apply a demographic post-stratification reweighting during 2020 and 2021 to account for panel composition shifts and changes in mobility behavior during the COVID-19 period (Section~\ref{supp:sec:post}).

Placer.ai does not release per-device records, so we cannot audit the CBG-level representativeness of the Placer panel as we do for Cuebiq. Placer states that it debiases its panel by age, geography, and mobile operating system distribution before aggregation~\cite{placer_our_data, placer_debiasing_v21}, but the calibration procedure is proprietary. We therefore rely on the vendor's post-stratification.

\subsection{Post-stratification of mobility data} \label{supp:sec:post}
Heterogeneous sampling rates across census tracts and income groups bias our estimates of visitation patterns to places, so we apply post-stratification, following previous work \cite{moro2021mobility,yabe2023behavioral}. Post-stratification is a standard sampling tool \cite{salganik2019bit} routinely used to correct sampling biases in mobile phone location data \cite{jiang2016timegeo} and (geolocated) social media data \cite{wang2018urban}.

Following Moro et al.\ \cite{moro2021mobility}, we define the expansion factor $w_g$ as the ratio between the
true population of census tract $g$ and the number of individuals from $g$ observed in our mobility
dataset (the distribution of which is shown in Fig~\ref{supp:fig:population_distribution}). We then reweight the visits from census tract $g$ to place $\alpha$ as
\[
\hat{n}_{g\alpha} = w_g n_{g\alpha},
\]
so that visit counts are adjusted inversely to the representativeness of mobile phone users in each census tract. With
this procedure, we effectively increase (decrease) the estimated visits from under-sampled (over-sampled)
census tracts. 

\subsubsection{Validation of visits data with MBTA bus ridership data}
\label{supp:sec:mbta_validation}
\begin{figure}[t!]
    \centering
    \includegraphics[width=\linewidth]{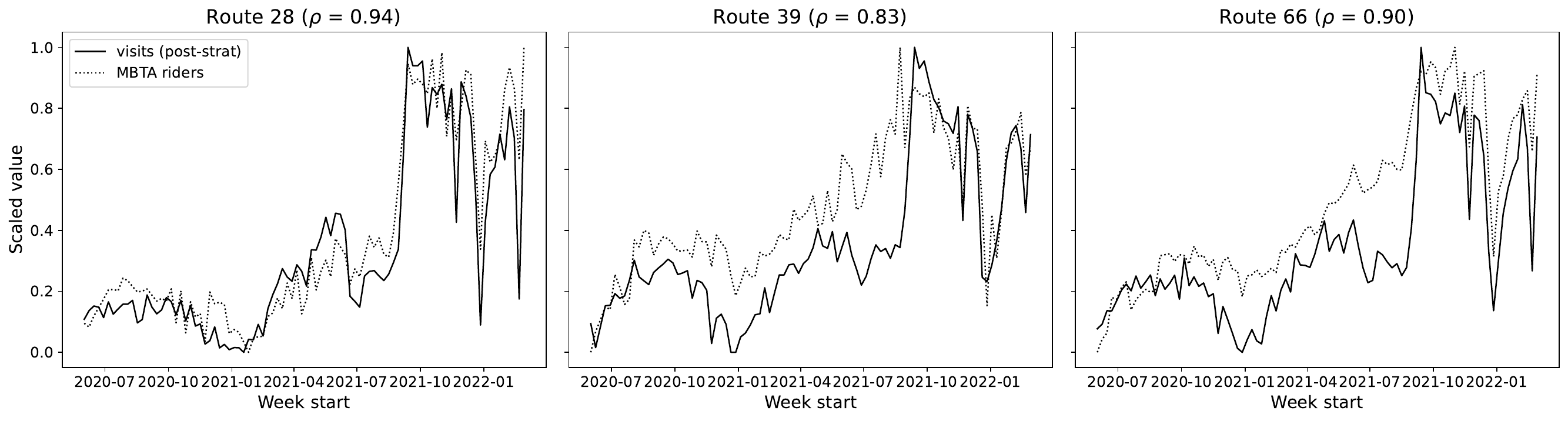}
    \caption[\textbf{Comparison of bus ridership between MBTA data and mobility data.}]{\textbf{Comparison of bus ridership between MBTA data and mobility data.} The bold lines are weekly visits to transportation POIs computed from the mobility data; the dotted lines are the corresponding MassDOT ridership series. For Routes 28, 39, and 66 the Pearson correlations are $0.94$, $0.83$, and $0.90$, respectively.}
    \label{supp:fig:correlation_data}
\end{figure}

We validate our visit counts against MBTA bus ridership data from MassDOT’s Mobility Dashboard~\cite{massdot_mobility_dashboard}. In addition to Route 28, we randomly select two other high-ridership bus routes, 39 and 66. For each of the three routes, we aggregate visits from our mobility data to transportation POIs along the route and compute their temporal correlation with the corresponding MassDOT ridership series. The two series are strongly correlated for all three routes (Fig.~\ref{supp:fig:correlation_data}). Small discrepancies are expected, as visits to transportation POIs do not necessarily translate directly into bus boardings and may also capture other transit-related activities occurring near the route.

\subsection{Demographic data} \label{supp:subsec:demographic}
Demographic data at the census-tract level were obtained from the American Community Survey (ACS) 5-year estimates published between 2019 and 2021~\cite{acs}. Each demographic characteristic corresponds to a specific ACS table. In our study, we use the following demographic \emph{estimates} at the census-tract level:

\begin{enumerate}
\item Proportion of the population by ethnicity (Table B03002);
\item Median value of owner-occupied housing units (Table B25077);
\item Proportion of workers commuting by public transportation (Table B08301);
\item Proportion of the employed population (Table B23025);
\item Proportion of people with a college degree or higher (Table B15002).
\end{enumerate}
We assign individuals the demographic characteristics of the census tract in which their home location is situated. For the Open Streets analysis, the demographic profile for each POI is defined as the mean demographic profile of the visitors to that POI.
\begin{figure}[t!]
    \centering
    \includegraphics[width=\linewidth]{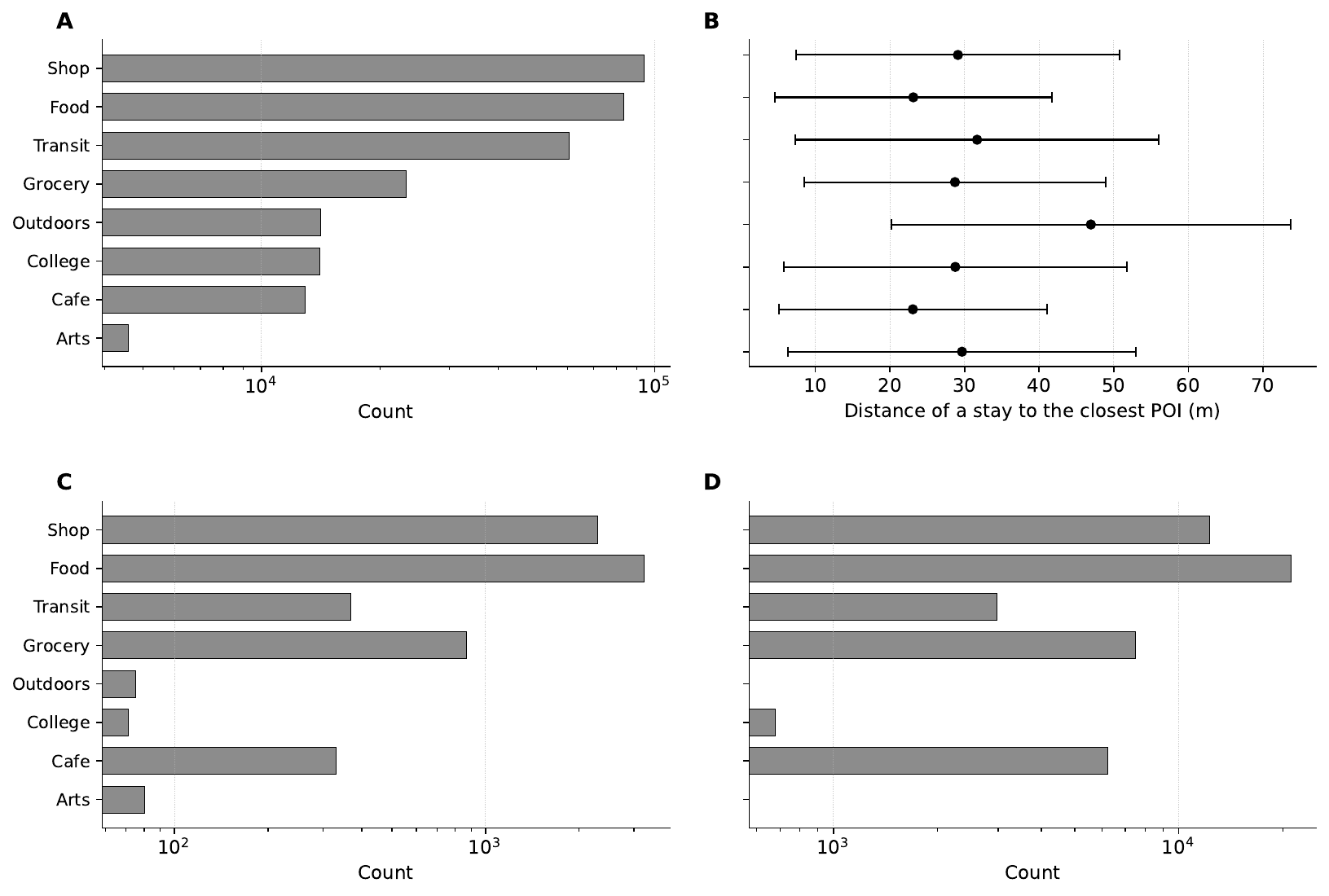}
    \caption[\textbf{POI inventory and stay-to-venue attribution across datasets.}]{\textbf{POI inventory and stay-to-venue attribution across datasets.}
    (A)~Distribution of points of interest across taxonomy categories in the Boston and New York CBSAs.
    (B)~Forest plot of stay-to-closest-POI distances per taxonomy; markers show the mean and horizontal bars the standard deviation of attribution distances.
    (C)~Placer POI inventory for the CRZ analysis, pooled across the CRZ treatment area and the 20\,km donor buffer.
    (D)~SafeGraph Spend Patterns POI inventory, summed across the Boston Free Fare window (2020-05 to 2022-03) and the CRZ + 20\,km donor buffer (2023-05 to 2025-07). Spend taxonomies do not separate Arts or Outdoors, so those categories are omitted from panel~D.}
    \label{supp:fig:poi_combined}
\end{figure}

\subsection{Weather data} \label{supp:subsec:weather}
Daily weather observations come from the Northeast Regional Climate Center's CLIMOD~2 station data~\cite{NRCC_CLIMOD2_2024}. We use the Boston Logan International Airport station for Boston Free Fare and the New York City Central Park station for Open Streets and the CRZ. We use daily average temperature ($^{\circ}$F), precipitation (inches), and snowfall (inches), with trace amounts set to zero. We aggregate the daily values to weekly resolution by averaging over seven-day windows, so precipitation and snowfall enter as mean daily amounts, and join the weekly series to the visit panels by week. The series cover the full study period of each intervention. The weekly weather variables enter the gap regression (Eq.~\ref{eq:gap_did}) as the period-level controls $\mathbf{x}_{t}$ for the visit panels.

\subsection{POI data} \label{supp:sec:poidata}
Our Point of Interest (POI) data contains information about the precise locations of venues and their corresponding activity category. We combine two sources to cover urban venues as completely as possible: Foursquare Open Source Places~\cite{foursquare} and SafeGraph Places Data, accessed through the Dewey Data platform (\url{https://www.deweydata.io/}) under SafeGraph's standard usage terms and conditions (\url{https://www.safegraph.com/}). 
The integrated dataset comprises approximately 2.5 million POIs across 11 CBSAs. We construct a custom taxonomy that groups these venues into 20 classes, of which we retain the eight that capture the activities relevant to our analysis (Fig.~\ref{supp:fig:poi_combined}A).
As for Placer.ai data, we use the POIs provided by the data provider. These POIs come with taxonomy, category, and subcategory tags assigned by the provider, and we manually map them to our custom taxonomies. The distribution of POIs in Boston and New York CBSA can be seen in Fig.~\ref{supp:fig:poi_combined}.

\subsubsection{Spatial attribution of stays to venues}\label{supp:poi_attribution}
A critical step in using mobility data is identifying the activity behind each stay, i.e., attributing each Cuebiq stay to a POI. Following established methods for location inference from sparse mobile phone data in research~\cite{cuttone2014inferring,moro2021mobility,saxena_poiformer_2025} and industry \cite{safegraph2022visit}, we assign each stay to its nearest POI, with both stays and POIs represented as discrete geographic points. We impose a maximum attribution distance of $d_{\text{max}} = 100$\,meters to avoid associating a stay with a distant venue; stays farther than this threshold from any venue are left unattributed. The median stay-to-venue distance is $20.6$\,meters, and the distribution is consistent across venue categories (Fig.~\ref{supp:fig:poi_combined}B). The validity of this attribution procedure is supported both by prior work validating similar approaches against ground-truth data~\cite{moro2021mobility, yabe2023behavioral,cuttone2014inferring} and by the strong correspondence between our mobility-derived transit activity measures and independent MBTA ridership records (Fig.~\ref{supp:fig:correlation_data}).

\subsection{Spend data} \label{supp:sec:spenddata}
To complement the visitation analyses with a monetary measure of activity, we use the SafeGraph Spend Patterns product, accessed through the Dewey Data platform (\url{https://www.deweydata.io/}) under SafeGraph's standard usage terms and conditions (\url{https://www.safegraph.com/}). Spend Patterns reports POI-level, monthly aggregates of card-transaction activity, where each record corresponds to one POI in a given calendar month. The POIs are tagged by NAICS code and we have three outcome fields that we use in this study: total dollars transacted at the venue during the month, number of card transactions, and number of unique cardholders. Data are aggregated and anonymized by the vendor prior to delivery, with no device- or cardholder-level records shared.

Spending records are manually mapped to our custom taxonomies, via a NAICS-to-taxonomy crosswalk. After spatial and temporal filtering, the Boston Free Fare spend panel contains 20{,}301 unique POIs (approximately \$987\,M in total spend summed across POI-months), and the CRZ spend panel contains 61{,}152 unique POIs within the Congestion Relief Zone and 20\,km donor buffer (approximately \$2.96\,B in total spend summed across POI-months). Figure~\ref{supp:fig:poi_combined}D shows the distribution of these POIs across the six taxonomy categories on which we focus the spend analyses (Food, Shop, Transit, Grocery, College, and Cafe). In both cities, Food dominates the POI inventory, with Shop and Grocery forming a secondary tier and Transit, Cafe, and College trailing on a logarithmic scale.

\subsection{Estimated spend outcome} \label{supp:sec:spend_methods}
To obtain a monetary outcome for the SCM--DiD analysis (Section~\ref{supp:subssection:scm}), we construct a panel of estimated weekly dollar spend per tract. The construction multiplies weekly visits at each POI by a monthly spend-per-customer figure derived from SafeGraph Spend Patterns (Section~\ref{supp:sec:spenddata}) and aggregates to the tract level. The visits and spend datasets must first be joined at the POI level; the linkage is direct in Free Fare and requires a fuzzy match in the CRZ.

\subsubsection{Linking visit-side and spend-side POIs.}
For Boston Free Fare, the visits panel is built by attributing Cuebiq stays to POIs from our combined inventory (Section~\ref{supp:poi_attribution}), which includes SafeGraph Places. These POIs carry the same unique identifier as SafeGraph Spend Patterns, so weekly visits at a POI merge with monthly spend at the same POI on the POI ID without further matching.

For the CRZ, the visits panel is built from Placer.ai data rather than SafeGraph, so the visit-side and spend-side POIs come from different inventories with different identifiers and must be linked by a one-time fuzzy-matching step. We index the SafeGraph POIs in a BallTree under the haversine (great-circle) metric and, for each Placer POI, retrieve every SafeGraph POI within $100$\,m. Among these spatial candidates we score each SafeGraph name against the Placer name with the rapidfuzz token-sort ratio~\cite{rapidfuzz}: the metric alphabetically sorts the word tokens of each name before computing a $0$--$100$ normalized similarity, so it is insensitive to word order. We keep the single highest-scoring candidate and retain the pair only if its score is at least $80$; otherwise the Placer POI is left unmatched and dropped from the spend panel. Finally, we enforce a one-to-one linkage by assigning each SafeGraph POI to at most one Placer POI, keeping the highest-scoring pair. This links $4{,}832$ of the $20{,}080$ Placer visit POIs ($24.1\%$; equivalently $7.9\%$ of the $61{,}152$ SafeGraph spend POIs); the remaining Placer POIs are left unmatched and excluded from the spend panel, so the spend analysis runs on the linked subset rather than the full visit inventory. 

\subsubsection{Visits-times-spend panel.}
For each linked POI $i$ and calendar month $m$, the SafeGraph Spend Patterns record reports the total transaction dollars $T_{i,m}$ and the number of unique cardholders $C_{i,m}$ observed at that POI in that month. We compute a per-customer spend
\[
s_{i,m} \;=\; T_{i,m}\,/\,C_{i,m},
\]
and assume this rate is stable within the month, so that every week $w$ in month $m$ inherits $s_{i,m(w)}$. The estimated weekly dollar volume at POI $i$ is then
\[
\widehat{D}_{i,w} \;=\; v_{i,w}\,\cdot\, s_{i,m(w)},
\]
where $v_{i,w}$ counts all weekly visits to POI $i$, irrespective of visitor origin. Tract-level outcomes are obtained by summing $\widehat{D}_{i,w}$ over POIs located in the tract, separately for each custom taxonomy and as a per-tract total. The resulting panel has the same shape as the visits panel, with rows indexed by (tract, week, taxonomy) and the outcome in dollars per week.

\subsection{Experienced segregation and exploration outcomes} \label{supp:sec:seg_exp_methods}
We construct two further outcomes from mobility data that summarize whom a tract's residents encounter and which venues they reach, rather than how many visits they make or how much they spend. The first, experienced income segregation, measures the income-group imbalance of the visits a tract participates in; the second, exploration, counts the diversity of distinct venues visited from the tract. Both are constructed at the (tract, week) level and serve as the dependent variable in the SCM--DiD analysis (Section~\ref{supp:subssection:scm}), leaving the donor pool, treatment date, and placebo inference unchanged. Segregation is reported for Free Fare and the CRZ; exploration is reported only for Free Fare, where tract-level per-POI weekly visits are available.

\subsubsection{Experienced income segregation.}
We follow Moro et al.~\cite{moro2021mobility} and define, per tract $i$ and week $w$, the index
\begin{equation}
S_{i,w} \;=\; \frac{2}{3}\sum_{q=1}^{4} \left| \tau^{q}_{i,w} - \tfrac{1}{4} \right|,
\label{eq:supp:seg_index}
\end{equation}
which lies in $[0,1]$.

For Boston Free Fare, $\tau^{q}_{i,w}$ is built up at the POI level. Each visiting tract is assigned to one of four equal-population quartiles by ACS median household income. For each POI $\alpha$ in week $w$ we form the visitor-quartile share $\tau^{q}_{\alpha,w}$ from the per-POI weekly visits obtained by attributing Cuebiq stays to POIs (Section~\ref{supp:poi_attribution}), and for each treated/donor tract $i$ we form $\tau^{\alpha}_{i,w}$, the share of tract $i$'s weekly visits going to POI $\alpha$. The tract-level mixture is then
\[
\tau^{q}_{i,w} \;=\; \sum_{\alpha} \tau^{\alpha}_{i,w}\, \tau^{q}_{\alpha,w}.
\]

$S_{i,w}$ captures the segregation of the income mix of co-visitors that residents of tract $i$ encounter through the venues they visit. When residents of tract $i$ encounter a representative sample of the city's income groups, $\tau_{i,w}^q = 1/4$ for all $q$ and $S_{i,w}=0$. Otherwise, $S_{i,w}$ increases with the concentration of encountered visits in a smaller number of income quartiles, reaching $S_{i,w}=1$ when all co-visitors encountered by residents of tract $i$ belong to a single income quartile.

For the CRZ, $\tau^{q}_{i,w}$ is taken directly from Placer.ai visitor-origin data: each visit is tagged with the income quartile of the visitor's home tract (NYC tract-level median income, quartiles taken across all NYC tracts), and $\tau^{q}_{i,w}$ is the share of week-$w$ visits to POIs located in tract $i$ whose visitor home falls in quartile $q$. We compute Eq.~\eqref{eq:supp:seg_index} three times: on all visits, on visits whose origin lies inside the CRZ tract pool (inside), and on those originating outside it (outside). This separates locally generated changes in visitor mix from changes driven by non-residents.

\subsubsection{Exploration.}
For Boston Free Fare, exploration is the count of distinct POIs visited from tract $i$ in week $w$,
\[
E_{i,w} \;=\; \big| \{ \alpha : v_{\alpha, i, w} > 0 \} \big|,
\]
where $v_{\alpha, i, w}$ is the count of weekly visits from tract $i$ to POI $\alpha$ obtained by attributing Cuebiq stays to POIs (Section~\ref{supp:poi_attribution}). Unlike $S_{i,w}$, exploration scales with the size of the observed visitor sample, so the count is post-stratified by the same census-population/observed-population ratio used for visit volumes (Section~\ref{supp:sec:post}) before being substituted into the SCM--DiD regression.

\subsection{Intervention data} \label{supp:section:interventiondata}
We obtained data for Massachusetts transit routes from the Mass.gov website~\cite{massgis_mbta_bus_routes_stops, massgis_mbta_rapid_transit}: these are geographic shapefiles giving the line geometry of each MBTA bus route and rapid-transit line, along with the locations of their stops and per-route attributes (route number, directional variant, trip headsign, and route description). Data on ticket validations for the Boston transit system were downloaded from MassDOT~\cite{massdot_mobility_dashboard}; these are route-level ridership (fare-validation) time series, which we use both to characterize the fare-free routes and to validate our mobility-derived visit counts (Fig.~\ref{supp:fig:correlation_data}). Information on the Fare Free Program was obtained from the City of Boston's mid-program report~\cite{boston_fare_free_mid_program_report}, which documents the pilot routes and dates, the reported change in ridership, and a rider survey of monthly savings and travel-behavior responses. For the New York Open Streets program we used the NYC DOT open-streets inventory~\cite{nyc_dot_open_streets}, which records, for each participating street, the managing partner organization, the street segment and its line geometry (borough and the on/from/to cross streets), the days of the week and opening and closing hours of operation, and the approved start and end dates. For the CRZ, we used MTA data~\cite{mta2025congestion_relief_zone_data} defining the Congestion Relief Zone boundary and its tolled entry points, together with the toll schedule by time of day and vehicle class.

\subsubsection{Boston Free Fare program}
The pilot for the free-fare bus program ran from August 29, 2021 to March 1, 2022. The City of Boston selected Route 28 for the pilot because it serves some of the city’s most socially vulnerable neighborhoods, with many zero-car households. The program allowed riders to board through all doors without paying a fare. After the pilot, the City expanded the program to include Routes 23 and 29 and extended it through 2026. According to a mid-term report produced by the City in March 2023~\cite{CityOfBoston_FareFree_2023}, most riders on the fare-free routes are transit-critical; 26\% of riders saved \$20 or more per month and reported using these savings to purchase food, save toward long-term goals, and build emergency funds. The report also notes that some riders now choose to take the bus rather than drive, which could reduce congestion and greenhouse gas emissions. Travel times on the routes remained essentially constant despite the influx of riders, and the City reports that ridership on the fare-free routes grew up to three times as fast as the rest of the MBTA bus system.

\subsubsection{New York Open Streets program}
New York City's Open Streets program closes selected street segments to through traffic and converts them into shared public space for pedestrians, cyclists, and local programming. Each open street is managed by a partner organization that applies to NYC DOT and acts as its steward: partners range from block associations, business improvement districts, and non-profits to schools, universities, merchant groups, and houses of worship, and the uses they host span outdoor dining, school programming, and cultural events. For our analysis we selected streets that operated on a recurring weekly schedule and closed to through traffic at least four days per week between June and December 2021, which narrowed the set to three sites: Community League of the Heights, NYCDOE/Washington Irving HS, and Downtown Brooklyn Partnership. 

\subsubsection{CRZ New York}
Beginning January 5, 2025, the Metropolitan Transportation Authority (MTA) implemented congestion pricing in Manhattan's Central Business District, formally designated the Congestion Relief Zone (CRZ). The policy imposes a toll on vehicles entering Manhattan south of 60th Street, with the aim of reducing traffic congestion, improving air quality, and generating revenue for public transit investment. The CRZ boundary is defined by a set of bridge and tunnel entry points, including the Lincoln Tunnel, Holland Tunnel, Queensboro Bridge, Williamsburg Bridge, Manhattan Bridge, and Brooklyn Bridge. Toll amounts vary by time of day and vehicle type: passenger vehicles pay \$9 during peak hours and \$2.25 overnight, while trucks and buses face higher charges. This intervention differs from the other two in that it increases rather than decreases the cost of accessing a central urban area, and it lets us test whether pricing mechanisms produce symmetric effects on urban activity.\\

\subsection{Global and Commuting mobility datasets}\label{supp:sec:metro_data}
To compare how well geographical distance proxies mobility network distance across metro areas, we used five independent origin-destination datasets across 138 cities in eight countries from three different continents. The datasets are:
\begin{itemize}
    \item {\bf United States, Mobility, Cuebiq}: anonymized aggregated device-based visit-to-visit mobility networks between tract pairs within Core-Base Statistical Areas (CBSAs). For statistical significance, we restrict the analysis to the most populated CBSAs that contained more than 100 tracts, which results in only 114 used in our analysis. Those CBSAs contained 222 million residents in the US. The mobility networks were constructed using 2019 mobility data. See Section \ref{supp:section:locationdata} for more details about this dataset.
    \item {\bf United States, Commuting, LODES}: Longitudinal Employer-Household Dynamics Origin-Destination Employment Statistics (LODES) for 2019 \cite{census_lodes2019}. The dataset contains home-work commuting flows from the U.S. Census Bureau aggregated at the tract level. We restrict our analysis to those CBSAs with more than 100 tracts, resulting again in 114 US metro areas analyzed.
    \item {\bf United Kingdom, Commuting, NOMIS}: Origin–destination workplace data for England and Wales from the U.K. Census 2021 (NOMIS table ODWP01EW), containing home–work commuting flows between Middle Layer Super Output Areas (MSOAs) \cite{ons_odwp01ew_2021}. We restrict to flows with place-of-work indicator 3 (working in the UK, not at/from home). Metropolitan areas are defined using Eurostat/ONS Urban Audit \emph{Functional Urban Areas} (FUAs, December 2016): each metro is the set of MSOAs whose Local Authority District (LAD) belongs to the corresponding FUA (or to several FUAs for poly-centric metros, e.g.\ Birmingham = West Midlands urban area + Coventry) \cite{ons_urban_audit_fua_2016}. Metro population is the sum of Census 2021 residents over those MSOAs. We only considered the 13 metro areas in England and Wales (London, Birmingham, Manchester, Leeds-Bradford, Liverpool-Birkenhead, Newcastle-Sunderland, Sheffield, Portsmouth-Southampton, Nottingham-Derby, Bristol, Brighton-Worthing-Littlehampton, Leicester, and Middlesbrough).               
    \item {\bf Spain, Mobility, MITMA}: District-level origin–destination trip matrices from Spain’s Ministerio de Transportes y Movilidad Sostenible (MITMA) open mobility study, based on anonymized mobile-network data \cite{mitms_opendata_mobility}. We use a representative weekday (1 March 2022) of all-purpose trips between MITMA \emph{distritos} (districts). Metropolitan areas are defined as MITMA \emph{Grandes Áreas Urbanas} (GAUs): each metro is the set of districts linked to a named GAU in the MITMA–INE crosswalk, retaining GAUs with population of at least 500{,}000. Geographic distances use district centroids, and metro population is the sum of MITMA population over those districts. We consider cities with more than 40 districts so we ended up with Madrid, Barcelona, Valencia, Seville, Bilbao, and Vigo only.  
   \item {\bf Latin America, Mobility, CEPAL/ECLAC}: Mobility Surveys (Origin-Destination Surveys) (\emph{Encuestas Origen--Destino}, EOD) harmonized by Economic Commission for Latin America and the Caribbean (ECLAC) through their Ciudades Inteligentes, Inclusivas y Sostenibles (CISI) mobility repository \cite{cepal_cisi_eod_platform,cepal_cisi_eod}. We use six metropolitan study areas: Bogotá (2019), Buenos Aires (2010), Mexico City (2017), Montevideo (2016), Santiago (2012), and São Paulo (2017). Zones are the official survey zoning units (e.g.\ ZATs, distritos, comunas), with centroids from the repository shapefiles. Trip counts come from pre-aggregated zone-to-zone matrices where available; for Buenos Aires and São Paulo we reconstruct weighted OD flows from trip microdata. Each metro is defined by its survey study area, geographic distances use zone centroids, and population is taken from published census or survey totals for that area. We consider areas with more than 40 zones and thus we only consider Bogotá (Colombia), Mexico City (Mexico), Montevideo (Uruguay), Santiago de Chile (Chile), and São Paulo (Brazil).
\end{itemize}

\noindent

\section{Causal inference methods} \label{supp:section:CIM}
Estimating the causal effect of an urban policy from observational mobility data requires two distinct choices: how to define which units are exposed to the intervention, and how to estimate the counterfactual outcome those units would have had in its absence. In our paper, we make these choices independently, and the central methodological contribution of this work concerns the first. This section sets out both. We begin with the estimator, which we hold fixed throughout: Difference-in-Differences (Section~\ref{supp:section:did}) applied to the gap between a treated unit and a synthetic counterfactual built by the Synthetic Control Method (Section~\ref{supp:subssection:scm}). Whichever exposure definition is used, donor units must also be demographically comparable to the treated ones; we quantify this comparability with a propensity score (Section~\ref{supp:sec:propensity}) that both methods rely on. 
We then assess significance with an empirical placebo test (Section~\ref{supp:sec:placebo_test}) rather than the regression standard error of the estimated effect, which is the standard practice in applied urban policy work. This criterion is more demanding, since it compares the estimated effect to the distribution of coefficients the same procedure produces on units that were never treated. With the estimator and this statistical significance criterion in place, we describe the two exposure definitions we compare: the conventional Geographical Distance Method (Section~\ref{supp:sec:GDM}), which assigns treatment and control by physical proximity, and the Network Distance Method (Section~\ref{supp:sec:NDM}), which assigns them by position in the mobility-flow network.

\subsection{Difference-in-Differences}
\label{supp:section:did}
Difference-in-Differences (DiD) estimates the effect of an intervention by comparing the pre-to-post change in a treated unit's outcome with the pre-to-post change in a comparable control~\cite{butts2021difference,callaway2021difference}. Identification rests on the parallel-trends assumption: absent the intervention, the treated and control series would have followed parallel trajectories.

We operationalize DiD on the gap between the observed treated outcome and a counterfactual outcome constructed from control units. Let $Y^{\text{tr}}_{t}$ denote the observed outcome of the treated unit at time $t$ and $Y^{\text{cf}}_{t}$ the counterfactual at the same time, and define the gap
\[
G_{t} \;=\; Y^{\text{tr}}_{t} - Y^{\text{cf}}_{t}.
\]
Because the synthetic counterfactual is fit to track the treated series over the pre-treatment window, working on the gap removes the variation common to treated and control units, so the residual reflects only treated-specific deviation and reduces the residual variance in the estimate of $\beta$. 

The treatment effect is the coefficient on the post-treatment indicator in the regression
\begin{equation}
G_{t} \;=\; \beta_{0} \;+\; \beta\, \mathbf{1}\{t \ge t^{\ast}\} \;+\; \mathbf{x}_{t}^{\top}\boldsymbol{\gamma} \;+\; \varepsilon_{t},
\label{eq:gap_did}
\end{equation}
where $t^{\ast}$ is the intervention date and $\mathbf{x}_{t}$ collects period-level controls. In the reported estimates $\mathbf{x}_{t}$ contains period-level weather variables (average temperature, precipitation, and snowfall), included for the visit and spend panels of all three interventions; the only exception is the Free Fare monthly-spend robustness panel, which uses no weather controls. Here $\beta_{0}$ absorbs any residual pre-period offset, and $\beta$ is the average treatment effect over the post-treatment window in outcome units.

For comparability across panels with different scales, we report the percentage gap
\[
\hat\beta_{\%} \;=\; 100\,\frac{\hat\beta}{\overline{Y^{\text{cf}}_{\text{post}}}},
\]
with $\overline{Y^{\text{cf}}_{\text{post}}}$ the mean of the counterfactual over the post-treatment window. 

\subsection{Synthetic Control Method (SCM)} \label{supp:subssection:scm}
DiD requires a control group, and nothing in the data identifies one. With a single treated set, any control group is a choice, and its parallel-trends assumption cannot be tested. The Synthetic Control Method (SCM) constructs the control group from a pool of untreated units, the \textit{donors}~\cite{abadie2010synthetic}. Donor weights are fit so that the weighted combination of donor series, the synthetic control, reproduces the treated unit's pre-treatment trajectory. Parallel trends then applies to the treated-versus-synthetic gap rather than to an assumed control.

Donors are census tracts in Free Fare and the CRZ, and individual businesses in Open Streets. The two exposure definitions determine which units enter the donor pool (Sections~\ref{supp:sec:GDM} and~\ref{supp:sec:NDM}). We fit one synthetic control per treated unit. 
We fit the weights by Lasso regression~\cite{tibshirani1996regression} with the coefficients constrained to be non-negative, so the synthetic control is a non-negative weighted combination of donor series, as in the original synthetic control formulation~\cite{abadie2010synthetic}. The $L_1$ penalty selects few donors and prevents overfitting the pre-treatment window when the donor pool is large. In the Boston Free Fare NDM total-visit fits a median of eight donors (range 1--18, out of roughly 45 candidates) receive nonzero weight. The synthetic control supplies $Y_t^{\text{cf}}$ in the gap regression, Eq.~\eqref{eq:gap_did}.

Two diagnostics measure how well each synthetic control reproduces its treated unit's pre-period trajectory. Both are computed over the pre-treatment window and averaged across treated units. Let $Y$ be a treated unit's observed pre-period series and $\hat{Y}$ its synthetic series. The \textit{pre-treatment correlation} is the Pearson correlation between $Y$ and $\hat{Y}$. The \textit{pre-treatment normalized root-mean-square error} (NRMSE) is the root-mean-square of the gap $Y-\hat{Y}$ divided by the standard deviation of $Y$,
\[
\text{NRMSE}=\sqrt{\overline{(Y-\hat{Y})^{2}}}\big/\operatorname{sd}(Y).
\]
A good match has a pre-treatment correlation close to $1$ and a low NRMSE. We refer to the two together as the \textit{pre-treatment fit}. They diagnose the credibility of the counterfactual and set the Leiden resolution (Section~\ref{supp:subsec:ndm_choices}). They play no part in estimating the treatment effect.

Finally, before fitting, we trim the estimation sample on mean visit volume. Units whose mean weekly visits over the study window fall in the extreme tails of the pooled treated-and-donor distribution are dropped: below the 10th or above the 90th percentile for Boston Free Fare, and below the 5th or above the 95th for the CRZ. The filter applies to treated and donor units alike, so it does not favour either pool, and it is applied identically under GDM and NDM. Its purpose is to keep the synthetic control from resting on units whose scale is incomparable to the treated series, since a donor with orders of magnitude more or fewer visits can dominate the fit or contribute nothing to it. No volume filter is applied in the Open Streets analysis, where the units are individual businesses and the volume range is narrower.

\subsection{Propensity score} \label{supp:sec:propensity}

A good counterfactual must respond to non-policy shocks as the treated unit does. Matching pre-treatment trajectories does not guarantee this, since two units can track each other before the intervention and diverge afterwards for reasons unrelated to it. Donors that resemble the treated units in income, transit access, and education are exposed to the same concurrent shocks and are more likely to keep tracking them. Both exposure methods therefore require donor units to be demographically comparable to the treated ones.

We quantify comparability with a propensity score, following matching methods that compare each treated unit with units of similar likelihood of receiving the treatment~\cite{rosenbaum1985constructing,austin2011introduction}. We fit a logistic regression predicting treatment status from standardized demographic covariates (Section~\ref{supp:subsec:demographic}), so $\hat p$ is the estimated probability of treatment given the observed covariates. We work on the log-odds scale, $\hat\ell=\mathrm{logit}(\hat p)=\log[\hat p/(1-\hat p)]$, which puts demographic similarity on an additive scale, and define the propensity distance of a candidate $c$ from the treated set $T$ as
\[
d_{\text{ps}}(c,T)\;=\;\bigl|\hat\ell_c-\overline{\hat\ell_T}\bigr|.
\]
A small $d_{\text{ps}}$ means the donor is as likely to be treated as the treated units are, given its covariates. It does not mean the donor matches them covariate by covariate, since the regression collapses the covariate vector into a single index. We therefore report per-covariate balance alongside $d_{\text{ps}}$ in the results tables (Section~\ref{supp:sec:gdm_ndm_results}).

The covariate set is intervention-specific. Boston Free Fare uses median household income, the public-transit commute share, and the Black and bachelor's-degree population shares. The CRZ uses median household income, the White and Black population shares, and the public-transit commute share. Open Streets, where the units are POIs, uses visitor-weighted income and the visitor-weighted White and Black population shares. The score filters the GDM donor ring (Section~\ref{supp:sec:GDM}) and serves as one of the two Pareto objectives in NDM (Section~\ref{supp:sec:NDM}).

\subsection{Placebo test}
\label{supp:sec:placebo_test}
Throughout our analysis, significance is conservatively assessed by an empirical placebo test~\cite{abadie2010synthetic,abadie2015comparative,abadie2021using}. In each placebo draw we relabel a random subset of the donor units as a pseudo-treated set and use the remaining units as the placebo donor pool, then refit the SCM and the gap regression on this pseudo design, recording the pseudo-coefficient $\hat \beta_{(b)}$. We repeat for $B=100$ independent draws and report the empirical $p$-value as the fraction of placebo coefficients whose magnitude reaches $|\hat\beta|$,
$$
\hat p \;=\; \frac{1 + \#\{b : |\hat\beta_{(b)}| \ge |\hat\beta|\}}{B+1}.
$$
The test is two-sided (on $|\hat\beta|$) in every analysis reported in this paper.\footnote{With $B=100$ the smallest attainable $p$-value is $1/101 \approx 0.0099$, so we report significance at the $p<0.05$ and $p<0.01$ levels only.}

This criterion is stricter than reading significance from the regression standard error of $\hat \beta$, which remains common in applied spatial and urban policy work. With a single treated unit and donor weights selected from the data, that standard error has no valid sampling interpretation~\cite{abadie2010synthetic,abadie2021using}. Reported standard errors in urban areas are often small \cite{zheng_effects_2024}, so a gap that the placebo distribution treats as ordinary can appear highly significant. The placebo test builds the null from the design itself and flags an effect only when it exceeds what the same procedure produces on units that were never treated.

The results tables additionally report, as a descriptive summary of each placebo null, the 5th--95th percentile range of the \textit{signed} placebo coefficients $\hat\beta_{(b)}$ (the ``placebo 5--95\%'' column). This range is not the decision rule: significance is always the two-sided test on the magnitude $|\hat\beta|$ defined above. Because the range is taken over signed values while the test compares magnitudes, an estimate can occasionally lie outside the reported $[5\%,95\%]$ range without being flagged significant. For example a one-sided placebo null that is itself shifted away from zero can place both percentiles on the same side of the origin. The asterisks therefore follow the $|\hat\beta|$ test, not visual comparison with the signed range.

\subsection{Geographical Distance Method (GDM)} \label{supp:sec:GDM}
SCM and DiD require a treatment and a donor group. The conventional choice is geographic distance from the intervention: units within a given radius are treated, and units in a surrounding ring form the donor pool~\cite{zheng_effects_2024,pollmann_causal_2023}. We treat census tracts within 1\,km of Route~28 in Free Fare, POIs within 50\,m of the open street in Open Streets, and all census tracts inside the zone boundary for the CRZ. Donors are drawn from a surrounding ring: a 2--8\,km ring for Free Fare, a 500--800\,m ring for Open Streets, and a 5--15\,km ring for the CRZ. The Open Streets buffers are the same for all three sites. Section~\ref{supp:sec:sensitivity_buffer} recomputes the estimates across alternative ring widths to check their robustness.

Distance is not the only criterion for choosing controls. To improve demographic comparability within the ring, we filter the donor pool by the propensity score of Section~\ref{supp:sec:propensity}, keeping donors whose logit propensity score lies within a fixed distance of the treated-pool mean. The distance is $6$ logit units for Boston Free Fare and the CRZ, and $0.5$, $1.0$, and $0.2$ logit units for the Community League of the Heights, Downtown Brooklyn Partnership, and Washington Irving HS Open Streets sites. The GDM control counts in Tables~\ref{supp:tab:free_fare_summary}, \ref{supp:tab:nyos_summary}, and~\ref{supp:tab:crz_summary} refer to this filtered pool. We additionally trim the estimation sample on visit volume, as described in Section~\ref{supp:subssection:scm}.

\subsection{Network Distance Method (NDM)}\label{supp:sec:NDM}
The flows that cross the buffer boundary also cross the boundary of any single unit. A tract's outcome is contaminated by the units it exchanges visitors with, so no individual unit is insulated and a single unit is the wrong scale at which to separate treated from control. We work instead at the level of mobility communities, groups of units within which flow is dense and across whose boundaries flow is sparse. Most movement stays inside a community, so spillover is largely contained within it and a community can be labeled treated or control as a near-closed unit. This is the urban analogue of cluster randomization in network experiments~\cite{ugander_graph_2013,eckles_design_2017}.

\subsubsection{Mobility network} 
\label{supp:sec:mobility_network}
Nodes in our network are spatial units, for example census tracts or businesses.
Edges record any behavioral dependence between units, that is, any measurable way in
which activity in one unit is tied to activity in another~\cite{yabe2025behaviour}.
The dependence we use is intervention-specific. For Boston Free Fare and the CRZ,
where nodes are census tracts, edges are visit-to-visit flows: each time a user's
trajectory contains two consecutive visits in tracts $u$ and $v$, the edge $u \to v$
gains one unit of weight. For Open Streets, where nodes are individual businesses,
edges are co-visitation: the edge between businesses $u$ and $v$ is weighted by the
number of users who visit both within the network-construction window, so two
businesses are connected when they draw on a shared customer base. The method applies
unchanged to other choices of dependence, such as home-to-work commuting
(Section~\ref{supp:sec:sensitivity_lodes}).

Summing edge weights across areas or POIs yields a weighted adjacency matrix $M$, directed
for visit-to-visit flows and symmetric for co-visitation. We zero its diagonal and
row-normalize it to obtain a row-stochastic matrix $W$. For visit-to-visit flows
$W_{uv}$ is the probability that a visit in $u$ is followed by a visit in $v$; for
co-visitation it is the share of unit $u$'s co-visitation weight that falls on $v$.

\subsubsection{Ensemble partitioning and consensus} \label{supp:subsec:consensus}
We detect communities in $W$ with the Leiden algorithm~\cite{traag_louvain_2019}, run for $N=100$ independent random seeds at a fixed resolution $\gamma$. Each run returns a hard partition, and the partitions differ across seeds because Leiden is stochastic. We consolidate them into one partition.

For each pair of units $(u,v)$ we compute the co-assignment frequency $C_{uv}\in[0,1]$, the fraction of the $N$ runs in which $u$ and $v$ sit in the same community. We treat $D_{uv}=1-C_{uv}$ as a distance, apply average-linkage hierarchical clustering, and cut the dendrogram at threshold $0.5$, so that pairs within a community are co-assigned in at least half of the runs. This yields a single consensus partition, together with a per-unit stability score equal to the mean co-assignment of the unit with its fellow community members. The consensus removes the dependence on any single seed and retains only community structure that is stable across the ensemble.

\subsubsection{Definition of treated and donor units} \label{supp:sec:treated_donor}
Let $T$ denote the set of directly treated units. A spatial unit belongs to $T$ if the intervention directly passes through it: census tracts within a 1\,km radius of Route~28, POIs along the pedestrianized street segments in the Open Streets policy, or all census tracts and POIs within the CRZ. Our Network Distance Method operates on mobility communities rather than individual units, so we declare a consensus community treated when at least 10\% of its units lie within $T$, and we include in the synthetic control all units from the treated community or communities.

For Boston Free Fare, this expands the treated set from the 40 tracts within 1\,km of Route~28 to the 159 tracts of the containing community, and for the CRZ from the 136 tracts inside the zone boundary to 162. For Open Streets, the treated communities are small and largely coincide with the street itself, so the treated POI count changes little relative to GDM (Table~\ref{supp:tab:nyos_summary}). NDM therefore changes both the treated set and the donor pool relative to GDM, and Section~\ref{supp:sec:decomposition} separates the two, showing that the difference between the designs comes from the donor pool.

The remaining communities form the donor pool, restricted to a study universe fixed per intervention. For Boston Free Fare the universe is the set of census tracts intersecting a 1\,km buffer around any MBTA bus route, so all candidate donors are bus-served tracts. For Open Streets it is the set of POIs within a 1.5\,km radius of the open street segments. For the CRZ it is the set of census tracts within 20\,km of the congestion zone boundary. The visit-volume trim of Section~\ref{supp:subssection:scm} applies to the treated and donor sets alike.

\subsubsection{Pareto selection of donor/control communities.} 
\label{supp:secc:pareto_selection}
Choosing controls in a spatial setting confronts two requirements that pull in opposite directions. Controls must be demographically comparable to the treated units, which is a requirement on covariates. Controls must also be insulated from the intervention, which is a requirement on mobility flows. Neither implies the other, and geography couples them.

Demographic comparability favors controls close to the intervention, because nearby areas tend to be more alike through residential segregation and planning history. Mobility spillovers, however, are strongest for nearby areas~\cite{pollmann_causal_2023,ugander_graph_2013,eckles_design_2017}, so the communities that are demographically most comparable are also the ones most exposed to the treatment. Minimizing demographic distance pulls controls inward, and escaping spillover pushes them outward. The two requirements cannot be satisfied at once.

Pushing the controls far away, or into other cities entirely, escapes spillover at a high cost. Distant neighborhoods and cities differ systematically in income, density, and infrastructure, and in commuting patterns and daily rhythms~\cite{pandit_behavioural_2026}. Even if pre-treatment trends appear to match, they may diverge for reasons unrelated to the policy~\cite{abadie2010synthetic,abadie2021using}. A control city also cannot register the substitutions the policy triggers within the treated city, because the mechanism that produces them does not propagate there. Cross-city designs trade a contaminated control group for one that is structurally incomparable.

We make the trade-off explicit through two scalars computed for every donor community $c$. The first measures exposure. A community is exposed to the intervention to the extent that people move between it and the treated region, so we measure its separation from the treated set by mobility flow rather than by space. Direct flows are not enough, since a community can be reached through an intermediate one, so we count chains of flows as well. We capture both with the truncated diffusion kernel~\cite{katz_new_1953,yabe2025behaviour} $W_{\text{diff}}=\sum_{k=1}^{K}\beta^{k} W^{k}$, with attenuation $\beta=0.8$ and depth $K=5$, so that $W_{\text{diff}}[u,v]$ is the total mobility connectivity between $u$ and $v$ across direct and indirect paths. Section~\ref{supp:sec:sensitivity_beta} shows that the donor selection does not depend on the specific choices of $\beta < 1$. The network distance is the inverse of the mean connectivity over the $c\times T$ pairs,
\[
d_{\text{net}}(c,T) \;=\; \bigl(\overline{W_{\text{diff}}[c,T]}\bigr)^{-1},
\]
which is small when $c$ receives many visits from the treated region and large when $c$ is mobility-isolated from it.

The second scalar measures demographic comparability. It is the propensity distance of Section~\ref{supp:sec:propensity}, computed here at the community level,
\[
d_{\text{ps}}(c,T) \;=\; \bigl|\hat\ell_{c}-\overline{\hat\ell_{T}}\bigr|,\qquad \hat\ell=\mathrm{logit}(\hat p),
\]
where $\hat p_c$ is the community-level propensity score. A community that resembles the treated set has $\hat\ell_c$ close to the treated mean $\overline{\hat\ell_T}$ and a small $d_{\text{ps}}$. Dissimilar communities have large $d_{\text{ps}}$.

Demographic similarity and mobility connection both decline with geographic distance, which couples the two objectives. The communities with the smallest $d_{\text{ps}}$ tend to have the smallest $d_{\text{net}}$, so no community is both maximally comparable and maximally insulated. An ideal control does not exist in a segregated city, and this is a structural property of urban space rather than a limitation of our data: any design that selects controls on one criterion alone accepts an unmeasured cost on the other. Buffer-based designs select on geographic distance and pay in spillover. Matching designs select on covariates and pay in the same currency, since the best-matched units are the most connected ones.

We therefore treat control selection as a two-objective problem and select along the Pareto frontier of $(d_{\text{ps}},d_{\text{net}})$, the set of communities for which no alternative is simultaneously closer in propensity and farther in network distance. These are the non-dominated trade-offs. The frontier does not remove the tension. It makes the position taken on it explicit and confines the choice to the communities that are not dominated on both criteria at once (Section~\ref{supp:subsec:ndm_choices}).

This two-objective formulation is what allows us to report an ATE. The distance analysis of Fig.~\ref{fig1}B varies network distance alone, using rings of donor tracts, and shows that the spillover bias vanishes beyond a characteristic distance. It locates where spillover stops mattering, and nothing more: the ring donors are internally incoherent (Section~\ref{supp:sec:exposure}) and are not matched to the treated region on observables. Selecting controls by network distance alone would therefore escape spillover at the cost of the comparability that the buffer design already fails on. For Boston Free Fare the far end of the Pareto front, community 3, carries a propensity distance of about 4, close to the 4.88 of the GDM ring and well above the 2.34 of the selected community (Table~\ref{supp:tab:gdm_demo_ff}). The Pareto front is what lets us escape spillover without paying that cost, and the position taken on it is fixed by the rule of Section~\ref{supp:subsec:ndm_choices} rather than by the size of the recovered effect.

We restrict the Pareto analysis to communities with at least $N_{\text{min}}=30$ units. Each community must carry enough data for a stable outcome series. The placebo test also relabels half of the donor pool as pseudo-treated on each draw and refits on the remainder, so the pool must be large enough for diverse placebo splits and stable synthetic-control fits.

\subsubsection{Why mobility communities, not network-distance rings} \label{supp:sec:exposure}

The Pareto step above ranks already-formed communities by their network distance to the treated set. A natural question is whether the communities themselves are needed at all: one could instead bin donor units directly by a network-distance score, forming concentric "rings" around the treated region and using a ring as the control pool. We show that this is inadequate, and that grouping units into mobility communities is what makes the donor pool valid.

A donor pool defined by distance from the intervention (using an outer ring on network distance, or binning spatial units by network distance score) does not control for the actual channel through which treatment contaminates controls: the mobility flow between donors and the treated region. Mobility communities do, for two reasons.

First, they contain spillover. Under network interference, the bias of a cluster-level estimator is bounded by the share of edge weight that crosses from the donor cluster back to the treated set, the quantity Ugander \emph{et al.}~\cite{ugander_graph_2013,eckles_design_2017} call $\varepsilon$. A good control group is therefore one whose boundary few flows cross. Community detection cuts the network where flow is sparse, which is the same criterion that keeps $\varepsilon$ small. A distance ring is drawn without regard to flow, so it can slice through a dense mobility corridor and admit the spillover it is meant to exclude. We do not estimate $\varepsilon$ directly. The coherence ratio defined below compares within-pool to across-boundary flow, so it measures the same boundary that bounds $\varepsilon$ and serves as our diagnostic for both conditions.

Second, they are internally coherent. Synthetic control weights are stable only when donor units resemble one another in their mobility, not only in their relationship to the treated set~\cite{abadie2021using}. An internally heterogeneous donor pool yields unstable weights and a noisy synthetic counterfactual. We diagnose mutual similarity with the within-pool diffusion-kernel coherence ratio of a group $B$,
\[
  \mathrm{coh}(B) \;=\; \frac{\overline{W_{\text{diff}}[u,v]}_{\,u,v \in B,\ u \ne v}}{\overline{W_{\text{diff}}[u,v]}_{\,u \in B,\ v \notin B}},
\]
with $W_{\text{diff}}$ the truncated diffusion kernel of Section~\ref{supp:secc:pareto_selection}. Leiden's objective rewards this ratio, so its communities are coherent by design. Distance bins are not.

\begin{figure}[t!]
\centering
\includegraphics[width=\textwidth]{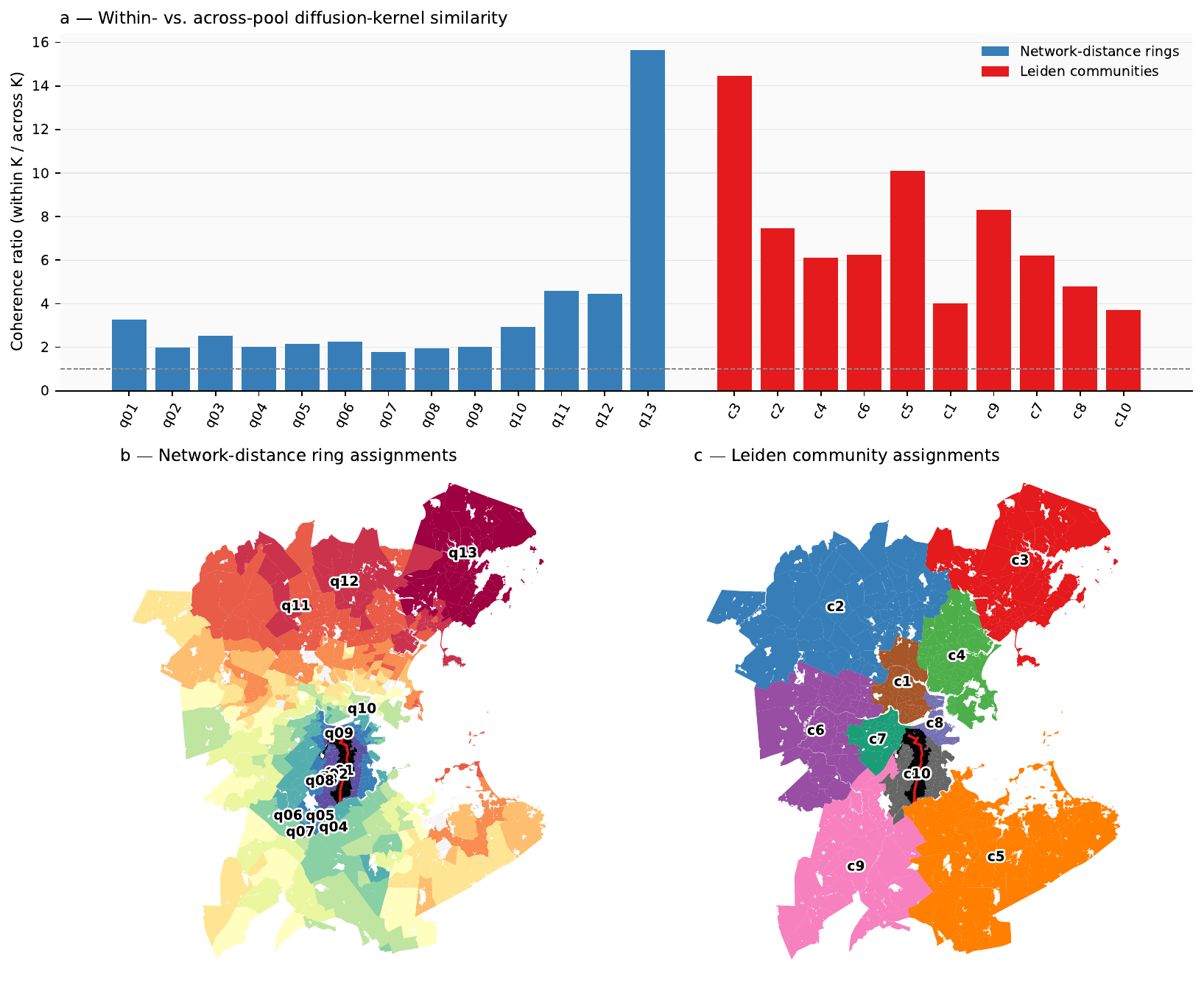}
\caption[{\bf Comparison of network-distance and Leiden-community donor partitions for the Boston Free Fare intervention.}]{{\bf Comparison of network-distance and Leiden-community donor partitions for the Boston Free Fare intervention.} (a) Within-pool coherence ratio $\text{coh}(B)$ for each candidate donor pool: the average within-pool mobility similarity divided by the across-boundary similarity, so higher values mean members are more mobility-similar to each other than to outsiders. Candidates are rings of different network distances (left) and communities (right). (b) Map of the network-distance ring assignments (q01--q13, low-to-high mean network distance) around Route 28. (c) Leiden community assignments ordered by decreasing mean network distance to treated; $c10$ contains the treated tracts and is excluded as a donor. Black tracts and red lines mark the treated set and Route 28.}
\label{supp:fig:exposure}
\end{figure}

Figure~\ref{supp:fig:exposure} confirms both points for Boston Free Fare. Communities reach coherence values between 4 and 14, indicating that pool members are several-fold more mobility-similar to each other than to outsiders. Network-distance rings cluster around $2$--$3$, with one accidental outlier at q13 where the rightmost ring happens to align with an isolated mobility cluster; only this outlier matches the coherence of an entire Leiden community. Panels b and c show why. The rings partition tracts by network-distance quantile from the treated set; they form roughly nested but irregular bands radiating outward from Route 28, but each band still mixes tracts that share little direct mobility. Communities instead group tracts by shared mobility flow, and so appear as discontiguous patches: tracts that exchange many visitors need not be geographically adjacent. Grouping by shared flow, not by distance rank, is what isolates the spillover channel.

Mobility communities therefore keep boundary-crossing flow low and are internally coherent, and distance bins are neither. We adopt them as the donor partition throughout NDM.

\subsubsection{Per-intervention donor choices of control communities}\label{supp:subsec:ndm_choices}
The Leiden resolution $\gamma$ and the donor community set used in the main-text analysis are fixed separately for each intervention. We choose $\gamma$ by how well the resulting donor communities reproduce the treated unit's pre-treatment trajectory, not by the size of the treatment effect. A good resolution is one that gives a low pre-treatment Normalized Root-Mean-Square-Error (NRMSE) and a high pre-treatment correlation in the synthetic control fit. We also require the communities to stay large enough to support placebo inference (Section \ref{supp:secc:pareto_selection}); very fine resolutions break the network into donor pools too small for that. The treatment effect itself is insensitive to this choice, so $\gamma$ affects the precision of the estimate rather than its value (Section \ref{supp:sec:sensitivity_community}).

At a fixed $\gamma$, we consider the Pareto front of candidate donor communities in $(d_{\text{ps}},d_{\text{net}})$ space, i.e.\ the set of communities for which no alternative is simultaneously closer in propensity distance and farther in network distance. Along this front the two objectives trade off by construction: ordering the communities by ascending $d_{\text{ps}}$ also orders them by ascending $d_{\text{net}}$, because near communities tend to be both demographically similar to the treated region and strongly connected to it in the mobility network (Section \ref{supp:secc:pareto_selection}). The leftmost community is thus the most demographically similar but at the highest risk of spillover, and the rightmost the most weakly connected but the most demographically dissimilar. We take the midpoint of this ordering as a neutral starting donor set since it sits farthest from both extremes. We take a single community when the front has odd length, the two central communities when it has even length. This starting set can be too small to support placebo inference, which relabels half the donor pool on each draw and so needs a donor count large enough for stable splits. Note that this is a separate constraint from the community-size floor that shapes $\gamma$ (Section \ref{supp:secc:pareto_selection}): there we require individual communities to be large; here we require the selected pool, at the already-fixed $\gamma$, to contain enough tracts. When the midpoint set falls short, we walk outward along the front, adding the next community at each step until the donor count is large enough to support stable placebo splits; the resulting NDM donor pools contain $64$ tracts for Boston Free Fare, $215$ tracts for the CRZ, and $50$, $110$, and $78$ POIs for the CLH, DBP, and WIHS Open Streets sites. We stop before propensity distance grows sharply, so the pool never drifts into demographically dissimilar donors. How far this allows us to walk differs by site: in Boston Free Fare the treated region already covers much of the comparable MBTA-served tract pool, so raising the target quickly pulls in high-propensity-distance donors; in the two NYC settings the donor pool is deeper and the treated regions smaller, so a larger pool improves placebo precision without degrading donor similarity. In both regimes the target is large enough that the synthetic weights spread across several donors (a median of eight in Free Fare; Section~\ref{supp:subssection:scm}) rather than resting on one or two.

Applying this rule fixes a specific donor set for each intervention, reported below. Unlike the width of a geographic buffer, the criterion is anchored to the channel that contaminates controls: a community's position on the front is set by its mobility flow to the treated region ($d_{\text{net}}$) and its demographic distance ($d_{\text{ps}}$), not by where it lies in space. The estimates are stable across the choice:  over donor subsets on the front, Leiden resolutions, and detection algorithm (See Section \ref{supp:sec:sensitivity_community}, and Fig.~\ref{supp:fig:comm_sensitivity}).

\begin{figure}[t!]
    \centering
    \includegraphics[width=\textwidth]{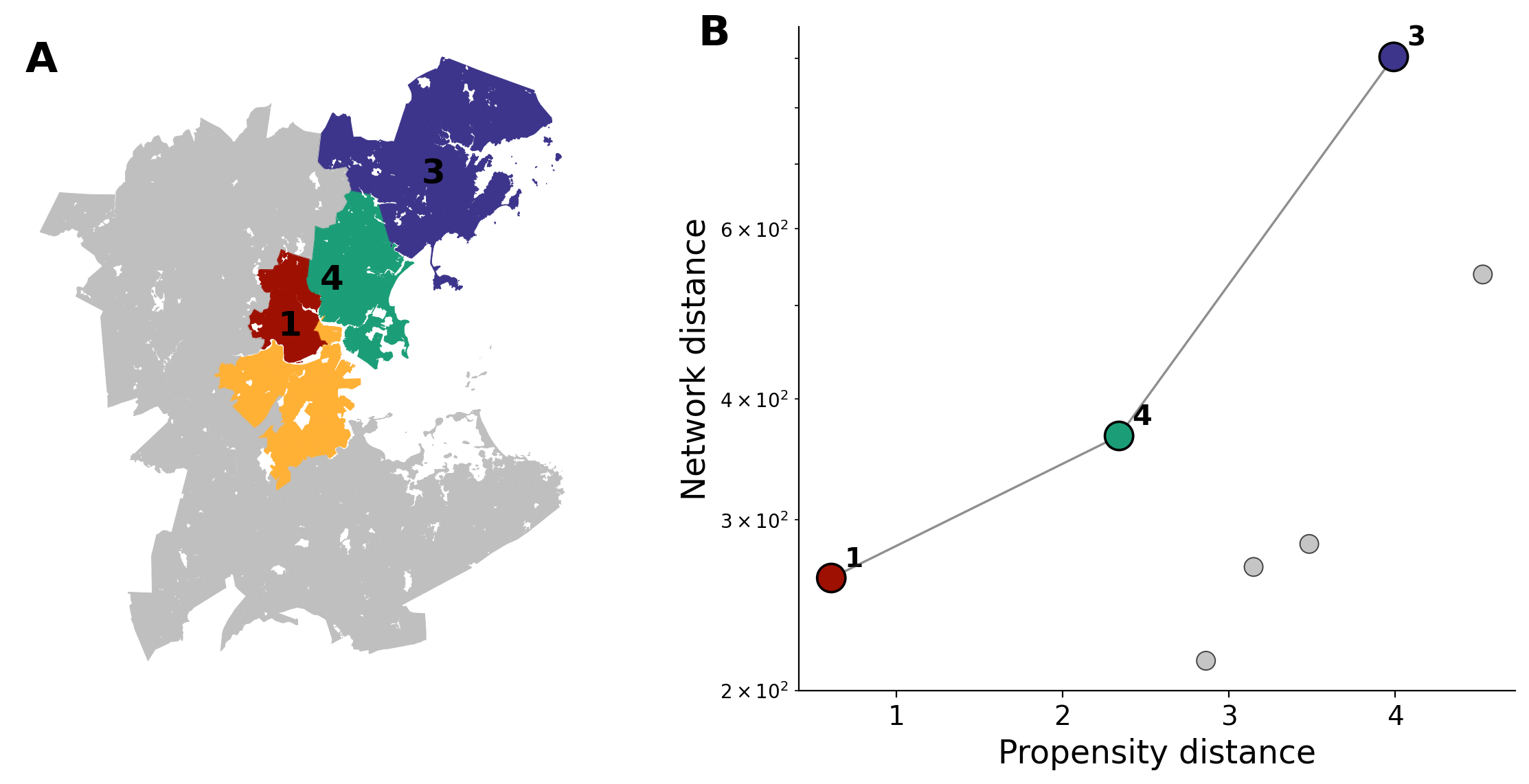}
    \caption[\textbf{Free Fare: selected donor community.}]{\textbf{Free Fare: selected donor community.} \textbf{(A)}~Consensus community map at Leiden $\gamma=0.0025$; the main-text donor is colored donor-green. \textbf{(B)}~Pareto scatter of candidate donor communities in (propensity distance, network distance) space, with the Pareto front connected; the chosen donor is green.}
    \label{supp:fig:free_fare_maps_pareto}
\end{figure}

\paragraph{Boston Free Fare.} We use Leiden $\gamma=0.0025$, which yields ten consensus communities. At this resolution the Pareto front over candidate donor communities is $\{1,\,4,\,3\}$, ordered by increasing network distance. Community 1 is closest in network distance (and in propensity distance), so it risks spillover; community 3 is farthest but the most demographically dissimilar. We therefore take community 4, the midpoint of the front, as the main-text donor: it carries a moderate network distance while keeping propensity distance well below community 3's. Figure~\ref{supp:fig:free_fare_maps_pareto} shows the map and Pareto scatter, with community 4 in donor-green. 

\begin{figure}[t!]
    \centering
    \includegraphics[width=\textwidth]{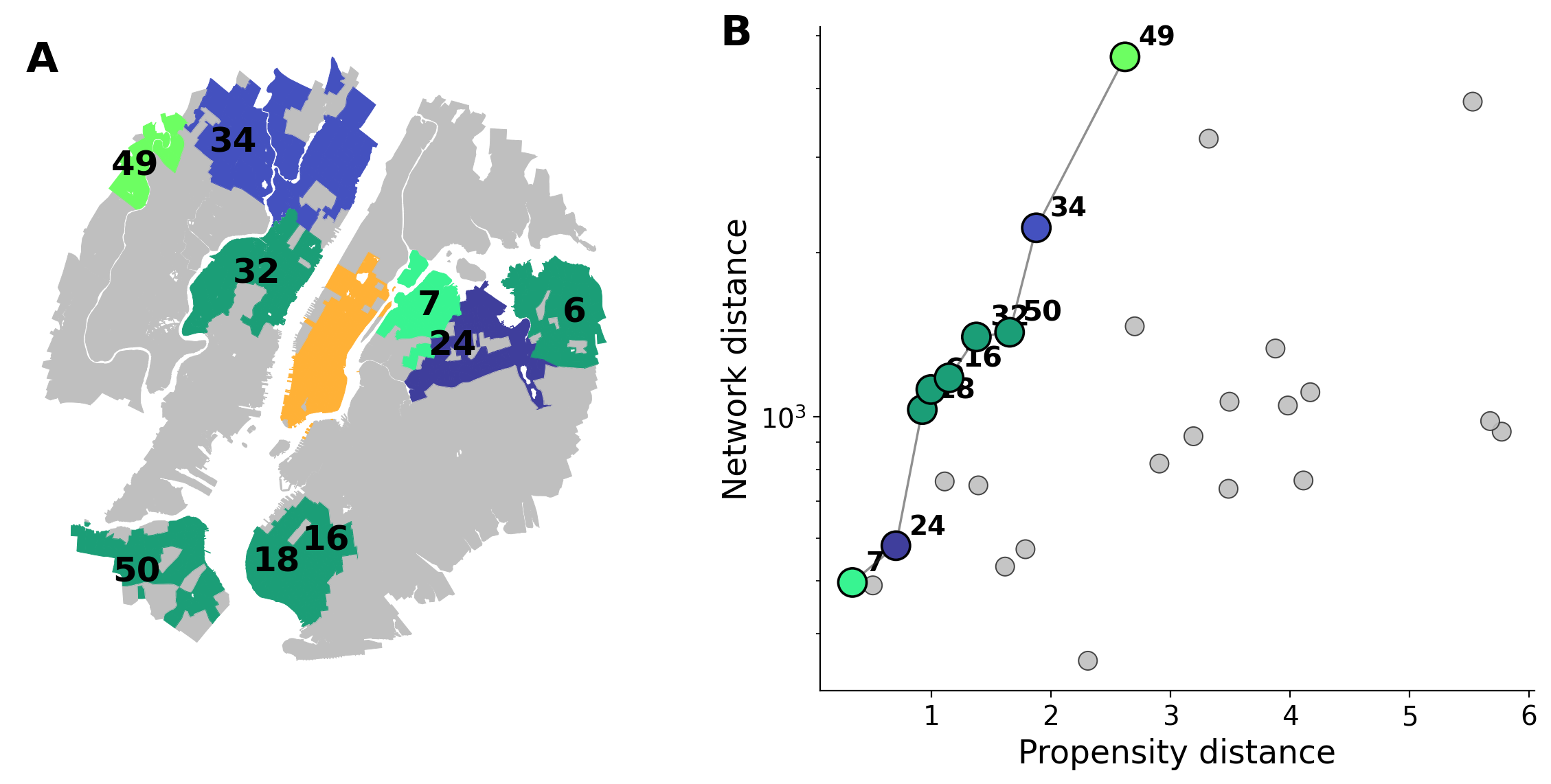}
    \caption[\textbf{CRZ: selected donor communities.}]{\textbf{CRZ: selected donor communities.} \textbf{(A)}~Consensus community map at Leiden $\gamma=0.002$; the five main-text donor communities $\{18,6,16,32,50\}$ are in donor-green. \textbf{(B)}~Pareto scatter of candidate donor communities; the chosen donors are green.}
    \label{supp:fig:crz_maps_pareto}
\end{figure}

\paragraph{CRZ: NYC Congestion Relief Zone.} We use Leiden $\gamma=0.002$, giving a nine-community Pareto front $\{7, 24, 18, 6, 16, 32, 50, 34, 49\}$, ordered by increasing network distance (and hence, along the front, increasing propensity distance). We retain the central five, $\{18, 6, 16, 32, 50\}$: extending outward from the midpoint reaches a workable donor count while trimming both extremes: communities 7 and 24, closest in network distance and at risk of spillover, and communities 34 and 49, farthest and demographically too dissimilar. Figure~\ref{supp:fig:crz_maps_pareto} shows the map and Pareto scatter.

\begin{figure}[t!]
    \centering
    \includegraphics[width=\textwidth]{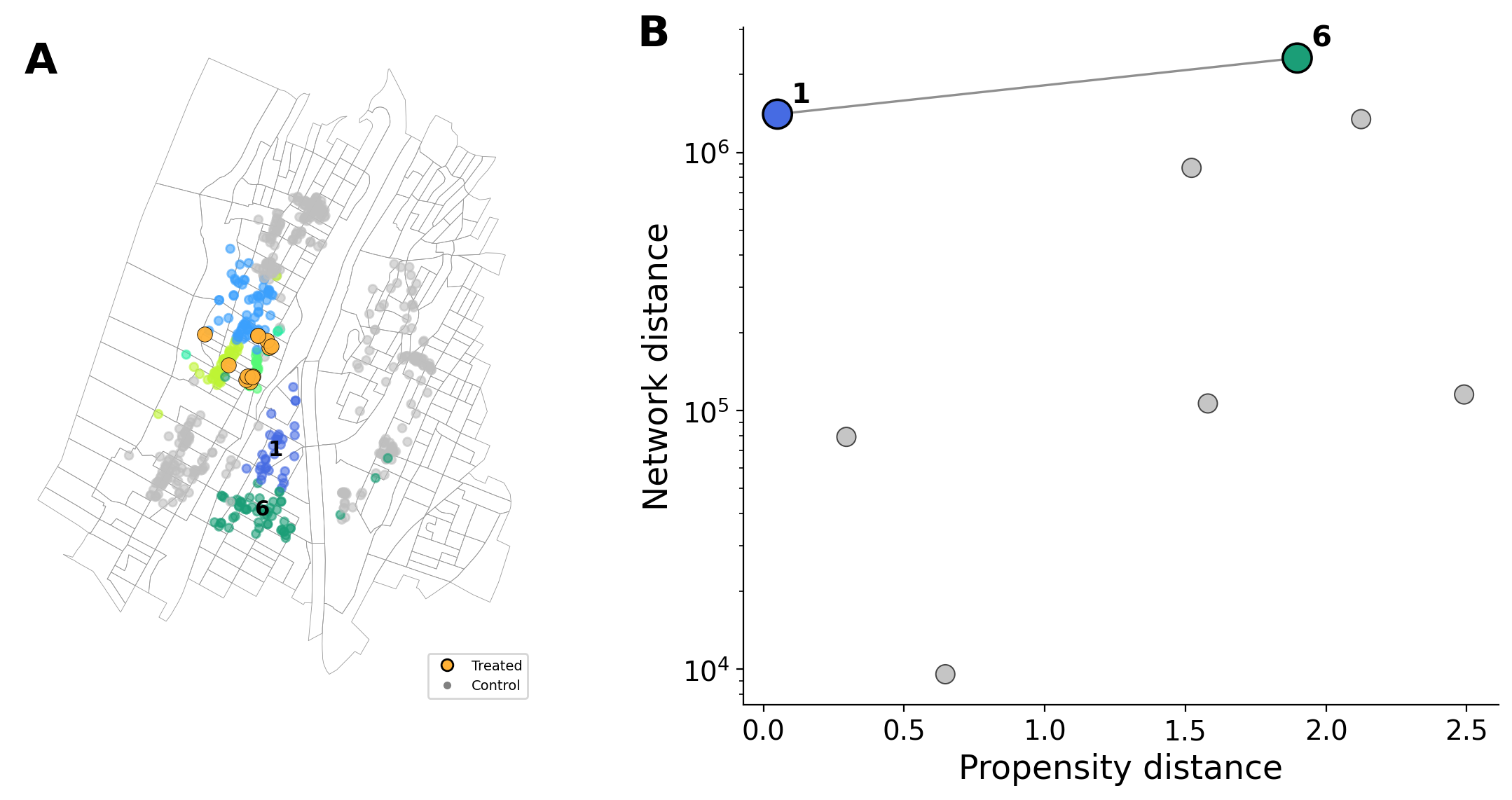}
    \caption[\textbf{CLH: selected donor community.}]{\textbf{CLH: selected donor community.} \textbf{(A)}~Consensus community map at Leiden $\gamma=2.0$; the main-text donor (community~6) is colored donor-green. \textbf{(B)}~Pareto scatter of candidate donor communities; the chosen donor is green.}
    \label{supp:fig:clh_maps_pareto}
\end{figure}

\paragraph{Community League of the Heights (CLH).} We use Leiden $\gamma=2.0$. At this resolution the Pareto front consists of only two communities, $\{1,\,6\}$, ordered by increasing network distance. With no interior community available, the midpoint rule does not select between them, so we choose on spillover risk: community 1 sits at the low-network-distance end and is the more exposed to the treated region, so we retain community 6. This trades demographic similarity for behavioral insulation, but it protects the control pool from the spillover that biases the buffer-based estimate, and the resulting effect is both significant and stable across resolutions and detection algorithm (Table \ref{supp:tab:gdm_os}, Fig. \ref{supp:fig:comm_sensitivity}). Figure~\ref{supp:fig:clh_maps_pareto} shows the map and Pareto scatter.

\begin{figure}[t!]
    \centering
    \includegraphics[width=\textwidth]{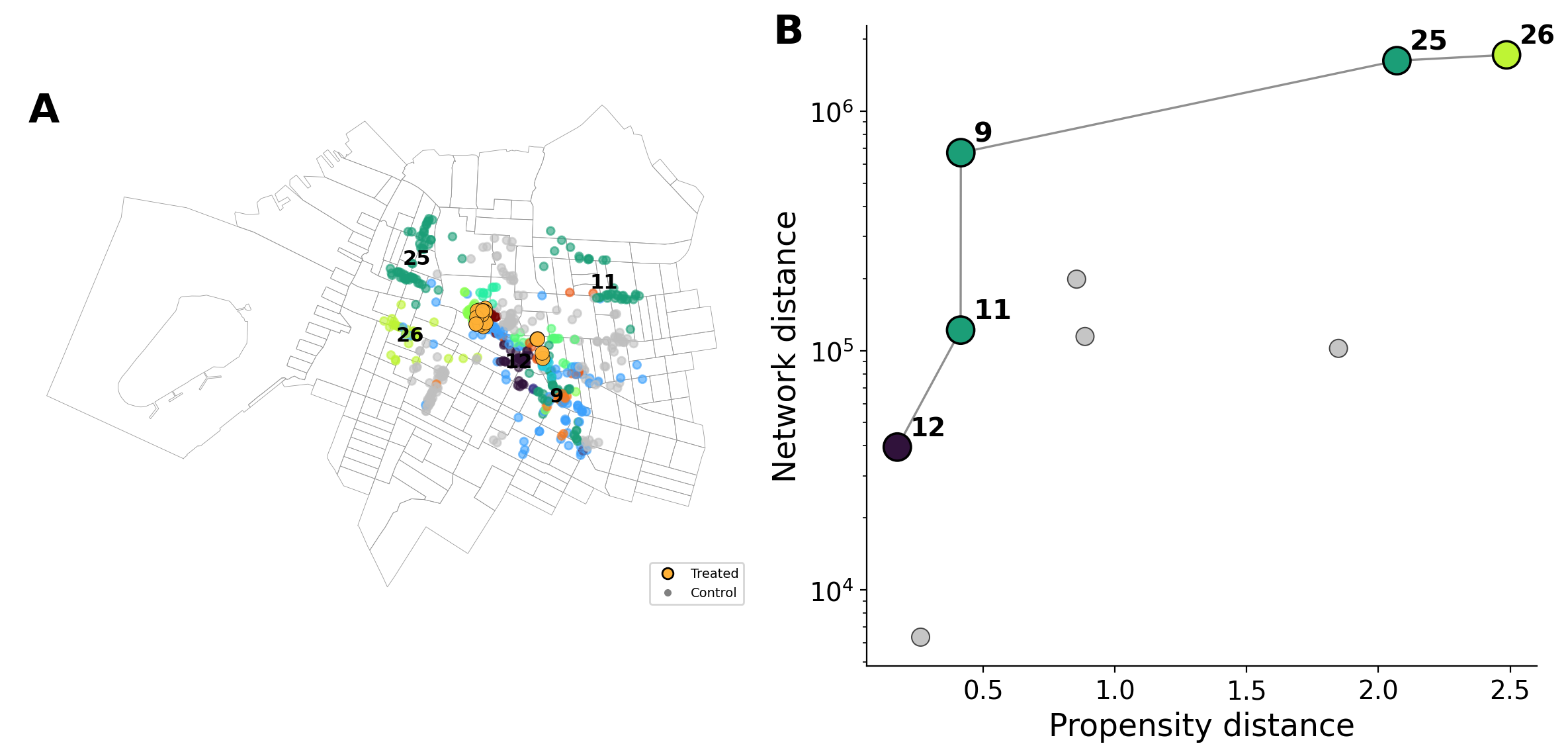}
    \caption[\textbf{DBP: selected donor communities.}]{\textbf{DBP: selected donor communities.} \textbf{(A)}~Consensus community map at Leiden $\gamma=2.0$; the main-text donors $\{9,11,25\}$ are in donor-green. \textbf{(B)}~Pareto scatter of candidate donor communities; the chosen donors are green.}
    \label{supp:fig:dbp_maps_pareto}
\end{figure}

\paragraph{Downtown Brooklyn Partnership (DBP).} We use Leiden $\gamma=2.0$. At this resolution the Pareto front is $\{12,\,11,\,9,\,25,\,26\}$, ordered by increasing network distance. Community~12 sits at the lowest network and propensity distance and risks spillover; community~26 is the most demographically dissimilar. We retain the central three $\{9,\,11,\,25\}$: extending outward from the midpoint (community 9) trims both ends while reaching a donor count large enough for stable placebo splits. Figure~\ref{supp:fig:dbp_maps_pareto} shows the map and Pareto scatter.

\begin{figure}[t!]
    \centering
    \includegraphics[width=\textwidth]{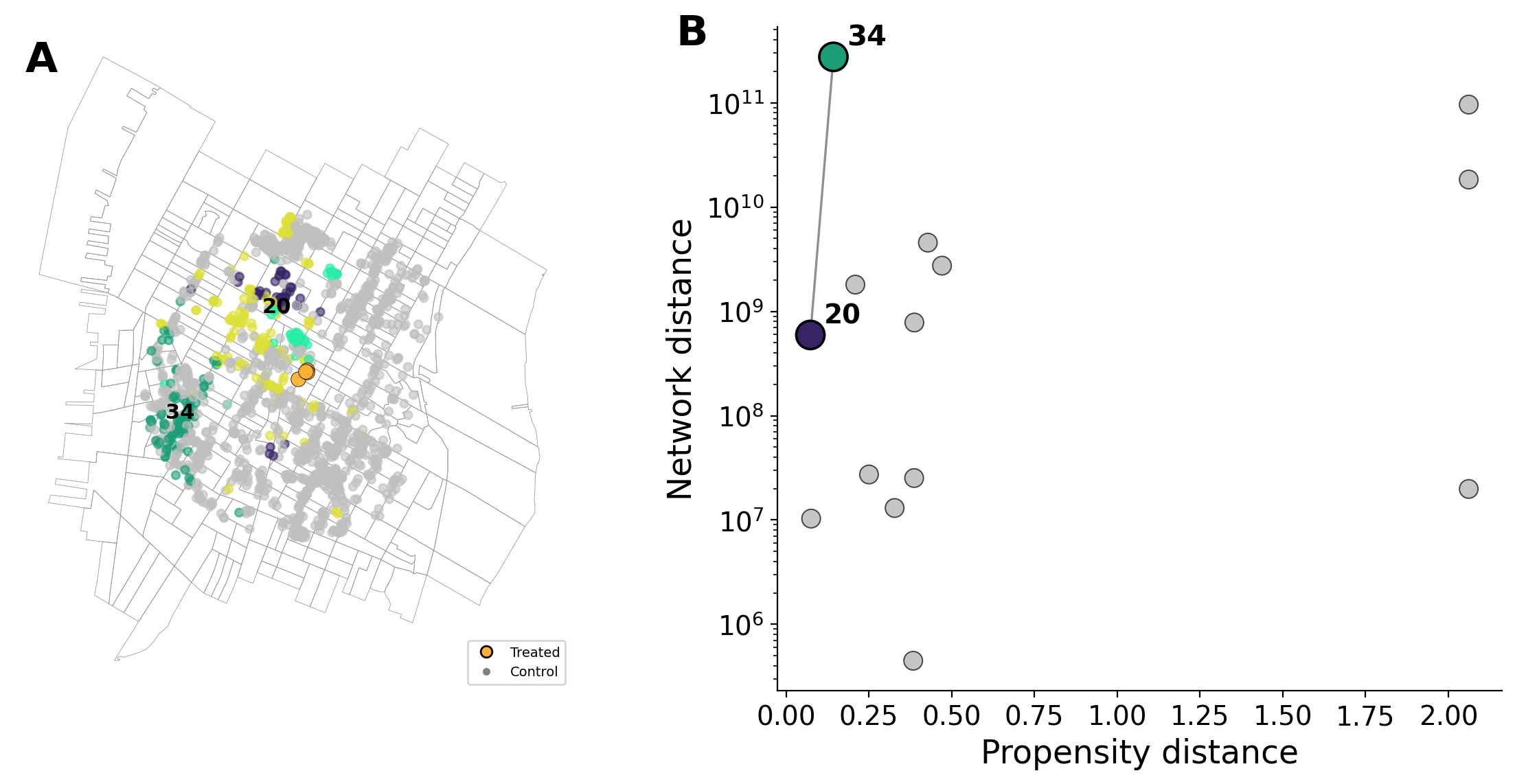}
    \caption[\textbf{WIHS: selected donor community.}]{\textbf{WIHS: selected donor community.} \textbf{(A)}~Consensus community map at Leiden $\gamma=2.0$; the main-text donor (community~34) is colored donor-green. \textbf{(B)}~Pareto scatter of candidate donor communities; the chosen donor is green.}
    \label{supp:fig:wihs_maps_pareto}
\end{figure}

\paragraph{NYCDOE / Washington Irving HS (WIHS).} We use Leiden $\gamma=2.0$. At this resolution the Pareto front consists of only two communities, $\{20,\,34\}$, ordered by increasing network distance. Community~20 has the lower propensity distance but the lower network distance, and is therefore at the higher risk of spillover. We retain community~34 as the main-text donor: it lies at the high-network-distance end of the front, with a propensity distance ($\approx 0.14$) only marginally larger than community~20's ($\approx 0.07$). Figure~\ref{supp:fig:wihs_maps_pareto} shows the map and Pareto scatter.

\section{Per-intervention GDM and NDM results and donor-pool balance} \label{supp:sec:gdm_ndm_results}

Tables~\ref{supp:tab:free_fare_summary}, \ref{supp:tab:nyos_summary}, and~\ref{supp:tab:crz_summary} summarize the estimation samples for each intervention: unit counts, POI-category cells, and mean weekly visits for each method's treatment and control pools. In this section, we report the per-intervention GDM and NDM ATEs side by side. Tables~\ref{supp:tab:gdm_ate_ff}--\ref{supp:tab:gdm_ate_crz} summarize, per intervention, the per-taxonomy SCM-DiD ATE and its standard error (in brackets) under both the Geographical Distance Method and the main-text Network Distance Method donor pool. ATEs for visits are reported as percent changes; ATEs for spend are reported in dollars per unit (treated unit-period); asterisks denote empirical placebo significance. For Boston Free Fare, the companion table (Table~\ref{supp:tab:gdm_demo_ff}) reports the donor-vs-treated demographic balance: the median and interquartile range (IQR) per pool, plus the propensity distance between treated and donor units in standardized demographic feature space; the same demographics block appears in the corresponding tables for NYC Open Streets and the CRZ. Propensity distance is the absolute difference between the donor-pool and treated-pool means of the logit propensity score (Section~\ref{supp:sec:propensity}); for NDM with a multi-community donor set we report the mean across constituent communities.

\begin{itemize}
\item Table~\ref{supp:tab:free_fare_summary} summarizes the Free Fare dataset. GDM and NDM operate on different treated tract sets (GDM treats the 40 tracts within 1\,km of Route~28, while NDM treats the 159 tracts of the Leiden community or communities that contain those treated tracts), so we list each method's treatment pool alongside its control pool.
\item Table~\ref{supp:tab:nyos_summary} summarizes the Open Streets dataset across the three sites. Units are individual POIs. GDM and NDM operate on different treated POI sets (GDM uses POIs within a fixed buffer of the open street; NDM uses the POIs in the Leiden community/communities that contain the open street), so we list each method's treatment pool alongside its control pool.
\item Table~\ref{supp:tab:crz_summary} summarizes the CRZ dataset. GDM and NDM operate on slightly different treated tract sets (GDM uses the 136 tracts inside the CRZ boundary; NDM treats the 162 tracts in the Leiden communities that contain those tracts), so we list each method's treatment pool alongside its control pool.
\end{itemize}

\begin{table}[tbp!]
\centering
\caption[\textbf{Free Fare: dataset summary.}]{\textbf{Free Fare: dataset summary.} Unit counts, weekly mean post-stratified visits (the SCM outcome), and POI-category cells for each method's treatment and control pools. POI-category cells count unique geoid and taxonomy pairs observed in the visits panel. GDM control counts refer to the ring donors retained after the demographic propensity-score filter described in Section~\ref{supp:sec:GDM}. Pre = 2020-06-01 to 2021-08-29, Post = 2021-08-29 to 2022-03-01.}
\label{supp:tab:free_fare_summary}
\small
\begin{tabular}{lrrrr}
\toprule
 & GDM Treatment & GDM control & NDM Treatment & NDM control \\
\midrule
Number of census tracts           & 40        & 62          & 159         & 64          \\
POI-category cells                & 956       & 1{,}483     & 3{,}167     & 1{,}277     \\
Mean weekly visits (pre)          & 415{,}771 & 625{,}419   & 1{,}554{,}894 & 1{,}185{,}864 \\
Mean weekly visits (post)         & 628{,}319 & 977{,}590   & 2{,}273{,}256 & 1{,}643{,}414 \\
\bottomrule
\end{tabular}
\end{table}

\begin{table}[tbp!]
\centering
\caption[\textbf{NYC Open Streets: dataset summary.}]{\textbf{NYC Open Streets: dataset summary.} Per-site POI counts and mean weekly post-stratified visits (the SCM outcome) for each method's treatment and control pools. Pre-period ends at each site's treatment start: CLH 2021-07-01, DBP 2021-06-17, WIHS 2021-09-13; post-periods span each site's documented open-street window: CLH through 2021-08-31, DBP through 2021-11-27, WIHS through 2021-12-23.}
\label{supp:tab:nyos_summary}
\small
\begin{tabular}{lrrrr}
\toprule
 & GDM Treatment & GDM control & NDM Treatment & NDM control \\
\midrule
\multicolumn{5}{l}{\textit{Community League of the Heights (CLH)}} \\
Number of POIs                  & 20      & 141      & 20      & 50      \\
Mean weekly visits (pre)        & 3{,}179 & 12{,}766 & 3{,}533 & 14{,}239 \\
Mean weekly visits (post)       & 4{,}500 & 15{,}997 & 4{,}922 & 17{,}354 \\
\midrule
\multicolumn{5}{l}{\textit{Downtown Brooklyn Partnership (DBP)}} \\
Number of POIs                  & 41      & 132       & 41       & 110      \\
Mean weekly visits (pre)        & 3{,}058 & 41{,}857  & 5{,}123  & 19{,}310 \\
Mean weekly visits (post)       & 5{,}161 & 65{,}162  & 10{,}341 & 28{,}474 \\
\midrule
\multicolumn{5}{l}{\textit{NYCDOE / Washington Irving HS (WIHS)}} \\
Number of POIs                  & 13      & 233       & 11      & 78       \\
Mean weekly visits (pre)        & 702     & 10{,}514  & 737     & 12{,}258 \\
Mean weekly visits (post)       & 2{,}101 & 18{,}503  & 2{,}156 & 21{,}470 \\
\bottomrule
\end{tabular}
\end{table}

\begin{table}[htbp]
\centering
\caption[\textbf{CRZ: dataset summary.}]{\textbf{CRZ: dataset summary.} Census-tract counts, POI-category cells, and mean weekly visits for each method's treatment and control pools. POI-category cells count unique geoid and taxonomy pairs observed in the visits panel. GDM control counts refer to the ring donors retained after the demographic propensity-score filter described in Section~\ref{supp:sec:GDM}. Pre = 2023-05-01 to 2025-01-05, Post = 2025-01-05 to 2025-07-21.}
\label{supp:tab:crz_summary}
\small
\begin{tabular}{lrrrr}
\toprule
 & GDM Treatment & GDM control & NDM Treatment & NDM control \\
\midrule
Number of census tracts           & 136       & 381       & 162       & 215       \\
POI-category cells                & 732       & 1{,}748   & 846       & 960       \\
Mean weekly visits (pre)          & 14.05M    & 13.64M    & 14.12M    & 6.61M     \\
Mean weekly visits (post)         & 13.98M    & 13.38M    & 14.11M    & 6.50M     \\
\bottomrule
\end{tabular}
\end{table}

\begin{table}[tbp]
\centering
\caption[\textbf{GDM and NDM results: Boston Free Fare.}]{\textbf{GDM and NDM results: Boston Free Fare.} (A) Average treatment effect (\%) and standard error (in brackets) from SCM-DiD under the Geographical Distance Method (GDM) and the Network Distance Method (NDM). ATEs for visits are reported as percent changes; ATEs for spend are reported in dollars per unit (treated unit-period). The placebo 5--95\% column reports the 5th--95th percentile range of the placebo ATE distribution (in the same units as the ATE); standard errors describe the precision of the gap regression and are not used for inference. Asterisks denote empirical placebo significance ($^{\ast}p<0.05$, $^{\ast\ast}p<0.01$). Dashes ($-$) mark income-quartile cells that cannot be estimated. A quartile ATE is computed only when the treated group contains units in that income quartile: the Route~28 treated area falls almost entirely in the lower income quartiles, so the highest quartile (Q4) is absent from the treated group under both designs, and Q2 is additionally absent under the smaller GDM treated set. Where a quartile is present but too sparse to support placebo relabeling, the empirical $p$-value is undefined and no placebo range is shown (NDM Q3). (B) Donor-vs-treated demographic balance: median value with interquartile range (IQR; in brackets, $\text{Q}_{3}-\text{Q}_{1}$) per pool. Propensity distance is the absolute difference between the donor-pool and treated-pool means of the logit propensity score (Section~\ref{supp:sec:propensity}); for NDM the value is the mean across constituent communities.}
\label{supp:tab:gdm_ff}

\renewcommand{\thesubtable}{\Alph{subtable}}

\begin{subtable}[t]{\linewidth}
\centering
\subcaption{ATE results}
\label{supp:tab:gdm_ate_ff}
\footnotesize
\setlength{\tabcolsep}{3.5pt}
\begin{tabular}{llrrrr}
\toprule
 & Taxonomy / Variable & GDM ATE (\%) [SE] & GDM placebo 5--95\% & NDM ATE (\%) [SE] & NDM placebo 5--95\% \\
\midrule
\multicolumn{6}{l}{\textit{Visits}} \\
 & Outdoors           & $-17.81\,[1.34]^{\ast\ast}$ & $[-12.6,\,13.8]$ & $1.22\,[0.92]$ & $[-12.5,\,15.8]$ \\
 & Cafe               & $2.90\,[1.12]$ & $[-6.8,\,8.0]$ & $2.27\,[0.63]$ & $[-7.0,\,7.1]$ \\
 & College            & $-24.30\,[1.44]^{\ast}$ & $[-15.4,\,21.9]$ & $11.87\,[1.10]$ & $[-12.8,\,16.1]$ \\
 & Food               & $-2.82\,[0.66]$ & $[-5.3,\,7.5]$ & $8.02\,[2.09]^{\ast}$ & $[-6.3,\,5.0]$ \\
 & Grocery            & $-6.75\,[1.29]$ & $[-7.5,\,8.0]$ & $-1.65\,[0.62]$ & $[-11.5,\,9.8]$ \\
 & Shop               & $-6.37\,[0.82]$ & $[-5.6,\,6.6]$ & $3.10\,[0.43]$ & $[-12.2,\,11.6]$ \\
 & Transit            & $0.55\,[0.82]$ & $[-9.1,\,7.8]$ & $16.39\,[0.51]^{\ast\ast}$ & $[-9.4,\,6.6]$ \\
 & Segregation        & $5.09\,[0.52]$ & $[-8.0,\,1.9]$ & $-3.74\,[0.35]$ & $[-5.9,\,6.6]$ \\
 & Exploration        & $-3.38\,[0.53]$ & $[-5.2,\,4.9]$ & $10.77\,[0.33]^{\ast}$ & $[-8.4,\,6.7]$ \\
 & Q1                 & $0.45\,[0.66]$ & $[-11.1,\,20.5]$ & $23.00\,[0.89]^{\ast\ast}$ & $[-12.2,\,14.6]$ \\
 & Q2                 & $-$ & $-$ & $-0.72\,[0.62]$ & $[-7.4,\,6.6]$ \\
 & Q3                 & $-13.16\,[1.84]^{\ast}$ & $[-5.2,\,8.1]$ & $-1.53\,[1.01]$ & $-$ \\
 & Q4                 & $-$ & $-$ & $-$ & $-$ \\
 & Total              & $1.94\,[0.63]$ & $[-5.9,\,6.4]$ & $15.65\,[0.55]^{\ast\ast}$ & $[-8.8,\,9.4]$ \\
\multicolumn{6}{l}{\textit{Spend (\$ per unit)}} \\
 & Food               & $-223\,[43]$ & $[-272,\,284]$ & $-50\,[67]$ & $[-526,\,161]$ \\
 & Grocery            & $-167\,[72]$ & $[-311,\,389]$ & $-33\,[36]$ & $[-408,\,287]$ \\
 & Shop               & $-404\,[172]$ & $[-540,\,176]$ & $-663\,[86]$ & $[-1{,}046,\,328]$ \\
 & Total              & $-478\,[57]$ & $[-435,\,406]$ & $-278\,[60]$ & $[-323,\,793]$ \\
\bottomrule
\end{tabular}
\end{subtable}

\vspace{1em}

\begin{subtable}[t]{\linewidth}
\centering
\subcaption{Donor-vs-treated demographic balance}
\label{supp:tab:gdm_demo_ff}
\small
\begin{tabular}{lrrrr}
\toprule
 & \multicolumn{2}{c}{GDM} & \multicolumn{2}{c}{NDM} \\
\cmidrule(lr){2-3} \cmidrule(lr){4-5}
Variable & Donor & Treated & Donor & Treated \\
\midrule
Median income (\$)         & $85{,}324\,[34{,}426]$  & $45{,}526\,[25{,}354]$ & $75{,}996\,[20{,}439]$ & $85{,}035\,[64{,}270]$ \\
Proportion Black           & $0.072\,[0.089]$        & $0.508\,[0.513]$       & $0.051\,[0.073]$        & $0.123\,[0.390]$ \\
Proportion Bachelor's+     & $0.585\,[0.418]$        & $0.269\,[0.302]$       & $0.317\,[0.204]$        & $0.559\,[0.514]$ \\
Public-transit share       & $0.244\,[0.128]$        & $0.309\,[0.104]$       & $0.233\,[0.187]$        & $0.267\,[0.127]$ \\
\midrule
Propensity distance        & \multicolumn{2}{c}{$4.88$} & \multicolumn{2}{c}{$2.34$} \\
\bottomrule
\end{tabular}
\end{subtable}
\end{table}

\begin{table}[htbp]
\centering
\caption[\textbf{GDM and NDM results: NYC Open Streets.}]{\textbf{GDM and NDM results: NYC Open Streets.} (A) Average treatment effect (\%) and standard error (in brackets) from SCM-DiD under GDM and NDM for each of the three sites (CLH, DBP, WIHS). The placebo 5--95\% column reports the 5th--95th percentile range of the placebo ATE distribution; for Open Streets the placebo draws are made on the total-visits panel of each site, so the range is shared across categories within a site. Standard errors describe the precision of the gap regression and are not used for inference. Asterisks denote empirical placebo significance ($^{\ast}p<0.05$, $^{\ast\ast}p<0.01$). (B) Donor-vs-treated demographic balance per pool: visitor-weighted POI-level median household income and visitor-weighted White and Black population shares, with interquartile ranges (IQR; in brackets, $\text{Q}_{3}-\text{Q}_{1}$). Propensity distance is the absolute difference between the donor-pool and treated-pool means of the logit propensity score on standardized visitor-weighted demographic features (Section~\ref{supp:sec:propensity}).}
\label{supp:tab:gdm_os}

\renewcommand{\thesubtable}{\Alph{subtable}}

\begin{subtable}[t]{\linewidth}
\centering
\subcaption{ATE results}
\label{supp:tab:gdm_ate_os}
\footnotesize
\setlength{\tabcolsep}{3.5pt}
\begin{tabular}{llrrrr}
\toprule
 & Taxonomy / Variable & GDM ATE (\%) [SE] & GDM placebo 5--95\% & NDM ATE (\%) [SE] & NDM placebo 5--95\% \\
\midrule
\multicolumn{6}{l}{\textit{CLH}} \\
 & Cafe               & $8.94\,[15.58]$ & $[-14.0,\,13.8]$ & $-$ & $-$ \\
 & Food               & $30.65\,[7.42]^{\ast\ast}$ & $[-14.0,\,13.8]$ & $41.94\,[7.75]^{\ast\ast}$ & $[-15.7,\,23.1]$ \\
 & Shop               & $8.76\,[4.51]$ & $[-14.0,\,13.8]$ & $26.24\,[6.24]^{\ast}$ & $[-15.7,\,23.1]$ \\
 & Total              & $15.72\,[3.89]$ & $[-14.0,\,13.8]$ & $32.60\,[4.86]^{\ast}$ & $[-15.7,\,23.1]$ \\
\multicolumn{6}{l}{\textit{DBP}} \\
 & Cafe               & $-5.93\,[4.26]$ & $[-18.0,\,10.5]$ & $18.52\,[5.60]^{\ast}$ & $[-14.9,\,14.0]$ \\
 & Food               & $6.17\,[2.88]$ & $[-18.0,\,10.5]$ & $24.74\,[3.11]^{\ast\ast}$ & $[-14.9,\,14.0]$ \\
 & Shop               & $-20.80\,[3.48]^{\ast}$ & $[-18.0,\,10.5]$ & $23.66\,[3.93]^{\ast\ast}$ & $[-14.9,\,14.0]$ \\
 & Total              & $-6.71\,[2.02]$ & $[-18.0,\,10.5]$ & $23.50\,[2.25]^{\ast\ast}$ & $[-14.9,\,14.0]$ \\
\multicolumn{6}{l}{\textit{WIHS}} \\
 & Cafe               & $6.61\,[16.49]$ & $[-38.6,\,12.6]$ & $-30.64\,[14.40]^{\ast\ast}$ & $[-19.7,\,16.5]$ \\
 & Food               & $18.39\,[3.91]$ & $[-38.6,\,12.6]$ & $31.92\,[5.11]^{\ast\ast}$ & $[-19.7,\,16.5]$ \\
 & Shop               & $-54.43\,[10.20]^{\ast\ast}$ & $[-38.6,\,12.6]$ & $-38.84\,[19.25]^{\ast\ast}$ & $[-19.7,\,16.5]$ \\
 & Total              & $12.42\,[4.15]$ & $[-38.6,\,12.6]$ & $27.27\,[5.25]^{\ast}$ & $[-19.7,\,16.5]$ \\
\bottomrule
\end{tabular}
\end{subtable}

\vspace{1em}

\begin{subtable}[t]{\linewidth}
\centering
\subcaption{Donor-vs-treated demographic balance}
\label{supp:tab:gdm_demo_os}
\small
\begin{tabular}{lrrrr}
\toprule
 & \multicolumn{2}{c}{GDM} & \multicolumn{2}{c}{NDM} \\
\cmidrule(lr){2-3} \cmidrule(lr){4-5}
Variable & Donor & Treated & Donor & Treated \\
\midrule
\multicolumn{5}{l}{\textit{CLH}} \\
Median income (\$)         & $53{,}367\,[8{,}898]$ & $55{,}320\,[6{,}536]$ & $63{,}954\,[6{,}719]$ & $55{,}395\,[6{,}822]$ \\
Proportion White           & $0.251\,[0.061]$ & $0.248\,[0.049]$ & $0.259\,[0.061]$ & $0.248\,[0.049]$ \\
Proportion Black           & $0.237\,[0.109]$ & $0.279\,[0.057]$ & $0.417\,[0.089]$ & $0.284\,[0.057]$ \\
Propensity distance        & \multicolumn{2}{c}{$0.02$} & \multicolumn{2}{c}{$1.90$} \\
\midrule
\multicolumn{5}{l}{\textit{DBP}} \\
Median income (\$)         & $77{,}522\,[11{,}220]$ & $69{,}721\,[13{,}962]$ & $84{,}467\,[27{,}156]$ & $71{,}855\,[17{,}072]$ \\
Proportion White           & $0.357\,[0.087]$ & $0.343\,[0.108]$ & $0.394\,[0.171]$ & $0.349\,[0.106]$ \\
Proportion Black           & $0.328\,[0.135]$ & $0.332\,[0.130]$ & $0.303\,[0.162]$ & $0.328\,[0.113]$ \\
Propensity distance        & \multicolumn{2}{c}{$0.27$} & \multicolumn{2}{c}{$0.92$} \\
\midrule
\multicolumn{5}{l}{\textit{WIHS}} \\
Median income (\$)         & $97{,}113\,[19{,}036]$ & $94{,}631\,[31{,}308]$ & $106{,}405\,[1{,}610]$ & $106{,}589\,[25{,}029]$ \\
Proportion White           & $0.483\,[0.055]$ & $0.488\,[0.160]$ & $0.489\,[0.007]$ & $0.535\,[0.135]$ \\
Proportion Black           & $0.162\,[0.027]$ & $0.161\,[0.099]$ & $0.174\,[0.035]$ & $0.134\,[0.082]$ \\
Propensity distance        & \multicolumn{2}{c}{$0.04$} & \multicolumn{2}{c}{$0.14$} \\
\bottomrule
\end{tabular}
\end{subtable}
\end{table}

\begin{table}[htbp]
\centering
\caption[\textbf{GDM and NDM results: NYC Congestion Relief Zone.}]{\textbf{GDM and NDM results: NYC Congestion Relief Zone.} (A) Average treatment effect (\%) and standard error (in brackets) from SCM-DiD under GDM and NDM. ATEs for visits are reported as percent changes; spend rows are reported in dollars per unit (treated unit-period) using visits $\times$ spend-per-customer (VxS). The placebo 5--95\% column reports the 5th--95th percentile range of the placebo ATE distribution (in the same units as the ATE); standard errors describe the precision of the gap regression and are not used for inference. Asterisks denote empirical placebo significance ($^{\ast}p<0.05$, $^{\ast\ast}p<0.01$). (B) Donor-vs-treated demographic balance: median value with interquartile range (IQR; in brackets, $\text{Q}_{3}-\text{Q}_{1}$) per pool. Propensity distance is the absolute difference between the donor-pool and treated-pool means of the logit propensity score (Section~\ref{supp:sec:propensity}); for NDM the value is the mean across constituent communities.}
\label{supp:tab:gdm_crz}

\renewcommand{\thesubtable}{\Alph{subtable}}

\begin{subtable}[t]{\linewidth}
\centering
\subcaption{ATE results}
\label{supp:tab:gdm_ate_crz}
\footnotesize
\setlength{\tabcolsep}{3.5pt}
\begin{tabular}{llrrrr}
\toprule
 & Taxonomy / Variable & GDM ATE (\%) [SE] & GDM placebo 5--95\% & NDM ATE (\%) [SE] & NDM placebo 5--95\% \\
\midrule
\multicolumn{6}{l}{\textit{Visits}} \\
 & Cafe               & $-4.76\,[0.84]$ & $[-7.5,\,7.8]$ & $-9.60\,[0.71]$ & $[-10.7,\,5.0]$ \\
 & Food               & $-5.00\,[0.42]^{\ast}$ & $[-2.2,\,2.8]$ & $-2.00\,[0.35]$ & $[-3.6,\,6.3]$ \\
 & Grocery            & $-7.95\,[0.84]^{\ast}$ & $[-3.5,\,2.6]$ & $-7.87\,[0.84]^{\ast\ast}$ & $[-3.4,\,5.0]$ \\
 & Shop               & $-7.73\,[0.86]^{\ast\ast}$ & $[-4.9,\,2.5]$ & $-8.54\,[0.76]^{\ast}$ & $[-6.3,\,4.1]$ \\
 & Transit            & $-0.62\,[1.71]$ & $[-5.9,\,4.2]$ & $-0.25\,[1.38]$ & $[-3.4,\,7.3]$ \\
 & Segregation        & $6.15\,[1.35]^{\ast\ast}$ & $[-0.7,\,0.8]$ & $4.20\,[1.11]^{\ast\ast}$ & $[-1.5,\,2.2]$ \\
 & Q1                 & $-3.40\,[0.33]$ & $[-3.5,\,2.1]$ & $-4.31\,[0.46]^{\ast}$ & $[-3.1,\,3.2]$ \\
 & Q2                 & $-11.31\,[0.39]^{\ast\ast}$ & $[-3.3,\,2.0]$ & $-0.29\,[0.44]$ & $[-2.4,\,2.9]$ \\
 & Q3                 & $-2.87\,[0.29]$ & $[-4.2,\,1.1]$ & $1.08\,[0.35]$ & $[-3.5,\,1.7]$ \\
 & Q4                 & $0.31\,[0.32]$ & $[-6.2,\,2.6]$ & $3.12\,[0.30]$ & $[-3.0,\,3.2]$ \\
 & Total              & $-6.46\,[0.39]^{\ast}$ & $[-5.2,\,1.0]$ & $-1.51\,[0.30]$ & $[-1.7,\,1.7]$ \\
\multicolumn{6}{l}{\textit{Spend (\$ per unit, VxS)}} \\
 & Food               & $-35{,}777\,[10{,}201]$ & $[-16{,}699,\,8{,}043]$ & $-20{,}181\,[7{,}717]$ & $[-40{,}363,\,13{,}369]$ \\
 & Grocery            & $-5{,}556\,[1{,}070]$ & $[-17{,}758,\,13{,}940]$ & $1{,}069\,[1{,}166]$ & $[-16{,}919,\,18{,}599]$ \\
 & Total              & $-466{,}241\,[34{,}560]^{\ast}$ & $[-82{,}102,\,63{,}089]$ & $70{,}902\,[13{,}662]$ & $[-64{,}897,\,43{,}806]$ \\
\bottomrule
\end{tabular}
\end{subtable}

\vspace{1em}

\begin{subtable}[t]{\linewidth}
\centering
\subcaption{Donor-vs-treated demographic balance}
\label{supp:tab:gdm_demo_crz}
\small
\begin{tabular}{lrrrr}
\toprule
 & \multicolumn{2}{c}{GDM} & \multicolumn{2}{c}{NDM} \\
\cmidrule(lr){2-3} \cmidrule(lr){4-5}
Variable & Donor & Treated & Donor & Treated \\
\midrule
Median income (\$)         & $64{,}310\,[34{,}354]$ & $137{,}882\,[68{,}982]$ & $74{,}167\,[33{,}597]$ & $139{,}294\,[65{,}432]$ \\
Proportion Black           & $0.124\,[0.396]$       & $0.038\,[0.056]$        & $0.019\,[0.054]$       & $0.030\,[0.046]$ \\
Proportion White           & $0.284\,[0.354]$       & $0.687\,[0.198]$        & $0.526\,[0.265]$       & $0.727\,[0.199]$ \\
Public-transit share       & $0.560\,[0.242]$       & $0.416\,[0.130]$        & $0.360\,[0.184]$       & $0.436\,[0.135]$ \\
\midrule
Propensity distance        & \multicolumn{2}{c}{$2.67$} & \multicolumn{2}{c}{$1.22$} \\
\bottomrule
\end{tabular}
\end{subtable}
\end{table}

\paragraph{Boston Free Fare.}
Table~\ref{supp:tab:free_fare_summary} summarizes the dataset. GDM and NDM operate on different treated sets: GDM treats the 40 tracts within 1\,km of Route~28, NDM the 159 tracts of the Leiden community or communities containing them, so each method's treatment pool is listed alongside its control pool. GDM draws controls from the 2--8\,km ring. Under GDM, SCM-DiD finds no significant change in total visitation ($+1.94\%\,[0.63]$, Table~\ref{supp:tab:gdm_ate_ff}; Fig.~\ref{fig2}B--D); the significant visit-side effects are wrong-signed declines in Outdoors ($-17.81\%$, $p<0.01$) and College ($-24.30\%$, $p<0.05$), the latter a category whose ring-based donor pool is dominated by university-area tracts rather than tracts comparable to those served by Route~28. The null result on total visits is consistent with spillover: tracts in the 2--8\,km ring remain indirectly exposed through mobility connections to the same destinations that Route~28 riders visit, attenuating the estimated effect.

NDM replaces the geographic ring with a mobility-network donor pool that is
behaviorally distant from Route~28. The donor pool is also closer to its treated
pool on observed demographics than the geographic ring is to its own
(Table~\ref{supp:tab:gdm_demo_ff}). Each design must be read against its own
treated set, since GDM treats the 40 tracts within 1\,km of Route~28 and NDM the
159 tracts of the containing community. Under GDM the donor median income exceeds
the treated median by nearly \$$40{,}000$ ($85{,}324$ vs $45{,}526$), the
Bachelor's-or-higher share by $0.32$ ($0.585$ vs $0.269$), and the Black-population
share falls short by $0.44$ ($0.072$ vs $0.508$). Under NDM the corresponding gaps
are $-\$9{,}039$ ($75{,}996$ vs $85{,}035$), $-0.24$ ($0.317$ vs $0.559$), and
$-0.07$ ($0.051$ vs $0.123$). The public-transit commute share is close to its
treated pool under both designs ($0.244$ vs $0.309$; $0.233$ vs $0.267$).
Aggregating these features into a single logit-propensity score, the propensity
distance to the treated pool falls from $4.88$ (GDM) to $2.34$ (NDM). Under this
donor pool, total visitation rises by $+15.65\%$ (SE~$0.55$, placebo $p<0.01$), with
positive effects on Transit ($+16.39\%$, $p<0.01$), Food ($+8.02\%$, $p<0.05$), and
Q1-quartile visitation ($+23.00\%$, $p<0.01$). The reversal between GDM and NDM is
the signature of network spillover combined with weak demographic matching in the
geographic ring; NDM mitigates both.

\paragraph{New York City Open Streets.}
Table~\ref{supp:tab:nyos_summary} summarizes the three sites; units are individual POIs. GDM and NDM operate on different treated POI sets — GDM uses POIs within a fixed buffer of the open street, NDM the POIs in the Leiden community or communities containing it. Controls are drawn from an outer ring under GDM. GDM produces no significant change in total visitation at any site (CLH $+15.72\%\,[3.89]$; DBP $-6.71\%\,[2.02]$; WIHS $+12.42\%\,[4.15]$, all non-significant under our placebo test, see Table~\ref{supp:tab:gdm_ate_os}). The per-category GDM estimates are sometimes significant but mutually inconsistent: Food rises significantly at CLH ($+30.65\%$, $p<0.01$) while Shop falls significantly at DBP ($-20.80\%$, $p<0.05$) and WIHS ($-54.43\%$, $p<0.01$). This is the same buffer-ring exposure problem as Free Fare: businesses in the outer ring share footfall with the pedestrianized stretch and are partially treated.

Switching to NDM donor sets selected on mobility-community structure recovers significant and consistent positive effects at all three sites: CLH $+32.60\%$ (SE~$4.86$, placebo $p<0.05$), DBP $+23.50\%$ (SE~$2.25$, placebo $p<0.01$), and WIHS $+27.27\%$ (SE~$5.25$, placebo $p<0.05$). The network-selected donors remain demographically close to the treated POIs while being behaviorally insulated from them. At DBP and WIHS the donor and treated White and Black shares differ by less than $0.05$, and at WIHS the median incomes are nearly identical (\$$106{,}405$ vs \$$106{,}589$); the propensity distances are $0.92$ and $0.14$. CLH shows the largest departure, a higher Black share among the donor POIs ($0.417$ vs treated $0.284$) at a propensity distance of $1.90$. The GDM pools, selected on propensity distance alone, are closer still by construction ($0.02$, $0.27$, and $0.04$). The mobility-community donors therefore deliver a behaviorally insulated control pool at all three sites while keeping it demographically comparable to the treated area.

\paragraph{New York City Congestion Pricing.}
We define census tracts inside the Congestion Relief Zone as treated and tracts within a 5--15\,km ring outside the zone boundary as potential controls. GDM yields a significant overall decline of $-6.46\%$ (SE~$0.39$, placebo $p<0.05$) in visits to POIs inside the zone, with the decline concentrated in the Q2 visitor-income quartile ($-11.31\%$, $p<0.01$; Q1 $-3.40\%$, Q3 $-2.87\%$, and Q4 $+0.31\%$ are not significant) and a significant $-\$466{,}241$ (SE~$34{,}560$, placebo $p<0.05$) drop in total weekly spend per treated unit (Table~\ref{supp:tab:gdm_ate_crz}). These estimates are misleading because the 5--15\,km buffer is itself part of the catchment that loses zone-bound trips: the controls are partially treated and the decline is overstated.

NDM resolves this bias by drawing donors from mobility communities that are distant from the CRZ in the network. The network-selected donor pool also moves closer to the treated tracts on every demographic dimension we measure (Table~\ref{supp:tab:gdm_demo_crz}): median income from \$$64{,}310$ (GDM) to \$$74{,}167$ (NDM) against treated \$$139{,}294$; the Black-share gap shrinks from $0.124{-}0.038$ to $0.019{-}0.030$; the White-share gap shrinks from $0.284{-}0.687$ to $0.526{-}0.727$; and the public-transit share moves from $0.560$ (GDM) to $0.360$ (NDM) against treated $0.436$. Propensity distance falls from $2.67$ to $1.22$. Under this donor pool, the overall visitation effect is no longer significant ($-1.51\%\,[0.30]$), the income-quartile pattern collapses to a single small Q1 effect ($-4.31\%$, $p<0.05$), and the total-spend coefficient is no longer significant. The contrast between GDM and NDM here is the cleanest in the paper: the geographic-ring estimate is inflated by spillover into the buffer, while the network donor pool is both more behaviorally insulated and a better demographic match.

\paragraph{Decomposition by visitor origin.} The main-text figure of a $-8.59\%$ change in visits \textit{originating from outside} the zone is one component of a split of the in-zone visitation effect by the visitor's origin. Table~\ref{supp:tab:crz_origin_split} reports this split under both designs. Under GDM the decline is carried almost entirely by out-of-zone visitors ($-8.59\%$, SE~$0.41$, placebo $p<0.01$), while the in-zone-origin component is smaller and not significant ($-3.51\%$, SE~$0.33$); under NDM neither component is significant ($-0.54\%$ outside, $-0.33\%$ inside). The GDM out-of-zone decline reflects the same spillover contamination as the overall GDM estimate: the 5--15\,km control ring is itself composed of out-of-zone origins that lose zone-bound trips, so the out-of-zone component is the most contaminated and the network-insulated NDM donor pool removes it.

\begin{table}[t]
\centering
\caption[\textbf{CRZ in-zone visitation split by visitor origin.}]{\textbf{CRZ in-zone visitation split by visitor origin.} Total-visit SCM-DiD ATE (\%) for visits to POIs inside the Congestion Relief Zone, split by whether the visitor's home origin lies outside or inside the zone, under GDM and NDM. Standard error [SE] in brackets; significance is the two-sided empirical placebo test ($^{\ast}p<0.05$, $^{\ast\ast}p<0.01$).}
\label{supp:tab:crz_origin_split}
\small
\begin{tabular}{lrr}
\toprule
Visitor origin & GDM ATE (\%) [SE] & NDM ATE (\%) [SE] \\
\midrule
Outside zone & $-8.59\,[0.41]^{\ast\ast}$ & $-0.54\,[0.33]$ \\
Inside zone  & $-3.51\,[0.33]$ & $-0.33\,[0.37]$ \\
\bottomrule
\end{tabular}
\end{table}

\section{Geographic/travel versus network distance across metro areas}
\label{supp:sec:divergence_local}
GDM and NDM select different donor pools because network distance diverges from geographical distance: units far apart in space can be strongly connected by mobility flows, and adjacent units can be weakly connected. This section measures how far the two distances diverge, and shows that the divergence holds in every metro area and data source we examine, so the mismatch is a general feature of urban mobility rather than a property of Boston, New York, or a single dataset.

\subsection{Definition of local divergence between geographical and network distance}
To test the divergence between geographical and network distance across metro areas we use the mobility networks described in Section~\ref{supp:sec:metro_data}, including the mobility data from Cuebiq. For each metro area we compute the row-normalized transition matrix $W$ between areas (census tracts, districts, MSOAs, etc.), obtain the network distance $d_{\text{net}}[u,v]$ for every pair of areas within the city, and compare it to the Euclidean distance $d_{\text{geo}}[u,v]$.

Across cities (Fig.~\ref{fig5}B), a given geographic distance corresponds to a wide range of network distances at all ranges, so geographic distance orders network distance imperfectly throughout. Rather than summarize this with a global correlation, we focus on the local scale ($<5$\,km). This is where buffer-based designs distinguish treated from control units, and therefore where a mismatch between spatial and network ordering most directly biases the design. For each origin area $u$, let $N_u$ be the set of other areas $v$ within a fixed geographic radius of $5$\,km from $u$ (the local scale noted above). We compute the Spearman rank correlation between $d_{\text{geo}}[u,v]$ and $d_{\text{net}}[u,v]$ over $v \in N_u$, discarding origins with fewer than 10 neighbors. The metro-level summary reported in Fig.~\ref{fig5}C is the mean over origins, $\rho = \overline{\rho_u}$. A value near 1 means that areas geographically close to $u$ are also close in the mobility network. Lower values indicate local mismatches, where areas nearby in space are distant in the network and vice versa, as in the example of Fig.~\ref{fig5}A.

\begin{figure}[t!]
    \centering
    \includegraphics[width=\textwidth]{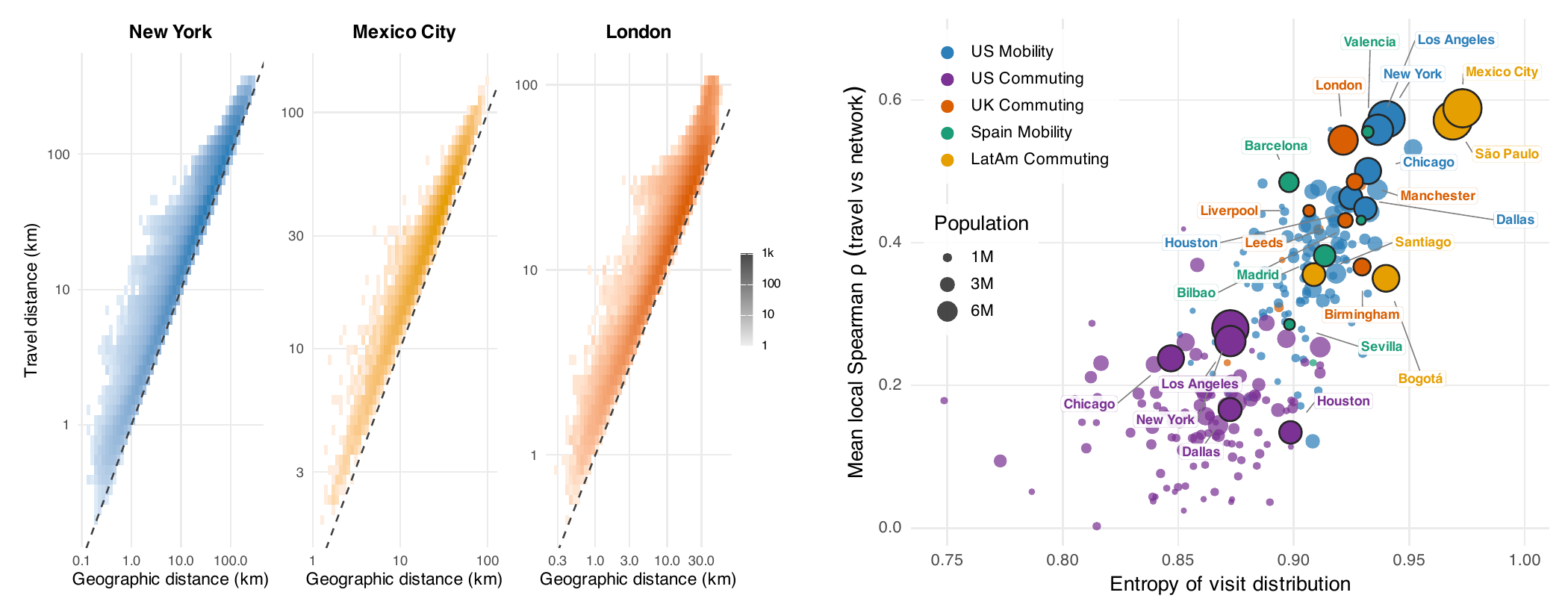}
    \caption[\textbf{Effect of geographical vs. travel distance on local divergence with network distance}]{\textbf{Effect of geographical vs. travel distance on local divergence with network distance.} Left: Scatter (binned) plots comparing travel (driving) distance between pairs of areas in different cities in our dataset. The plots show the strong correlation at different ranges of geographical distance. Right: comparison between the entropy of visits and the average local divergence $\rho$ of different metro areas in different countries using traveling distance instead of geographical distance as in Fig.\ \ref{fig5}C.}
    \label{supp:fig:geo_vs_travel}
\end{figure}

\subsection{Robustness against travel distance}
\label{supp:sec:travel_distance}
Geographic distance between centroids ignores physical routing: two areas separated by a river or a highway may be close as the crow flies but far by road. To test whether the geographic-network divergence reflects behavior rather than physical routing barriers, we repeat the analysis using travel distance in place of Euclidean distance. Travel distance $d_{\text{travel}}[u,v]$ is the driving route distance between area centroids, computed with the Open Source Routing Machine (OSRM) \cite{luxen-vetter-2011} on OpenStreetMap road networks \cite{openstreetmap} using country/region extracts from Geofabrik \cite{geofabrik}. Figure~\ref{supp:fig:geo_vs_travel} compares the two spatial proxies. The left panel shows travel distance against geographic distance for NYC, Mexico City, and London: the two are tightly related, so road routing accounts for little of the gap between spatial and network distance. The right panel reproduces the cross-metro summary of Figure~\ref{fig5}C using local Spearman $\rho$ from $d_{\text{travel}}$ instead of $d_{\text{geo}}$. The entropy--$\rho$ relationship and the ordering across data sources are unchanged, confirming that the local divergence between spatial and network distance is behavioral and not an artifact of physical routing.

\subsection{Correlates of the local divergence}
\label{supp:sec:rho_correlates}
Finally, we test statistically the relationship between the local divergence $\rho$ and properties of the urban areas (Table~\ref{supp:table:rho-models-all}). Using the mobility flows we define the entropy of visitations as
$$
\text{Entropy} = - \frac{1}{\log N}\sum_{u=1}^N p_{u} \log p_{u},
$$
where $p_u = n_u / \sum_{v} n_v$ is the fraction of visits to area $u$, and $N$ is the total number of areas in the city. For the US data we also use US Census variables associated with mobility at the metro-area level, including median household income, car share, and average commuting time. 

Table~\ref{supp:table:rho-models-all} regresses each metro's mean local $\rho$ on visit-distribution entropy and population, fit separately by data source; for the US sources we add ACS covariates (car share, commuting time, median income). Entropy is positively associated with $\rho$ in every source and significantly so in all but the small Latin-American sample, confirming that geographic distance approximates the network ordering more closely where activity is dispersed across destinations. The association is estimated most precisely in the two large US samples ($n=114$); the international models ($n = 5-13$) are consistent in sign but rest on few cities and should be read as corroborating rather than independently establishing the relationship. Population enters positively throughout and is the larger of the two structural predictors in the US data. Finally, we see a significant (although not universal across datasets) dependence on commuting time (more commuting time, more divergence between geographical and network distance) and car share (more car share, more divergence between geographical and network distance.

In any case, the values of $\rho$ are well below one across all cities and specifications, so while these covariates explain part of the variation in the divergence, none of them eliminates it: geographic distance remains a poor local proxy for network distance everywhere.

\begin{table}[t] \centering 
  \caption[\textbf{Correlates of the local divergence between geographic and network distance.}]{\textbf{Correlates of the local divergence between geographic and network distance.} Ordinary least squares regressions of each metro's mean local Spearman $\rho$ on visit-distribution entropy and population, fit separately by data source. The two US columns add ACS covariates (car share, commuting time, median household income), which are unavailable for the other sources. All regressors are standardized, so coefficients give the change in $\rho$ per standard deviation. Standard errors in parentheses.} 
  \label{supp:table:rho-models-all} 
\begin{tabular}{@{\extracolsep{5pt}}lccccc} 
\\[-1.8ex]\hline 
\hline \\[-1.8ex] 
 & \multicolumn{5}{c}{\textit{Dependent variable: $\rho$}} \\ 
\cline{2-6} 
 & US & US & UK & Spain & LatAm \\
 & Mobility & Commuting & Commuting & Mobility & Mobility \\ 
 \hline \\[-1.8ex] 
 Population (log) & 0.055$^{***}$ & 0.022$^{***}$ & 0.021 & 0.051$^{**}$ & 0.115$^{*}$ \\ 
  & (0.009) & (0.008) & (0.013) & (0.012) & (0.036) \\ 
 Entropy & 0.014$^{**}$ & 0.007$^{*}$ & 0.052$^{***}$ & 0.075$^{**}$ & 0.026 \\ 
  & (0.007) & (0.004) & (0.015) & (0.018) & (0.030) \\ 
  & & & & & \\ 
 Car share (log) & $-$0.007 & $-$0.019$^{**}$ &  &  &  \\ 
  & (0.008) & (0.007) &  &  &  \\ 
 Commuting time (log) & $-$0.022$^{***}$ & $-$0.005 &  &  &  \\ 
  & (0.008) & (0.008) &  &  &  \\ 
 Median income (log) & 0.004 & 0.003 &  &  &  \\ 
  & (0.007) & (0.007) &  &  &  \\ 
 Constant & 0.352$^{***}$ & 0.156$^{***}$ & 0.355$^{***}$ & 0.430$^{***}$ & 0.062 \\ 
  & (0.005) & (0.010) & (0.012) & (0.016) & (0.062) \\ 
\hline \\[-1.8ex] 
Observations & 114 & 114 & 13 & 6 & 5 \\ 
R$^{2}$ & 0.501 & 0.317 & 0.758 & 0.901 & 0.941 \\ 
Adjusted R$^{2}$ & 0.478 & 0.286 & 0.709 & 0.834 & 0.883 \\ 
\hline 
\hline \\[-1.8ex] 
\textit{Note:}  & \multicolumn{5}{r}{$^{*}$p$<$0.1; $^{**}$p$<$0.05; $^{***}$p$<$0.01} \\ 
\end{tabular} 
\end{table} 

\section{Free Fare effects on activity across visit subsamples}
\label{supp:sec:ff_subsamples}
The main text estimates the Free Fare effect on all visits made by residents of the treated communities. This section narrows the activity measure in three steps and re-estimates at each one. We first measure visits received by POIs in the treated tracts rather than visits made by their residents (Section~\ref{supp:sec:ff_poi_visits}). We then restrict that set to the POIs within 100\,m of any MBTA bus route (Section~\ref{supp:sec:ff_near_bus}), and finally to visits made by treated residents to POIs along Route~28 itself (Section~\ref{supp:sec:ff_samegroup_tr28}). Each step brings the outcome closer to the mechanism the program acts through. Under NDM the total-visit estimate is positive at every step and significant at the first and the third, and the Transit effect is positive and significant at all three. Under GDM the total is negative at the first two steps and never significant, and the per-taxonomy estimates do not share a sign. The main-text result is therefore not an artifact of how broadly activity is measured.

\subsection{Effect of visits to POIs in the treated area due to the Free Fare program}
\label{supp:sec:ff_poi_visits}
The main-text results measure the program from the residents' side. The outcome of each census tract is the weekly post-stratified visits made by its residents, wherever those visits take place. Here we measure the program from the destination side. The treated and control tracts are unchanged, and the outcome of each tract is now the weekly post-stratified visits received by the POIs located in it, regardless of where the visitors live. The rest of the pipeline is unchanged. Visits are split by POI taxonomy, a synthetic control is fitted for each treated tract from the same donor pool, and the post-treatment gaps enter the gap regression of Eq.~\eqref{eq:gap_did}. NDM uses the main-text community. The two measures should move together. If the additional trips treated residents make concentrate within their own communities, the POIs located there will register the extra visits. Table~\ref{supp:tab:ff_poi_visits} reports the resulting SCM-DiD ATEs under both GDM and NDM.

\begin{table}[t]
\centering
\caption[\textbf{Free Fare: effect on the visits to POIs in the treated area.}]{\textbf{Free Fare: effect on the visits to POIs in the treated area.} GDM and NDM ATE (\%) on the number of visits to POIs in the treated area of the Free Fare policy (total and per taxonomy). Standard error [SE] in brackets; the placebo 5--95\% column is the 5th--95th percentile range of the empirical placebo ATE distribution; significance is the two-sided empirical placebo test on $|\hat\beta|$ (Section~\ref{supp:sec:placebo_test}; $^{\ast}p<0.05$, $^{\ast\ast}p<0.01$).}
\label{supp:tab:ff_poi_visits}
\small
\setlength{\tabcolsep}{3.5pt}
\begin{tabular}{llrrrr}
\toprule
 & Taxonomy & GDM ATE (\%) [SE] & GDM placebo 5--95\% & NDM ATE (\%) [SE] & NDM placebo 5--95\% \\
\midrule
\multicolumn{6}{l}{\textit{Visits}} \\
 & City / Outdoors & $-15.27\,[1.82]$ & $[-16.6,\,12.2]$ & $6.21\,[1.17]$ & $[-19.5,\,24.7]$ \\
 & Cafe & $2.40\,[2.11]$ & $[-9.5,\,15.0]$ & $8.46\,[0.99]$ & $[-19.9,\,13.4]$ \\
 & College & $17.62\,[3.67]$ & $[-22.2,\,24.9]$ & $17.40\,[1.93]$ & $[-38.3,\,64.4]$ \\
 & Food & $-1.59\,[0.93]$ & $[-5.5,\,13.5]$ & $7.72\,[0.57]$ & $[-9.3,\,8.7]$ \\
 & Grocery & $-5.97\,[1.23]$ & $[-10.1,\,9.2]$ & $-6.07\,[0.74]$ & $[-11.0,\,10.6]$ \\
 & Shop & $9.80\,[2.22]$ & $[-9.3,\,8.2]$ & $5.34\,[0.80]$ & $[-10.1,\,13.4]$ \\
 & Transit & $4.59\,[0.97]$ & $[-6.8,\,9.6]$ & $14.09\,[0.60]^{\ast\ast}$ & $[-7.7,\,7.0]$ \\
 & Total & $-3.68\,[0.63]$ & $[-9.3,\,13.8]$ & $16.82\,[0.94]^{\ast\ast}$ & $[-3.9,\,8.2]$ \\
\bottomrule
\end{tabular}
\end{table}

Under GDM no per-taxonomy effect is significant and the total is null and slightly negative ($-3.68\%\,[0.63]$). Under NDM total POI visits rise by $+16.82\%$ (SE~$0.94$, placebo $p<0.01$), with a significant gain in Transit ($+14.09\%$, SE~$0.60$, placebo $p<0.01$). Food ($+7.72\%$), Cafe ($+8.46\%$), College ($+17.40\%$), Outdoors ($+6.21\%$) and Shop ($+5.34\%$) are positive but not significant. Grocery is the only category on which both designs agree, with a small negative effect under each ($-5.97\%$ GDM, $-6.07\%$ NDM, neither significant).

The signs and the ordering of the per-taxonomy effects match the tract-level NDM estimates in the main text, with the largest gains in transit-adjacent and food categories. The destination-side and resident-side measures therefore agree. This outcome counts all visits received, so it does not by itself separate additional trips by treated residents from additional arrivals by outsiders. Section~\ref{supp:sec:ff_samegroup_tr28} imposes that restriction directly.

\subsection{Visits to venues near any bus routes}
\label{supp:sec:ff_near_bus}
The main text finds that the Free Fare program raised city-wide activity among
treated-area residents. If this increase operates through bus travel, it should be
visible at venues along the bus network itself. We therefore refit SCM-DiD under both GDM and NDM after redefining only the outcome.
The units of analysis and the treated and control sets are exactly those of the
main-text Free Fare analysis under each design: GDM treats the 40 census tracts
within 1\,km of Route~28 and draws donors from the 2--8\,km ring, while NDM treats
the 159-tract mobility community containing Route~28 and draws donors from its
network-selected donor community. What changes is the outcome assigned to each of
these same tracts: as in Section~\ref{supp:sec:ff_poi_visits} each tract's outcome is
the weekly post-stratified visits \textit{received} by the POIs located in it, but
now counting only those POIs whose centroid lies within 100~m of any MBTA bus route
(not only Route~28). This near-bus POI set is a single city-wide whitelist, fixed
once and applied identically to treated and control tracts, so a tract with no
near-bus POIs contributes a near-zero outcome under both designs.
Table~\ref{supp:tab:ff_near_bus}
reports the resulting ATEs. All $p$-values refer to the empirical placebo test
(Section~\ref{supp:sec:placebo_test}).

\begin{table}[t]
\centering
\caption[\textbf{Free Fare: visits to POIs within 100\,m of MBTA bus routes.}]{\textbf{Free Fare: visits to POIs within 100\,m of MBTA bus routes.} GDM and NDM ATE (\%) on the number of visits to POIs close to any MBTA bus route (total and per taxonomy). Standard error [SE] in brackets; the placebo 5--95\% column is the 5th--95th percentile range of the empirical placebo ATE distribution; significance is the two-sided empirical placebo test on $|\hat\beta|$ (Section~\ref{supp:sec:placebo_test}; $^{\ast}p<0.05$, $^{\ast\ast}p<0.01$).}
\label{supp:tab:ff_near_bus}
\small
\setlength{\tabcolsep}{3.5pt}
\begin{tabular}{llrrrr}
\toprule
 & Taxonomy & GDM ATE (\%) [SE] & GDM placebo 5--95\% & NDM ATE (\%) [SE] & NDM placebo 5--95\% \\
\midrule
\multicolumn{6}{l}{\textit{Visits}} \\
 & City / Outdoors & $3.20\,[4.85]$ & $[-28.7,\,33.6]$ & $-7.52\,[1.74]$ & $[-30.8,\,32.1]$ \\
 & Cafe & $13.16\,[2.51]^{\ast}$ & $[-11.3,\,8.2]$ & $5.64\,[1.11]$ & $[-11.9,\,8.1]$ \\
 & Food & $-11.50\,[1.13]$ & $[-7.6,\,11.4]$ & $4.37\,[0.48]$ & $[-9.6,\,11.3]$ \\
 & Grocery & $-11.01\,[1.37]$ & $[-11.4,\,11.0]$ & $-4.87\,[0.81]$ & $[-11.2,\,11.8]$ \\
 & Shop & $-7.14\,[1.06]$ & $[-10.5,\,7.4]$ & $8.37\,[0.99]$ & $[-12.1,\,10.4]$ \\
 & Transit & $5.08\,[0.93]$ & $[-8.0,\,8.4]$ & $13.89\,[0.64]^{\ast}$ & $[-10.2,\,7.1]$ \\
 & Total & $-6.27\,[0.58]$ & $[-6.5,\,9.2]$ & $6.25\,[0.46]$ & $[-5.1,\,7.1]$ \\
\bottomrule
\end{tabular}
\end{table}

Under GDM the estimates are mutually inconsistent. The total is null ($-6.27\%$, not significant), Food is negative and not significant ($-11.50\%$), and the only significant effect is a positive one on Cafe ($+13.16\%$, $p<0.05$). A single program effect does not produce one significant positive category against negative Food, Grocery, and Shop estimates.
 
Under NDM the total turns positive ($+6.25\%$), though it is not significant ($p=0.109$), and Transit is positive and significant ($+13.89\%$, $p<0.05$). Shop ($+8.37\%$) is positive but not significant. Food and Cafe re-align with the main-text findings, Food moving from $-11.50\%$ (GDM) to $+4.37\%$ (NDM) and Cafe from $+13.16\%$ (GDM) to $+5.64\%$ (NDM), neither significant under NDM. Grocery is negative and not significant under both designs ($-11.01\%$ GDM, $-4.87\%$ NDM). Outdoors is the only category that turns negative under NDM ($+3.20\%$ GDM, $-7.52\%$ NDM), and neither estimate is significant.
 
The NDM total is smaller than the $+16.82\%$ recovered on all POIs in the treated tracts and is not individually significant. The restriction keeps a small share of the POIs in each tract, so the outcome series rests on fewer venues and the placebo distribution widens. It also selects on proximity to any MBTA route rather than to the fare-free route, so most retained venues are not on the Route~28 corridor. The direction of the main-text result survives, with a positive total and a significant positive Transit effect, which is what the fare-free mechanism predicts at the venues the bus network reaches. The transit result is further supported by the correlation between our visit counts to bus-stop POIs and the corresponding MBTA ridership series (Fig.~\ref{supp:fig:correlation_data}).

\subsection{Treated-resident visits along Route 28} 
\label{supp:sec:ff_samegroup_tr28}
The previous subsection measures visits from anywhere in the city. Here we take the narrowing to its endpoint and isolate the mechanism the program targets, treated residents riding the free Route~28 bus to destinations along it.

On the treated side a visit is kept only if the visitor lives in a treated tract, the POI is located in a treated tract, and the POI lies within 100\,m of the Route~28 alignment. On the control side we keep the within-area restriction, retaining visits made by donor-tract residents to POIs located in donor tracts, but not the corridor restriction, since there is no treated route in the control communities to condition on. Estimation is unchanged. The outcome of each tract is the weekly post-stratified visits received by its retained POIs, fitted with the same synthetic controls and gap regression as before. Table~\ref{supp:tab:ff_samegroup_tr28} reports ATEs under both GDM and NDM.

\begin{table}[t]
\centering
\caption[\textbf{Free Fare: visits to POIs near Route~28 by residents in that area.}]{\textbf{Free Fare: visits to POIs near Route~28 by residents in that area.} GDM and NDM ATE (\%) on the number of visits to POIs around Route~28 of residents in that area (total and per taxonomy). Standard error [SE] in brackets; the placebo 5--95\% column is the 5th--95th percentile range of the empirical placebo ATE distribution; significance is the two-sided empirical placebo test on $|\hat\beta|$ (Section~\ref{supp:sec:placebo_test}; $^{\ast}p<0.05$, $^{\ast\ast}p<0.01$).}
\label{supp:tab:ff_samegroup_tr28}
\small
\setlength{\tabcolsep}{3.5pt}
\begin{tabular}{llrrrr}
\toprule
 & Taxonomy & GDM ATE (\%) [SE] & GDM placebo 5--95\% & NDM ATE (\%) [SE] & NDM placebo 5--95\% \\
\midrule
\multicolumn{6}{l}{\textit{Visits}} \\
 & Total & $9.54\,[2.31]$ & $[-13.9,\,11.0]$ & $18.76\,[3.59]^{\ast\ast}$ & $[-6.1,\,8.9]$ \\
 & Cafe & $1.22\,[5.40]$ & $[-14.0,\,10.7]$ & $36.42\,[5.58]^{\ast\ast}$ & $[-16.9,\,8.9]$ \\
 & Food & $-20.53\,[2.36]^{\ast}$ & $[-8.9,\,18.2]$ & $1.47\,[2.14]$ & $[-6.6,\,9.3]$ \\
 & Grocery & $-3.15\,[3.07]$ & $[-13.6,\,12.2]$ & $-9.51\,[2.68]$ & $[-9.5,\,14.4]$ \\
 & Shop & $9.16\,[2.15]$ & $[-13.7,\,10.6]$ & $14.72\,[2.66]$ & $[-12.2,\,12.5]$ \\
 & Transit & $15.33\,[2.09]^{\ast}$ & $[-11.5,\,11.4]$ & $26.67\,[3.10]^{\ast\ast}$ & $[-8.7,\,9.3]$ \\
\bottomrule
\end{tabular}
\end{table}

Under GDM no coherent uplift emerges. The only significant effects are a negative Food estimate ($-20.53\%$, $p<0.05$) and a positive Transit one ($+15.33\%$, $p<0.05$). The negative Food effect is the ring-contamination signature diagnosed in Section~\ref{supp:sec:ff_near_bus}, since control POIs near bus stops share visitors with the treated POIs, so geographical distance alone does not isolate a clean control group.
 
Under NDM the total is significant ($+18.76\%$, $p<0.01$), with large significant gains in Cafe ($+36.42\%$, $p<0.01$) and Transit ($+26.67\%$, $p<0.01$). Shop ($+14.72\%$) is positive but not significant. This is consistent with treated residents concentrating newly induced trips on the Route~28 corridor.
 
This narrowest definition gives the same answer as the two preceding subsections and as the main text, a significant positive total uplift under NDM and no such effect under GDM. It is also the sharpest version of that result, concentrating the effect exactly where the program's mechanism predicts, among treated residents making newly induced trips along the Route~28 corridor.

\section{Validation on synthetic data} \label{supp:sec:simulation}

Across all three interventions, GDM and NDM give different estimates of the policy effect (Section~\ref{supp:sec:gdm_ndm_results}). The true effect is unknown in every case, so the observed data cannot tell which method is closer to it. We therefore construct synthetic data for Boston Free Fare and New York congestion pricing in which the ground-truth effect is fixed by construction, and apply both methods to it.

\paragraph{Data-generating process.} The same construction serves both interventions. We place a direct effect on the truly treated tracts only, and let it diffuse to the rest of the city through the same pre-period mobility network $W$ that defines the network distance (Section~\ref{supp:sec:NDM}). Let $\mathbf{d}\in\{0,1\}^{n}$ indicate the directly treated tracts, the tracts within 1\,km of Route~28 for Free Fare and the tracts inside the zone boundary for the CRZ. The simulated outcome for tract $i$ is
\begin{equation}
Y_i \;=\; \frac{s_{\text{inc}}}{\text{income}_i} \;+\; d_i\,\Delta \;+\; \kappa\,(1-d_i)\,\bigl(W_{\text{diff}}\,\mathbf{d}\bigr)_i \;+\; \varepsilon_i,
\label{eq:supp:dgp}
\end{equation}
where $W_{\text{diff}}=\sum_{k=1}^{K}\beta^{k}W^{k}$ is the truncated diffusion kernel ($\beta=0.8$, $K=5$, identical to the main-text $d_{\text{net}}$ construction). The first term is a baseline that varies inversely with tract income, with $s_{\text{inc}}=1000$. The second places the direct effect $\Delta$ on the treated tracts. The third spreads the effect to the untreated tracts in proportion to their diffusion-kernel proximity to the treated set, with spillover coefficient $\kappa=5$. The last is i.i.d.\ Gaussian noise with standard deviation $\sigma=0.05$.

The two interventions differ only in $\Delta$. For Free Fare we set $\Delta=1$, a unit increase in activity on the treated tracts. For congestion pricing we set $\Delta=0$, so the policy has no direct effect on the treated tracts and anything an estimator recovers there comes from the donor pool.
 
The ground-truth ATE on the treated tracts is therefore $\Delta$ itself, $+1$ for Free Fare and $0$ for congestion pricing. The income baseline enters the treated and counterfactual outcomes identically and cancels in the difference, and the spillover term is switched off on treated tracts by the factor $(1-d_i)$, so $\Delta$ is the whole of the true effect. Any estimate away from $\Delta$ is method bias, not a property of the simulation. The outcome $Y$ is built once per intervention and held fixed, so GDM and NDM are applied to a single common ground truth rather than to two separately simulated outcomes.
 
The bias has a single source in both cases. A donor tract carries the spillover term $\kappa\,(W_{\text{diff}}\mathbf{d})_j$, which is positive for any donor with flow to the treated set, while a treated tract does not. Donors therefore sit above their own baseline, the fitted counterfactual inherits that inflation, and the recovered gap is pushed downward. How far it goes depends on how much spillover the donor pool absorbs, which the two analyses below vary.

\begin{figure}[t!]
    \centering
    \includegraphics[width=0.8\textwidth]{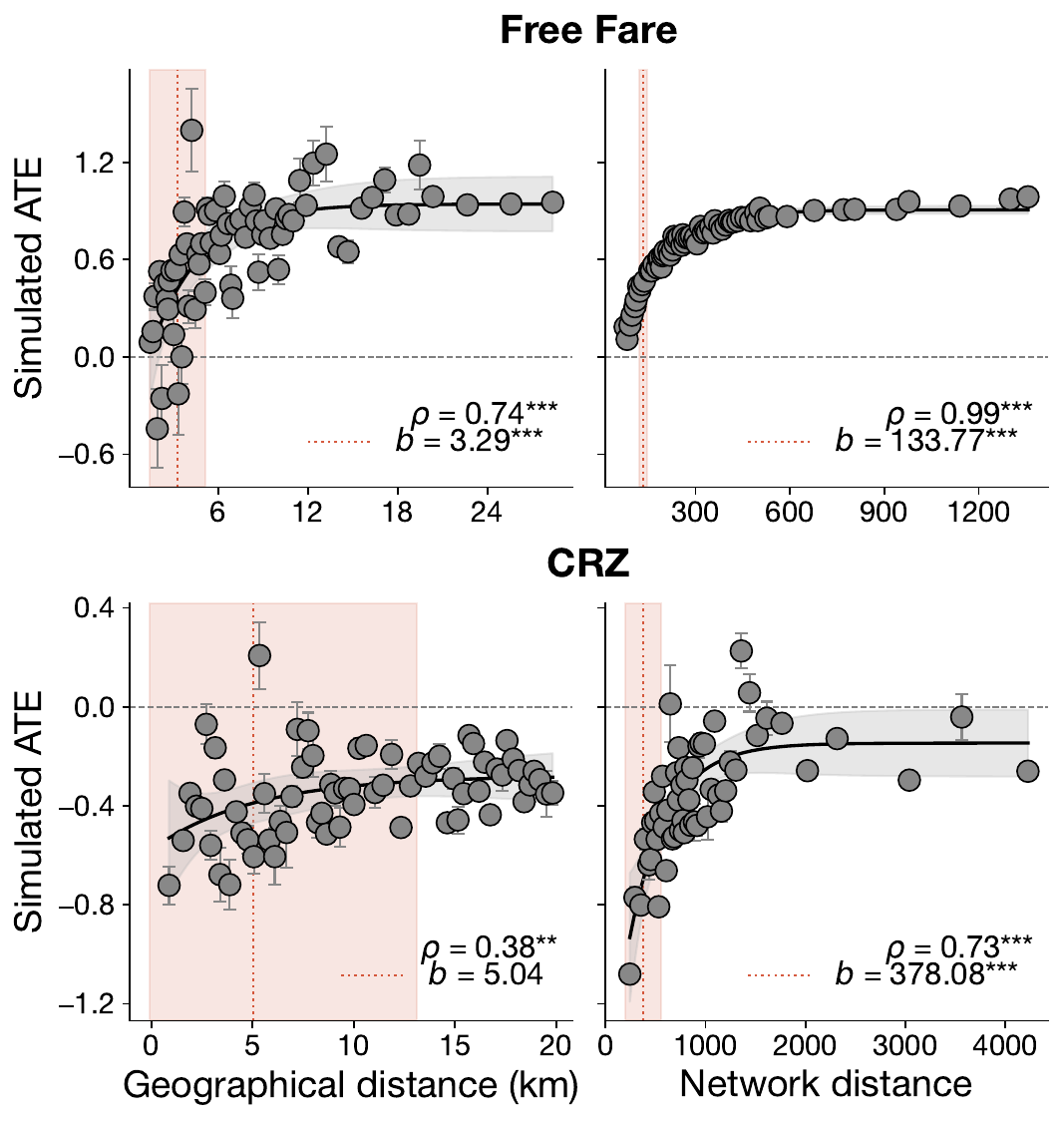}
    \caption[\textbf{Simulated validation: ATE versus distance.}]{\textbf{Simulated validation: ATE versus distance.} Per-ring simulated ATE from the diffusion DGP for Boston Free Fare (top row, planted effect $\Delta=1$) and New York congestion pricing (bottom row, $\Delta=0$), with donor tracts sorted into equal-count rings by geographical distance (left) and network distance (right) to the treated set; points are plotted at each ring's midpoint distance with $\pm1.96$ regression SE. Solid curves are saturating fits $c + a\,(1-e^{-d/b})$; the dotted red line marks the fitted characteristic distance $b$ with its $\pm1.96$ SE band shaded, and each panel is annotated with its Spearman $\rho$ (permutation-test stars) and $b$ ($t$-test stars; $^{\ast}p<0.05$, $^{\ast\ast}p<0.01$, $^{\ast\ast\ast}p<0.001$; Table~\ref{supp:tab:distance_fit_stats}). Against network distance the recovered ATE converges on the planted effect in both interventions, rising to $+1$ for Free Fare (RMSE $0.03$, $b=133.8^{\ast\ast\ast}$) and to $0$ for congestion pricing (RMSE $0.16$, $b=378.1^{\ast\ast\ast}$). Against geographical distance it scatters and does not saturate (Free Fare RMSE $0.25$, $b=3.29^{\ast\ast\ast}$; congestion pricing RMSE $0.16$, $b$ not significant).}
    \label{supp:fig:sim_2panel_ff_crz}
\end{figure}

\paragraph{Distance decay of the simulated effect.} Because the direct effect sits on the treated tracts and spreads outward through $W$, a donor pool drawn closer to the treated set absorbs more spillover, so the recovered ATE should vary systematically with distance from the treated set, as in the observed-data exposure analysis (Fig.~\ref{fig1}B). We sort the donor universe into equal-count rings ($\approx 5$ tracts per ring) by geographical and by network distance to the treated set, estimate the ATE for each ring with a per-treated-unit counterfactual (an OLS of the donor outcomes on $1/\text{income}$, evaluated at each treated tract's income), and fit a saturating curve $c + a\,(1-e^{-d/b})$ to the per-ring ATEs (Fig.~\ref{supp:fig:sim_2panel_ff_crz}).
 
Against network distance the per-ring ATE converges on the planted effect in both interventions. For Free Fare it rises along a tight curve toward $+1$ (RMSE $0.03$, $b=133.8^{\ast\ast\ast}$); for congestion pricing it rises from a spurious decline toward $0$ (RMSE $0.16$, $b=378.1^{\ast\ast\ast}$). The bias therefore vanishes once the donors are network-distant, which the figure validates: the estimator recovers the truth when the donor pool is uncontaminated, and misses it by an amount that tracks contamination. Against geographical distance, the same points scatter and do not saturate (Free Fare RMSE $0.25$, $b=3.29^{\ast\ast\ast}$; congestion pricing RMSE $0.16$, $b$ not significant). Network distance gives the cleaner fit in part by construction, since the effect propagates through $W$ and network distance is itself $\propto 1/(W_{\text{diff}}\mathbf{d})$, so it tracks the spillover directly where geographical distance only approximates it.

\paragraph{GDM vs.\ NDM under the known ground truth.} We now apply each method to the common simulated outcome using its real treatment and donor pools from the main analysis. For each model we fit a per-treated-unit cross-sectional synthetic control, regressing the donor outcomes on $1/\text{income}$ and predicting at each treated unit's income to obtain $\text{effect}_i=Y_i-\hat{Y}_i$, and combine across treated units (ATE $=$ mean effect, with the confidence interval taken from the spread across treated units). Significance is assessed against a placebo null of 100 draws that relabel half of each model's donors as pseudo-treated.
 
\begin{figure}[t!]
    \centering
    \includegraphics[width=0.62\textwidth]{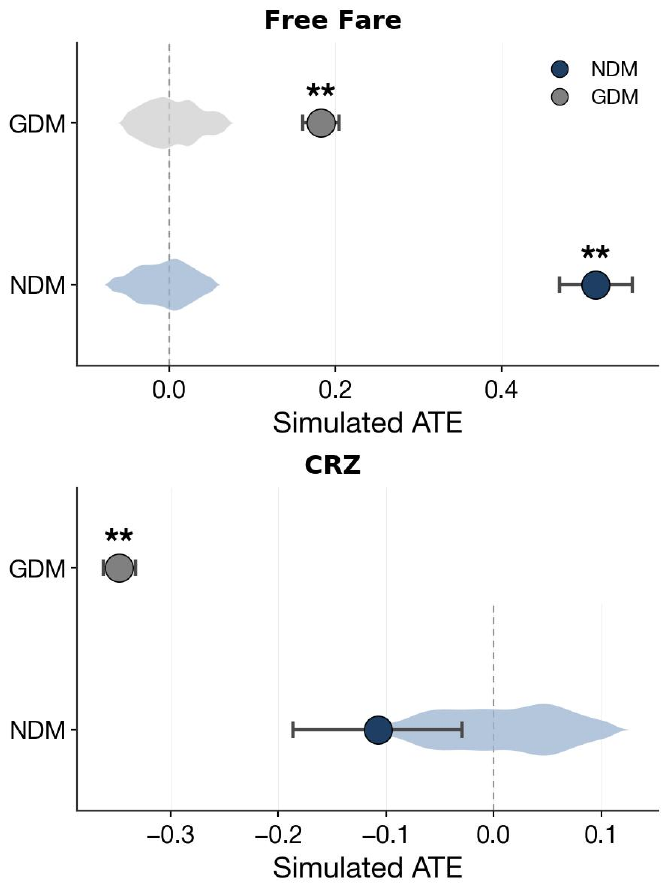}
    \caption[\textbf{Simulated validation: GDM vs.\ NDM under a known ground truth.}]{\textbf{Simulated validation: GDM vs.\ NDM under a known ground truth.} Total-visit ATE recovered from the diffusion DGP by each method for Boston Free Fare (top) and New York congestion pricing (bottom), using each method's real treatment and donor pools. Points are the combined per-treated-unit synthetic-control ATE (Free Fare: NDM $+0.51$, GDM $+0.18$; congestion pricing: GDM $-0.35$, NDM $-0.11$); horizontal bars span the spread across treated units; violins show the placebo-null distribution (100 draws relabeling half the donors as pseudo-treated). Asterisks denote empirical placebo $p$-values ($^\ast p<0.05$, $^{\ast\ast} p<0.01$). For Free Fare both estimators are significant, with NDM recovering the larger (correct) uplift; for congestion pricing GDM registers a significant spurious decline while NDM's effect is small and not significant---in both cases NDM lands closer to the truth because its donor pool is less contaminated by network spillover.}
    \label{supp:fig:sim_forest_ff_crz}
\end{figure}

The two interventions show the same bias acting in opposite directions (Fig.~\ref{supp:fig:sim_forest_ff_crz}). With a planted effect of $+1$ (Free-Fare), contaminated donors inflate the counterfactual and both estimators attenuate, GDM to $+0.18$ and NDM to $+0.51$. With a planted effect of $0$ (CRZ), the same inflation manufactures a decline, $-0.35$ under GDM and $-0.11$ under NDM. In both cases the bias scales with how much spillover the donor pool absorbs, and the network-distant NDM pool absorbs less. The GDM decline for congestion pricing is entirely an artifact of donor contamination, since the true effect there is zero, and NDM removes roughly two thirds of it.
 
NDM reduces the bias but does not eliminate it: on Free Fare it recovers about half the planted effect. This matches the distance analysis above. The bias vanishes only beyond a characteristic network distance, and the donor community selected for the main text sits at the midpoint of the Pareto front by the rule of Section~\ref{supp:subsec:ndm_choices} rather than at its far end, so it retains some exposure to the treated region. Moving further out along the front would reduce the residual bias at the cost of demographic comparability, which is the trade-off the front makes explicit.

\section{Robustness checks}

\subsection{Decomposing the GDM--NDM difference: treated group versus donor pool} \label{supp:sec:decomposition}
Moving from GDM to NDM changes two things at once. It changes the \textit{treated group}. GDM treats the tracts inside a fixed geographic buffer from the intervention, i.e., the 40 tracts within 1\,km of Route~28 for Free Fare and the tracts inside the zone for the CRZ. NDM treats the full mobility community or communities that contain them, i.e., 159 tracts for Free Fare. But it also changes the \textit{donor pool}. GDM draws controls from a surrounding geographic ring, NDM from mobility-distant communities. The GDM and NDM numbers reported in the main text do not separate the two. They leave open whether the estimate changes because the donor pool controls spillover or because the treated group is defined differently.

We separate the two by crossing each method's treated group with each method's donor pool, holding all other SCM settings fixed, so that the four cells differ only in their pair of unit sets (Table~\ref{supp:tab:decomp_treated_vs_donor}). The two off-diagonal cells change exactly one factor. We use identical SCM settings in every cell and disable the propensity score filter throughout, so the cells are strictly comparable. Under this common setting the NDM cell reproduces the main-text NDM estimate exactly for both interventions, $+15.65\%$ for Free Fare and $-1.51\%$ for the CRZ. The GDM cell keeps the sign and significance of the main-text GDM estimate but not its exact magnitude, since the main-text GDM also applies the propensity score filter of Section~\ref{supp:sec:GDM} that we switch off here. The filter barely moves the Free Fare GDM estimate and tightens the CRZ GDM estimate substantially.

\begin{table}[t]
\centering
\caption[\textbf{Decomposing the GDM--NDM difference.}]{\textbf{Decomposing the GDM--NDM difference.} Total-visit SCM-DiD ATE for each combination of treated group (GDM geographic buffer or NDM mobility community) and donor pool (GDM geographic ring or NDM mobility-distant community), for Boston Free Fare and the NYC Congestion Relief Zone. The diagonal cells are the main-text GDM and NDM designs. The off-diagonal cells change one factor at a time. Standard error [SE] in brackets. The placebo 5--95\% column is the 5th--95th percentile range of the empirical placebo ATE distribution, and significance is the two-sided empirical placebo test on $|\hat\beta|$ (Section~\ref{supp:sec:placebo_test}; $^{\ast}p<0.05$, $^{\ast\ast}p<0.01$). All cells use identical SCM settings with the propensity score filter disabled, so the four cells differ only in their unit sets. The NDM cell reproduces the main-text NDM estimate exactly.}\label{supp:tab:decomp_treated_vs_donor}
\small
\setlength{\tabcolsep}{5pt}
\begin{tabular}{lllrr}
\toprule
Intervention & Treated group & Donor pool & ATE (\%) [SE] & Placebo 5--95\% \\
\midrule
\multicolumn{5}{l}{\textit{Boston Free Fare}} \\
 & GDM buffer     & GDM ring     & $2.48\,[0.64]$             & $[-5.9,\,6.4]$ \\
 & GDM buffer     & NDM distant community  & $17.34\,[1.05]^{\ast\ast}$ & $[-7.1,\,6.8]$ \\
 & NDM treated community  & GDM ring     & $-6.02\,[0.31]^{\ast}$     & $[-4.4,\,7.6]$ \\
 & NDM treated community  & NDM distant community  & $15.65\,[0.55]^{\ast\ast}$ & $[-8.7,\,9.4]$ \\
\midrule
\multicolumn{5}{l}{\textit{NYC Congestion Relief Zone}} \\
 & GDM zone                & GDM ring               & $-10.85\,[0.37]^{\ast\ast}$ & $[-5.2,\,1.0]$ \\
 & GDM zone                & NDM distant community  & $-1.17\,[0.30]$             & $[-2.1,\,1.9]$ \\
 & NDM treated community   & GDM ring               & $-11.01\,[0.33]^{\ast\ast}$ & $[-4.0,\,1.8]$ \\
 & NDM treated community   & NDM distant community  & $-1.51\,[0.30]$             & $[-1.7,\,1.7]$ \\
\bottomrule
\end{tabular}
\end{table}

The two off-diagonal cells locate the source of the difference, and in both interventions it is the donor pool. For Free Fare, swapping only the donor pool raises the estimate from $+2.48\%$ (not significant) to $+17.34\%$ (placebo $p<0.01$). This keeps the geographic buffer treated group and draws controls from the mobility-distant community, and it recovers essentially the full NDM effect of $+15.65\%$. The donor swap is large and positive at either treated group ($+2.48\%\to+17.34\%$ at the GDM treated set, $-6.02\%\to+15.65\%$ at the NDM treated set). The treated-group swap depends on which donors it is paired with. With the mobility-distant NDM donors it barely moves the estimate ($+17.34\%$ vs $+15.65\%$). With the contaminated geographic ring it flips the sign to a significant $-6.02\%$. Once the donor pool controls for spillover, the positive effect is robust to how the treated group is defined. Redefining the treated group without fixing the donor pool does not recover it.

For the CRZ the pattern is sharper. Swapping only the donor pool removes the effect at either treated group ($-10.85\%\to-1.17\%$ at the GDM treated set, $-11.01\%\to-1.51\%$ at the NDM treated set), turning a significant decline into a non-significant near-null. Swapping only the treated group does essentially nothing ($-10.85\%$ vs $-11.01\%$ with the geographic ring, $-1.17\%$ vs $-1.51\%$ with the mobility-distant community). The outcome here is visits to POIs inside the zone, which are largely fixed however the surrounding treated tracts are grouped, so the whole GDM--NDM difference falls on the donor pool. Across both interventions the same conclusion holds. The difference between the GDM and NDM estimates comes from the donor pool, and therefore from controlling spillover, rather than from the redefinition of the treated group.

\subsection{Sensitivity to geographical buffer size in GDM} \label{supp:sec:sensitivity_buffer}
GDM draws its control units from a fixed-width buffer ring around the treated area. The width of that ring is arbitrary. We recompute the GDM ATE for each intervention across several concentric rings and report it with its empirical placebo $p$-value. Each ring is filtered on the propensity score as in the main-text GDM (Section~\ref{supp:sec:GDM}), so the sweep tests the full design rather than the unfiltered donor rings of Fig.~\ref{fig1}B.

\begin{figure}[t!]
    \centering
    \includegraphics[width=\textwidth]{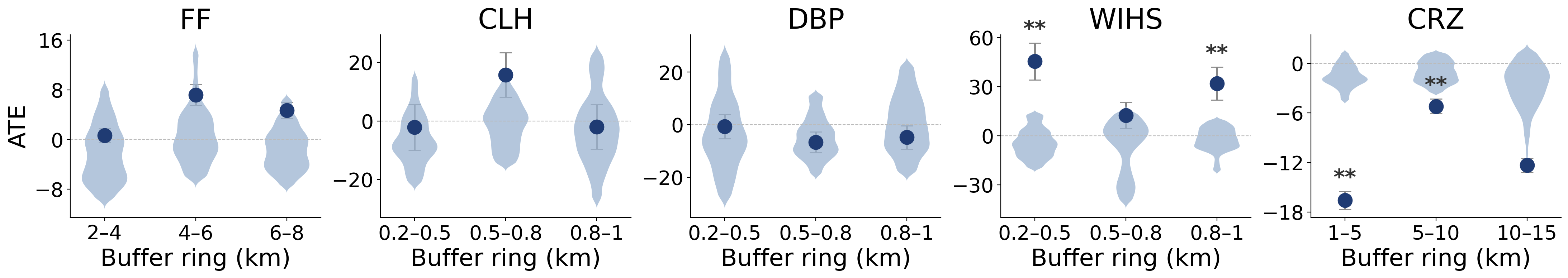}
    \caption[\textbf{GDM robustness: ATE across geographic buffer rings.}]{\textbf{GDM robustness: ATE across geographic buffer rings.} For each intervention (Free Fare, the three Open Streets sites, and CRZ), SCM estimates are recomputed using control units drawn from successive distance rings around the treated area. Error bars show $\pm 1.96\,\mathrm{SE}$; asterisks denote empirical placebo $p$-values ($^\ast p<0.05$, $^{\ast\ast} p<0.01$).}
    \label{supp:fig:gdm_buffer_robustness}
\end{figure}

At three of the five sites the ATE is not significant at any ring (Free Fare, Community League of the Heights, and Downtown Brooklyn Partnership; Fig.~\ref{supp:fig:gdm_buffer_robustness}). The control group is contaminated by spillover at every distance we test, and widening the ring does not recover the effect NDM finds.

For the other two, the estimate depends on where the ring is drawn. For the CRZ the ATE is negative at every ring, but its magnitude varies threefold, from $-16\%$ at 1--5\,km to $-5\%$ at 5--10\,km and $-12\%$ at 10--15\,km, and the outermost ring is not significant despite being larger in magnitude than the middle one. The reported size of the economic loss therefore turns on a choice the design does not justify, and Section~\ref{supp:sec:gdm_ndm_results} shows the decline itself is driven by spillover into the buffer. At NYCDOE/Washington Irving HS the estimate is significantly positive at the inner and outer rings and not significant at the middle ring, which is the main-text one (Table~\ref{supp:tab:gdm_ate_os}): significance appears and disappears with the width rather than tracking a stable effect. The per-category estimates at this site are also mutually inconsistent under GDM, with a significant decline in Shop ($-54.43\%$, $p<0.01$) alongside a positive Food estimate ($+18.39\%$) and a non-significant total. WIHS is a site where the two designs agree in direction, with NDM recovering $+27.27\%$ ($p<0.05$) against GDM's $+12.42\%$; the buffer design is not pointing the wrong way there, it is unable to resolve the effect at a width chosen in advance.

The buffer width is therefore not a neutral implementation detail. Where GDM finds nothing, it finds nothing at every width; where it finds something, the size and significance depend on the ring.

\subsection{Sensitivity to an alternative definition of proximity distance: travel time} \label{supp:sec:sensitivity_traveltime}
The main-text exposure analysis (Fig.~\ref{fig1}B) contrasts two ways of ordering control communities by their separation from the treated region: (Euclidean) geographical distance, which shows no clear saturation of the recovered ATE, and network distance, which does. 

However, at short range especially, Euclidean distance does not measure real distance between areas: a river, highway, or rail barrier can leave two areas that are close geographically far apart by road, while areas that are farther in straight-line distance may be quicker to reach. Travel time removes this problem, so we repeat the analysis to test whether the failure of geographical distance is really a routing artifact.

We add road travel time as a third distance metric alongside geographical and network distance, for both Boston Free Fare and New York congestion pricing. For each intervention we build a tract-to-tract driving-time matrix with the Open Source Routing Machine (OSRM, car profile,~\cite{luxen-vetter-2011}) on the OpenStreetMap road network~\cite{openstreetmap}, take each donor tract's minimum driving time to the treated set, and sort the donor universe into equal-count travel-time bins. Free Fare uses sixteen bins, and congestion pricing uses rings of $\approx 120$ tracts. Each bin serves as the SCM donor pool and its total-visit ATE is estimated as in that intervention's geographical and network panels (Section~\ref{supp:sec:gdm_ndm_results}).

\begin{figure}[t!]
    \centering
    \includegraphics[width=\textwidth]{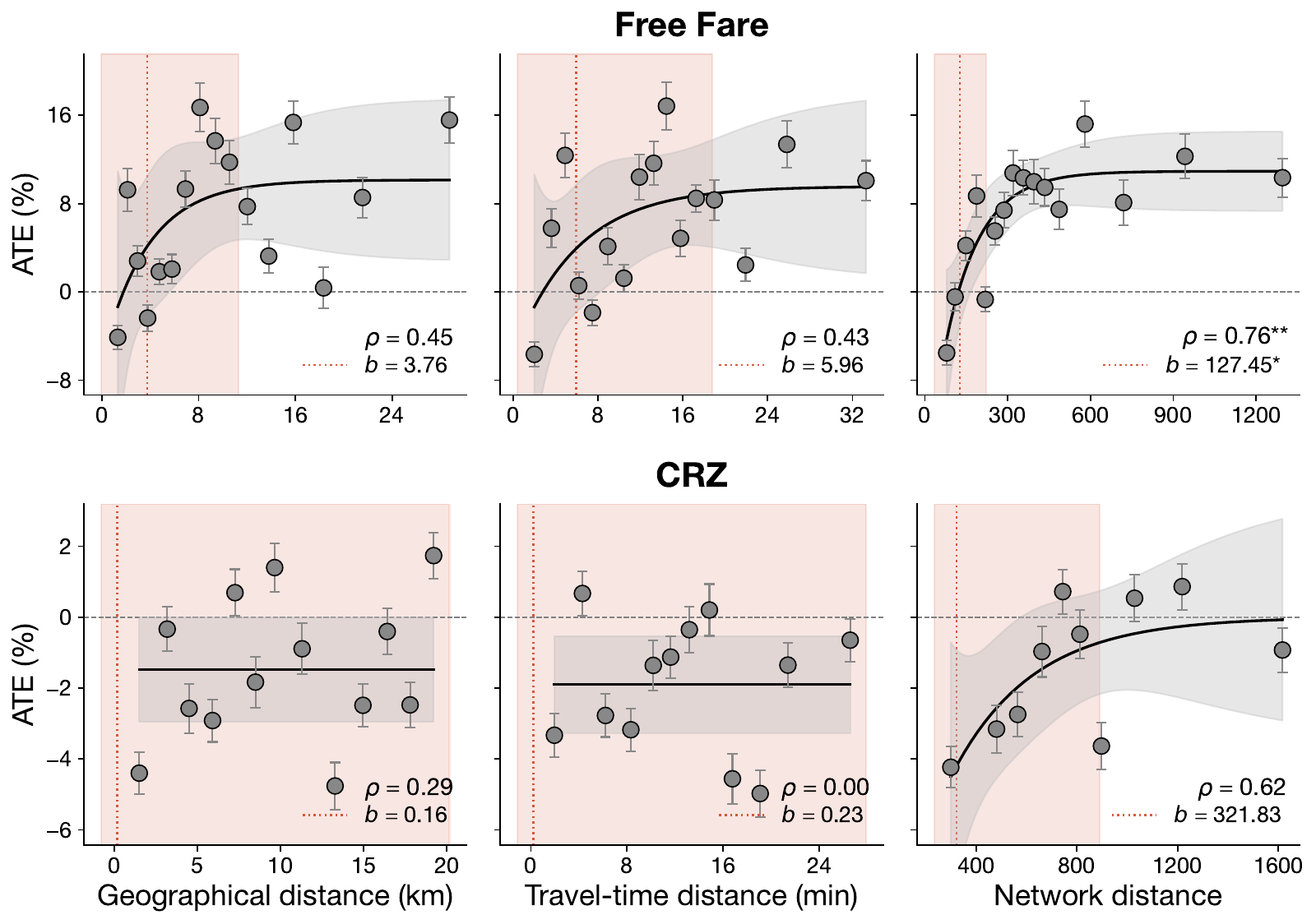}
    \caption[\textbf{Recovered effect versus geographic, travel-time, and network distance.}]{\textbf{Recovered effect versus geographic, travel-time, and network distance.} Per-ring SCM-DiD total-visit ATE for Boston Free Fare (top row) and New York congestion pricing (bottom row), with donor tracts sorted into equal-count rings by geographical distance (left), minimum road driving time (OSRM, car profile; middle), and network distance (right) to the treated set. Free Fare uses sixteen equal-count bins; congestion pricing uses rings of $\approx 120$ tracts, and its network panel excludes the single farthest ring, which is composed of New Jersey tracts outside the metropolitan mobility core. Points are plotted at each ring's midpoint distance and error bars span $\pm1.96$ SE; solid curves are saturating fits $c + a\,(1-e^{-d/b})$, the dotted red line marks the fitted characteristic distance $b$ where the fit is non-degenerate, and each panel is annotated with its Spearman $\rho$ (permutation-test stars) and $b$ ($t$-test stars; $^{\ast}p<0.05$, $^{\ast\ast}p<0.01$). Its Free Fare geographical and network panels are the two panels of the main-text Fig.~\ref{fig1}B. In both cities, only network distance orders the recovered effect.}
    \label{supp:fig:distance_saturation_6panel}
\end{figure}

Only network distance orders the recovered ATE into a clean saturating rise (Fig.~\ref{supp:fig:distance_saturation_6panel}). For Free Fare (top row) the per-ring ATE rises and saturates with network distance ($\rho = 0.76^{\ast\ast}$, characteristic distance $b = 127.45^{\ast}$). Against geographical distance ($\rho = 0.45$, $b = 3.76$) and travel time ($\rho = 0.43$, $b = 5.96$) the same points scatter more widely, and neither correlation is significant. Congestion pricing (bottom row) shows the contrast more sharply. Network distance produces a saturating rise toward zero ($\rho = 0.62$, $b = 321.83$), while against geographical distance ($\rho = 0.29$) and travel time ($\rho = 0.00$) the per-ring ATEs scatter around a flat line and the saturating fit degenerates to a constant ($b = 0.16$ and $b = 0.23$). Road travel time therefore behaves like geographical distance rather than like network distance. It is a better proxy for physical separation than straight-line distance, but it remains a physical metric and does not reproduce the rise and plateau that the mobility network recovers. Exposure is governed by behavioral connectivity rather than by any notion of physical proximity.

Table~\ref{supp:tab:distance_fit_stats} reports the fit statistics of every panel with a saturating fit, in Fig.~\ref{fig1}B and Figs.~\ref{supp:fig:sim_2panel_ff_crz} and~\ref{supp:fig:distance_saturation_6panel}. The figures annotate each panel with the Spearman $\rho$ and the characteristic distance $b$ and their significance; the table adds their exact $p$-values and the RMSE, AIC, and $R^2$ of each fit.

\begin{table}[t]
\centering
\caption[\textbf{Fit statistics of the ATE-versus-distance panels.}]{\textbf{Fit statistics of the ATE-versus-distance panels.} One row per panel of Fig.~\ref{supp:fig:distance_saturation_6panel} (observed data) and Fig.~\ref{supp:fig:sim_2panel_ff_crz} (simulated data); the Free Fare observed geographical and network rows are also the two panels of Fig.~\ref{fig1}B. $n$ is the number of rings; the CRZ observed network row excludes the farthest ring, as in Fig.~\ref{supp:fig:distance_saturation_6panel}. $\rho$ is the Spearman correlation between ring midpoint distance and ring ATE, and $p_{\rho}$ is its two-sided permutation $p$-value from 20{,}000 random permutations (the smallest attainable value is $1/20{,}001\approx5.0\times10^{-5}$). $b$ is the characteristic distance of the saturating fit $c + a\,(1-e^{-d/b})$, in km (geographical), minutes (travel time), or network-distance units, with its SE in brackets, and $p_{b}$ is the two-sided $t$-test $p$-value for $b=0$ on $n-3$ degrees of freedom. RMSE is $\sqrt{\mathrm{RSS}/(n-3)}$ in units of the ATE (percent for the observed data), AIC is $n\ln(\mathrm{RSS}/n)+6$, and $R^2=1-\mathrm{RSS}/\mathrm{TSS}$ is the goodness of fit. The significance stars in the figures follow these $p$-values ($^{\ast}p<0.05$, $^{\ast\ast}p<0.01$, $^{\ast\ast\ast}p<0.001$). $^{\dagger}$The best saturating fit is a step at the nearest ring, so $b$ lies below the nearest ring distance and is not identified; the figure draws the constant model instead.}
\label{supp:tab:distance_fit_stats}
\small
\setlength{\tabcolsep}{4pt}
\begin{tabular}{lrrrrrrrr}
\toprule
Distance & $n$ & $\rho$ & $p_{\rho}$ & $b$ [SE] & $p_{b}$ & RMSE & AIC & $R^2$ \\
\midrule
\multicolumn{9}{l}{\textit{Boston Free Fare, observed data (Fig.~\ref{supp:fig:distance_saturation_6panel}, top row)}} \\
Geographical & 16 & $0.45$ & $0.080$ & $3.76\,[3.85]$ & $0.35$ & $5.95$ & $59.7$ & $0.30$ \\
Travel time & 16 & $0.43$ & $0.096$ & $5.96\,[6.53]$ & $0.38$ & $5.55$ & $57.5$ & $0.28$ \\
Network & 16 & $0.76$ & $0.0010$ & $127.45\,[47.66]$ & $0.019$ & $2.97$ & $37.5$ & $0.74$ \\
\multicolumn{9}{l}{\textit{CRZ, observed data (Fig.~\ref{supp:fig:distance_saturation_6panel}, bottom row)}} \\
Geographical & 13 & $0.29$ & $0.33$ & $0.16^{\dagger}\,[167.79]$ & $1.0$ & $2.04$ & $21.1$ & $0.18$ \\
Travel time & 12 & $0.00$ & $1.0$ & $0.23^{\dagger}\,[497.53]$ & $1.0$ & $1.97$ & $18.8$ & $0.06$ \\
Network & 10 & $0.62$ & $0.062$ & $321.83\,[290.58]$ & $0.30$ & $1.53$ & $11.0$ & $0.49$ \\
\midrule
\multicolumn{9}{l}{\textit{Boston Free Fare, simulated data (Fig.~\ref{supp:fig:sim_2panel_ff_crz}, top row)}} \\
Geographical & 67 & $0.74$ & $5.0\times10^{-5}$ & $3.29\,[0.94]$ & $8.7\times10^{-4}$ & $0.25$ & $-182.8$ & $0.53$ \\
Network & 67 & $0.99$ & $5.0\times10^{-5}$ & $133.77\,[6.50]$ & $6.4\times10^{-30}$ & $0.03$ & $-447.9$ & $0.97$ \\
\multicolumn{9}{l}{\textit{CRZ, simulated data (Fig.~\ref{supp:fig:sim_2panel_ff_crz}, bottom row)}} \\
Geographical & 66 & $0.38$ & $0.0015$ & $5.04\,[4.12]$ & $0.23$ & $0.16$ & $-240.9$ & $0.14$ \\
Network & 54 & $0.73$ & $5.0\times10^{-5}$ & $378.08\,[92.28]$ & $1.5\times10^{-4}$ & $0.16$ & $-197.3$ & $0.58$ \\
\bottomrule
\end{tabular}
\end{table}

We also repeat the Geographical Distance Method itself with control tracts selected by travel time. The treated set is unchanged, the 40 tracts within 1\,km of Route~28, and the donor pool is redefined as the tracts whose minimum driving time to the treated set falls in a 5 to 10 minute ring (89 tracts), in place of the 2 to 8\,km geographic ring (188 tracts). We apply the same estimator, propensity-score donor filter, and placebo inference used for the geographic GDM, across the visit, social, income, and spend outcomes. Table~\ref{supp:tab:traveltime_gdm} reports the result. The total-visit effect is $0.6\%$, smaller than the $1.9\%$ recovered with the geographic ring and not distinguishable from its placebo distribution. The total spend effect is also not significant, and the per-category effects are scattered and mostly consistent with their placebo nulls. Selecting the GDM control pool by driving time therefore does not recover a cleaner program effect.

In summary, our main-text results are not an artifact of using Euclidean distance as the measure of physical proximity.

\begin{table}[t]
    \centering
    \caption[\textbf{Travel-time GDM: Free Fare and congestion pricing.}]{\textbf{Travel-time GDM: Free Fare and congestion pricing.} GDM average treatment effects with control tracts chosen by driving time instead of a straight-line buffer (Free Fare: 5--10 min ring, 89 donors; congestion pricing: 6--18 min ring, 1254 donors). Visit and social outcomes in percent; spend in dollars (Free Fare: per-POI weekly spend; congestion pricing: VxS spend per treated unit---scales not directly comparable). ATE with standard error [SE]; placebo 5--95\% is the 5th--95th percentile of the empirical placebo distribution ($^\ast p<0.05$, $^{\ast\ast} p<0.01$). Dashes mark outcomes not reported for that intervention.}
    \label{supp:tab:traveltime_gdm}
    \small
    \setlength{\tabcolsep}{4pt}
    \begin{tabular}{lrrrr}
        \toprule
         & \multicolumn{2}{c}{Free Fare} & \multicolumn{2}{c}{Congestion pricing} \\
        \cmidrule(lr){2-3} \cmidrule(lr){4-5}
        Outcome & ATE [SE] & placebo 5--95\% & ATE [SE] & placebo 5--95\% \\
        \midrule
        \multicolumn{5}{l}{\textit{Visits (\%)}} \\
        GDM (Total) & $0.6\,[0.7]$ & $[-7.3,\,17.2]$ & $-2.6\,[0.3]$ & $[-4.4,\,0.4]$ \\
        Cafe & $-2.4\,[1.0]$ & $[-8.9,\,14.2]$ & $-1.4\,[0.8]$ & $[-8.3,\,4.3]$ \\
        Food & $-2.0\,[0.7]$ & $[-10.3,\,12.0]$ & $-14.3\,[0.5]^{\ast\ast}$ & $[-2.5,\,3.4]$ \\
        Grocery & $-8.8\,[1.2]$ & $[-9.4,\,16.0]$ & $-8.2\,[0.8]^{\ast\ast}$ & $[-3.6,\,2.3]$ \\
        Shop & $-3.6\,[0.9]$ & $[-9.5,\,10.5]$ & $-13.1\,[1.1]^{\ast\ast}$ & $[-4.8,\,3.4]$ \\
        Transit & $0.0\,[0.9]$ & $[-9.5,\,10.7]$ & $-7.8\,[1.6]^{\ast}$ & $[-5.9,\,3.0]$ \\
        Outdoors & $-18.6\,[1.6]$ & $[-17.7,\,20.2]$ & -- & -- \\
        College & $-31.8\,[1.5]^{\ast}$ & $[-22.3,\,19.5]$ & -- & -- \\
        \midrule
        \multicolumn{5}{l}{\textit{Social (\%)}} \\
        Segregation & $14.3\,[0.6]^{\ast}$ & $[-6.9,\,9.0]$ & $6.4\,[1.3]^{\ast\ast}$ & $[-0.2,\,1.7]$ \\
        Exploration & $-6.3\,[0.5]$ & $[-6.4,\,6.6]$ & -- & -- \\
        Q1 & $-3.7\,[0.8]$ & $[-17.5,\,44.5]$ & $-3.8\,[0.3]$ & $[-5.2,\,2.4]$ \\
        Q2 & -- & -- & $-3.0\,[0.3]$ & $[-2.9,\,2.2]$ \\
        Q3 & $-2.8\,[2.2]$ & $[-14.5,\,11.2]$ & $-3.2\,[0.3]$ & $[-4.2,\,0.8]$ \\
        Q4 & -- & -- & $-1.6\,[0.3]$ & $[-3.8,\,1.6]$ \\
        \midrule
        \multicolumn{5}{l}{\textit{Spend (\$)}} \\
        GDM (Total) & $-155\,[159]$ & $[-357,\,661]$ & $-244{,}438\,[23{,}971]^{\ast}$ & $[-75{,}533,\,54{,}508]$ \\
        Food & $121\,[49]$ & $[-229,\,396]$ & $-49{,}391\,[3{,}664]^{\ast}$ & $[-16{,}547,\,9{,}206]$ \\
        Grocery & $-18\,[64]$ & $[-419,\,865]$ & $-1{,}932\,[1{,}013]$ & $[-10{,}820,\,9{,}168]$ \\
        Shop & $-354\,[178]$ & $[-517,\,729]$ & -- & -- \\
        \bottomrule
    \end{tabular}
\end{table}

For congestion pricing we repeat the same exercise, replacing the geographic control ring with a driving-time ring (donors 6 to 18 minutes from the treated set, 1254 tracts, matched to the geographic GDM's donor count). Because the treated unit is the entire Congestion Relief Zone, driving time to the nearest treated tract is close to distance from the zone boundary, so the travel-time ring largely overlaps the geographic buffer. Table~\ref{supp:tab:traveltime_gdm} shows that this control selection does not recover a cleaner effect: the overall visit effect is small and not distinguishable from its placebo distribution ($-2.6\%\,[0.3]$), while several categories (Food, Grocery, Shop, Transit) and total spend remain significantly negative, the same spillover-driven pattern seen under the geographic GDM (Section~\ref{supp:sec:gdm_ndm_results}). As for Free Fare, selecting controls by travel time rather than straight-line distance does not resolve the buffer-contamination problem.

\subsection{Sensitivity to the diffusion kernel parameters} \label{supp:sec:sensitivity_beta}
The network distance $d_{\text{net}}(c,T)=\bigl(\overline{W_{\text{diff}}[c,T]}\bigr)^{-1}$ is built from the truncated diffusion kernel $W_{\text{diff}}=\sum_{k=1}^{K}\beta^{k}W^{k}$, which depends on the attenuation $\beta < 1$ that discounts longer flow chains geometrically. The main-text analyses fix $\beta=0.8$. To show that the donor selection does not depend on this choice, we recompute the Pareto front of $(d_{\text{ps}},\,d_{\text{net}})$ for the Boston Free Fare consensus communities at four values of $\beta$, holding the kernel depth at $K=5$, and recompute $d_{\text{net}}$ at four values of $K$ with $\beta=0.8$.

\begin{figure}[t!]
    \centering
    \includegraphics[width=\textwidth]{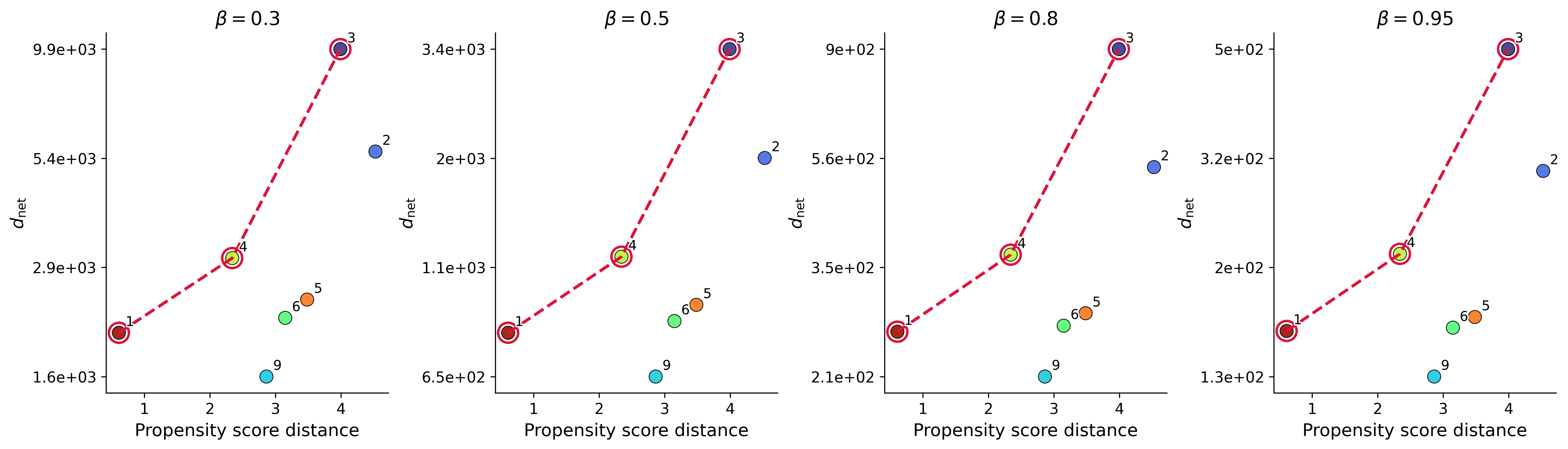}
    \caption[\textbf{Free Fare: Pareto front sensitivity to the diffusion attenuation $\beta$.}]{\textbf{Free Fare: Pareto front sensitivity to the diffusion attenuation $\beta$.} Each panel shows the candidate donor communities (consensus partition at Leiden $\gamma=0.0025$) in $(d_{\text{ps}},\,d_{\text{net}})$ space for a different value of $\beta$. The Pareto front (red, dashed) and its members are computed independently at each $\beta$. Numbers label the community id. The propensity distance $d_{\text{ps}}$ does not depend on $\beta$; only the vertical $d_{\text{net}}$ scale rescales across panels.}
    \label{supp:fig:ff_pareto_beta_sensitivity}
\end{figure}

The Pareto front contains the same three communities $\{1,4,3\}$ at every $\beta\in\{0.3,0.5,0.8,0.95\}$, in the same order along the front (Fig.~\ref{supp:fig:ff_pareto_beta_sensitivity}). Only the absolute $d_{\text{net}}$ values rescale. Larger $\beta$ amplifies the contribution of longer flow chains and shrinks $d_{\text{net}}$ on the log scale, but the relative ordering of communities is preserved, and that ordering is the input to the donor selection step. The selected donor community is therefore the same for any $\beta<1$ we test

The depth $K$ behaves the same way. Holding $\beta=0.8$ and the consensus partition fixed, we recompute $d_{\text{net}}(c,T)$ for every control community at $K\in\{4,5,6,7\}$. The Spearman correlation with the $K=5$ distances is exactly 1 at every $K$, and the Pearson correlation exceeds 0.9998. Absolute distances fall as $K$ grows, since deeper kernels accumulate more walks and $1/\overline{W_{\text{diff}}[c,T]}$ shrinks. The ordering of control communities is identical at every $K$, and the selected donor community is the same. Donor selection depends on the ranking in network distance, not on its level, so neither $\beta$ nor $K$ affects the design.

\subsection{Sensitivity to the community detection method and resolution} \label{supp:sec:sensitivity_community}
The main-text analysis detects mobility communities with one algorithm at one resolution, and selects a specific donor community or combination for each intervention. Here we show that the estimates do not depend on that choice. For every intervention we recompute the community assignments across several Leiden resolutions and with Infomap as an alternative algorithm. At each variant we compute the Pareto front, select the donor group by the procedure of Section~\ref{supp:subsec:ndm_choices}, and report the resulting ATE.

\begin{figure}[p]
    \centering
    \includegraphics[width=\textwidth]{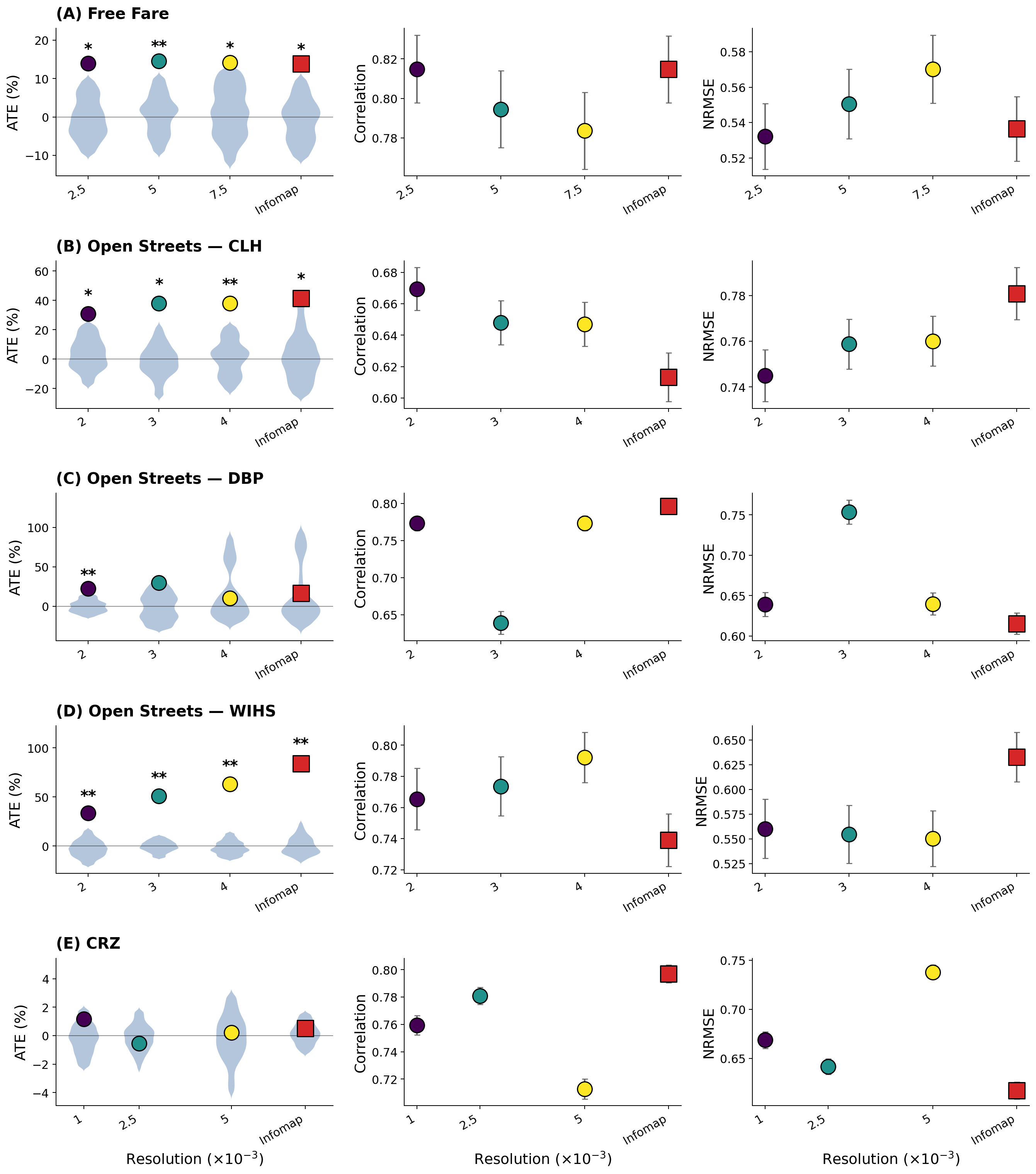}
    \caption[\textbf{Community-detection sensitivity across interventions.}]{\textbf{Community-detection sensitivity across interventions.} Each row is one intervention (top to bottom: Boston Free Fare; Community League of the Heights, CLH; Downtown Brooklyn Partnership, DBP; NYCDOE\,/\,Washington Irving HS, WIHS; and the NYC Congestion Relief Zone, CRZ). Within each row the donor community set is reselected by the procedure of Section~\ref{supp:subsec:ndm_choices} under several Leiden resolutions and under Infomap, and three quantities are plotted against the detection variant: \textbf{(left)} the SCM-DiD ATE, with the placebo-null distribution shown as a violin and asterisks marking empirical placebo significance ($^\ast p<0.05$, $^{\ast\ast} p<0.01$); \textbf{(center)} the pre-treatment correlation; and \textbf{(right)} the pre-treatment NRMSE. For every intervention the ATE is stable across detection variants, while pre-treatment fit varies and guides the main-text resolution choice (Section~\ref{supp:subsec:ndm_choices}).}
    \label{supp:fig:comm_sensitivity}
\end{figure}
\paragraph{Boston Free Fare.} The main-text analysis uses Leiden at $\gamma=0.0025$. Fig.~\ref{supp:fig:comm_sensitivity}A compares this against $\gamma=0.005$, $\gamma=0.0075$, and Infomap. The ATE is stable across all variants, so the estimate does not depend on the resolution. Pre-treatment fit does depend on it. NRMSE is lowest and pre-treatment correlation highest at $\gamma=0.0025$, and both degrade as $\gamma$ increases. We therefore adopt $\gamma=0.0025$ as the main partition.

\paragraph{New York Open Streets.} We report the three sites separately, since each was evaluated against its own local mobility network and Pareto pool.

\begin{itemize}
\item \textbf{Community League of the Heights (CLH).} The main-text analysis uses Leiden $\gamma=2.0$. Fig.~\ref{supp:fig:comm_sensitivity}B compares this against $\gamma=3$, $\gamma=4$, and Infomap. The ATE is stable across all variants, as for Free Fare.
\item \textbf{Downtown Brooklyn Partnership (DBP).} The main-text analysis uses Leiden $\gamma=2.0$. Fig.~\ref{supp:fig:comm_sensitivity}C compares this against $\gamma=3$, $\gamma=4$, and Infomap. The point estimate is positive at every variant, but significance is reached only at the main-text resolution. Finer resolutions fragment the communities into smaller donor pools, which inflates placebo dispersion and weakens the test while leaving the central tendency unchanged. The resolution was selected on pre-treatment fit (Section~\ref{supp:subsec:ndm_choices}), not on significance.
\item \textbf{NYCDOE / Washington Irving HS (WIHS).} The main-text analysis uses Leiden $\gamma=2.0$. Fig.~\ref{supp:fig:comm_sensitivity}D compares this against $\gamma=3$, $\gamma=4$, and Infomap. The ATE is robust to both the resolution and the detection algorithm.
\end{itemize}

\paragraph{CRZ: NYC Congestion Relief Zone.} The main-text analysis uses Leiden $\gamma=0.002$. Fig.~\ref{supp:fig:comm_sensitivity}E shows the ATE at $\gamma=0.001$, $\gamma=0.0025$, and $\gamma=0.005$, which bracket the main-text resolution, and under Infomap. As in the other interventions, the ATE is unaffected by the resolution and the detection algorithm.

\subsection{Sensitivity to the definition of mobility flows} \label{supp:sec:sensitivity_flows}
The mobility network is the input from which the whole NDM design follows, since the communities, the network distances, and the donor pool are all derived from it. We therefore test whether the design survives changes to how the flows are defined. Section~\ref{supp:sec:sensitivity_flows_visits} perturbs the visit-flow specification, and Section~\ref{supp:sec:sensitivity_lodes} rebuilds the network from administrative commute data instead.

\subsubsection{Alternative visit-flow specifications}
\label{supp:sec:sensitivity_flows_visits}
The NDM partition used for the Free Fare analysis is derived from visit-to-visit mobility flows spanning 2021-01-01 to 2021-07-01, strictly before the intervention begins on August 29, 2021. Two concerns motivate testing its stability. The partition could be an artifact of this particular flow definition rather than of genuine mobility structure. Mobility panels are also sampled from opted-in devices and are therefore noisy, so the partition should not hinge on the specific set of visits observed.

\begin{figure}[t!]
    \centering
\includegraphics[width=\textwidth]{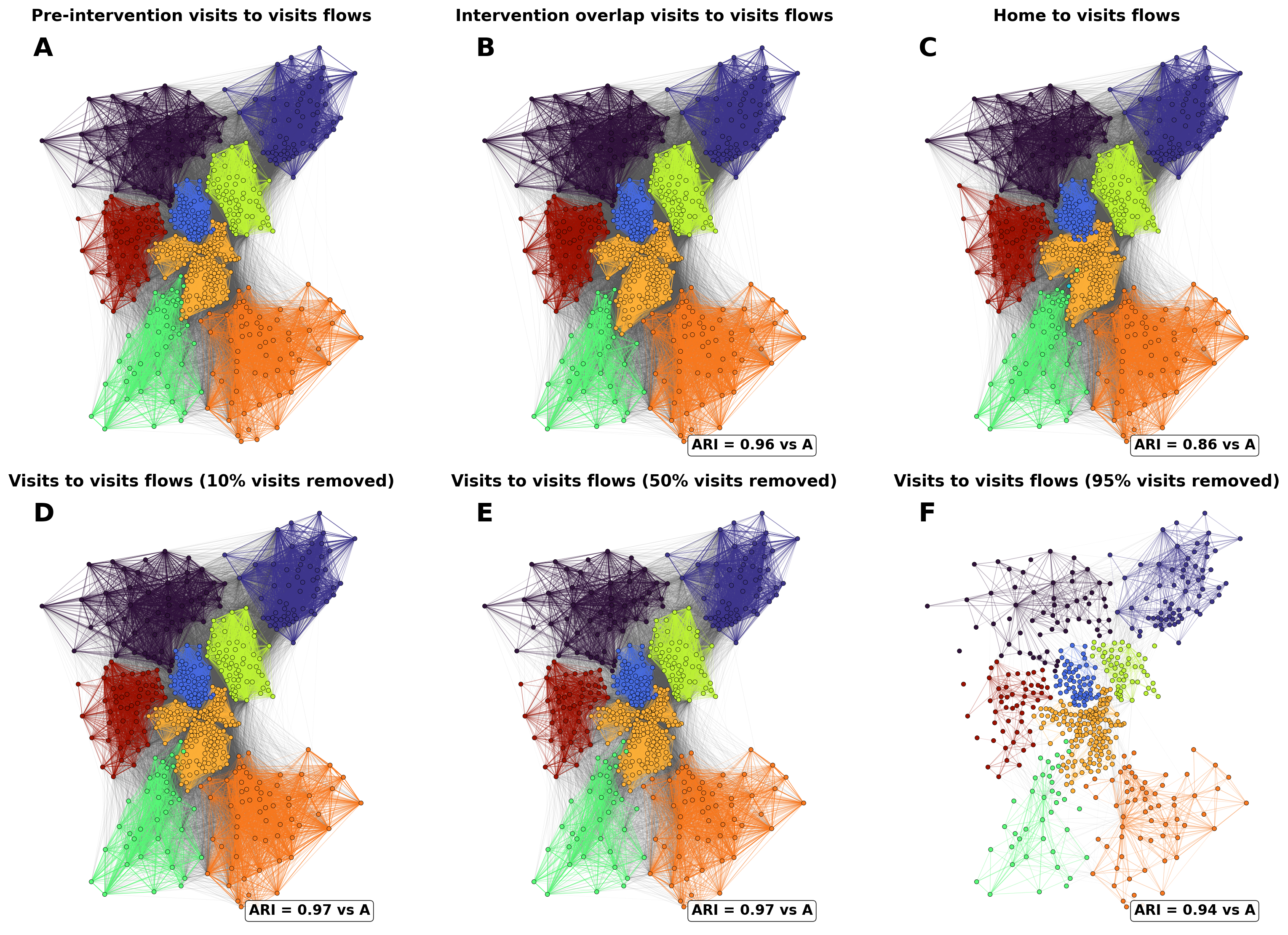}
    \caption[\textbf{Free Fare: sensitivity of NDM partition to the mobility-flow definition.}]{\textbf{Free Fare: sensitivity of NDM partition to the mobility-flow definition.} \textbf{(A)} Consensus partition built from pre-intervention visit-to-visit flows (2021-01-01 to 2021-07-01), the flow used in the main analysis. \textbf{(B)} Partition built from post-intervention visit-to-visit flows (2021-09-01 to 2022-03-01). \textbf{(C)} Partition built from home-to-visit flows over the pre-intervention window. \textbf{(D)} Partition built from pre-intervention visit-to-visit flows with 10\% of visits removed uniformly at random before aggregation. \textbf{(E)} Same with 50\% of visits removed. \textbf{(F)} Same with 95\% of visits removed. Community IDs are aligned across panels via Hungarian matching on the co-assignment overlap matrix; orange shading marks communities that intersect the treated tracts. Each non-reference panel is annotated with its Adjusted Rand Index (ARI) against panel~A. All six partitions recover the same community structure and identify the same treated communities.}
    \label{supp:fig:free_fare_flows_sensitivity}
\end{figure}

To address both, we re-derive the partition under five alternative flow specifications, holding all other parameters fixed (Leiden at $\gamma=0.0025$, $N=100$ ensemble runs, same tract universe): (i)~post-intervention visit-to-visit flows spanning 2021-09-01 to 2022-03-01; (ii)~home-to-visit flows over the pre-intervention window, where edges are home-tract~$\to$~visited-tract; (iii)~pre-intervention visit-to-visit flows with 10\% of visits removed uniformly at random before aggregation; (iv)~with 50\% of visits removed; and (v)~with 95\% of visits removed. Fig.~\ref{supp:fig:free_fare_flows_sensitivity} shows the six consensus partitions, with community labels aligned across panels via Hungarian matching on the co-assignment overlap matrix so that matched communities share colors. Each non-reference panel is annotated with its Adjusted Rand Index (ARI) against panel~A. The ARI measures the agreement between two partitions of the same units, corrected for the agreement expected by chance~\cite{hubert_comparing_1985}. It equals 1 for identical partitions and 0 for a random labeling, so values near 1 indicate that the two partitions assign tracts to communities in nearly the same way.

All six partitions recover the same community structure. Each yields 10 communities, except the home-to-visits partition, which splits community~10 into a further community~11, and all six identify the same treated communities (those with $\geq 10\%$ treated-tract share). Pairwise ARI against the main-text partition ranges from $0.86$ for home-to-visits to $0.97$ for 10\% and 50\% random thinning, and 95\% random thinning still yields $0.94$. The only visible differences are small boundary reassignments between adjacent communities. The community structure used for NDM is therefore stable under these perturbations of the mobility network.

The post-intervention partition agrees with the pre-intervention one at ARI $0.96$ (Fig.~\ref{supp:fig:free_fare_flows_sensitivity}B), which matters beyond robustness. The program changed the volume of travel within the existing mobility communities, not the community structure itself. The partition that defines treated and control units is a stable property of the city that the intervention operates within rather than reshapes, which is consistent with the magnitude of the mobility change the policy produces.

\subsubsection{LODES commute flows} \label{supp:sec:sensitivity_lodes}

As an independent check on the NDM network, we rebuild the Free Fare donor partition from LEHD LODES 2021 home-to-work commute flows \cite{census_lodes2021} instead of the Cuebiq visit-to-visit flows used in the main text. LODES encodes residence-to-workplace ties from administrative records, so it provides a mobility signal disjoint from the device-stay panels. We apply the same Leiden and hierarchical-consensus pipeline as the main analysis (Section~\ref{supp:sec:NDM}) and compute network and propensity distances as before. Because commute flows yield sparser and smaller communities than visit flows, we relax the community-size floor of Section~\ref{supp:secc:pareto_selection} and keep communities with at least 10 active tracts.

\begin{figure}[t!]
    \centering
    \includegraphics[width=\textwidth]{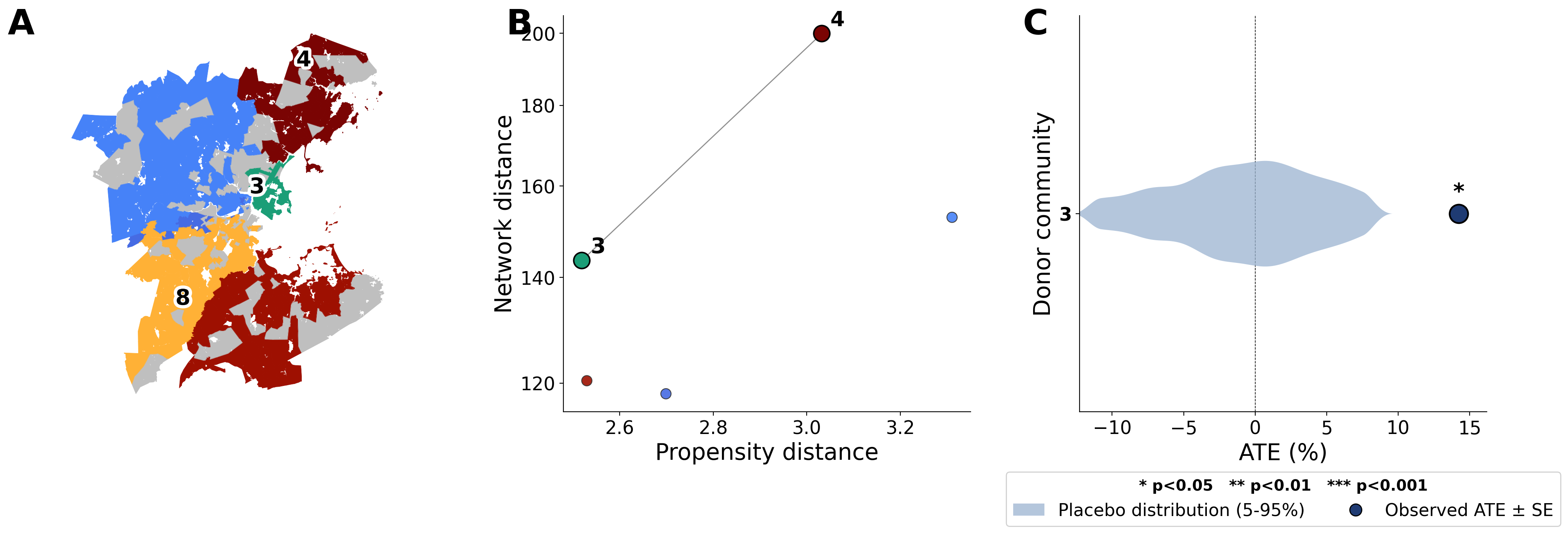}
    \caption[\textbf{Free Fare: sensitivity to LODES commute flows.}]{\textbf{Free Fare: sensitivity to LODES commute flows.} \textbf{(A)}~Consensus community partition built from LEHD LODES 2021 home-to-work flows on the Boston tracts. Treated communities are orange, the selected donor (community~3) green, other Pareto-front communities in their map colors, and zero-outflow or too-small ($<10$ tract) communities in grey. \textbf{(B)}~Candidate donor communities in the (propensity distance, network distance) space; the Pareto front $\{3,4\}$ is connected and labeled by community id, with the selected donor in green. \textbf{(C)}~Observed total-visits ATE for the selected donor community (dark navy, $\pm 1$ regression SE) against the placebo-null distribution (violin, 5th--95th percentile of 100 placebo draws, each pseudo-treating half of the donor tracts). Significance is assessed against the placebo band, not the regression SE; the asterisks denote empirical placebo $p$-values ($^\ast p<0.05$, $^{\ast\ast} p<0.01$).}
    \label{supp:fig:free_fare_lodes_pareto}
\end{figure}

The Pareto front of donor candidates contains two communities (Fig.~\ref{supp:fig:free_fare_lodes_pareto}A,B): community~3 (36 donor tracts), at the lower propensity distance, and community~4 (57 tracts), farther in network distance but the more demographically dissimilar. We retain community~3 as the donor pool and fit the same SCM-DiD as the main analysis, with an empirical $p$-value from a 100-draw placebo test.

The observed total-visits ATE under the community~3 donor pool is $+14.24\%$ (SE $0.47$, placebo $p=0.020$), outside the 5th--95th percentile band of the placebo distribution, and the pre-treatment fit remains tight (NRMSE $0.25$, correlation $0.97$). The farther community~4 yields a smaller $+6.38\%$ [$0.38$] ($p=0.059$), consistent with commute ties capturing only part of the leisure-and-errand mobility through which the program operates. A commute-only administrative network therefore recovers the same conclusion as the main-text NDM estimate of $+15.65\%$: a significant positive uplift in visits among treated communities, from a mobility signal collected independently of the device panels.

\subsection{Sensitivity to random treated-unit removal} \label{supp:sec:sensitivity_remove_tracts}
To check that the NDM ATE does not hinge on a small number of influential treated units, we re-estimate SCM after randomly removing $\{10, 25, 50, 75, 90\}\%$ of the treated pool. For each removal percentage we draw 10 independent random subsets of the treated units and refit SCM on each. We apply the procedure to all five treated settings: Boston Free Fare, the CRZ, and the three Open Streets sites. The donor pool is held fixed, so the placebo null does not change as treated units are removed, and we reuse the full-pool placebo coefficients to score every sub-sample.

\begin{figure}[t!]
    \centering
    \includegraphics[width=\textwidth]{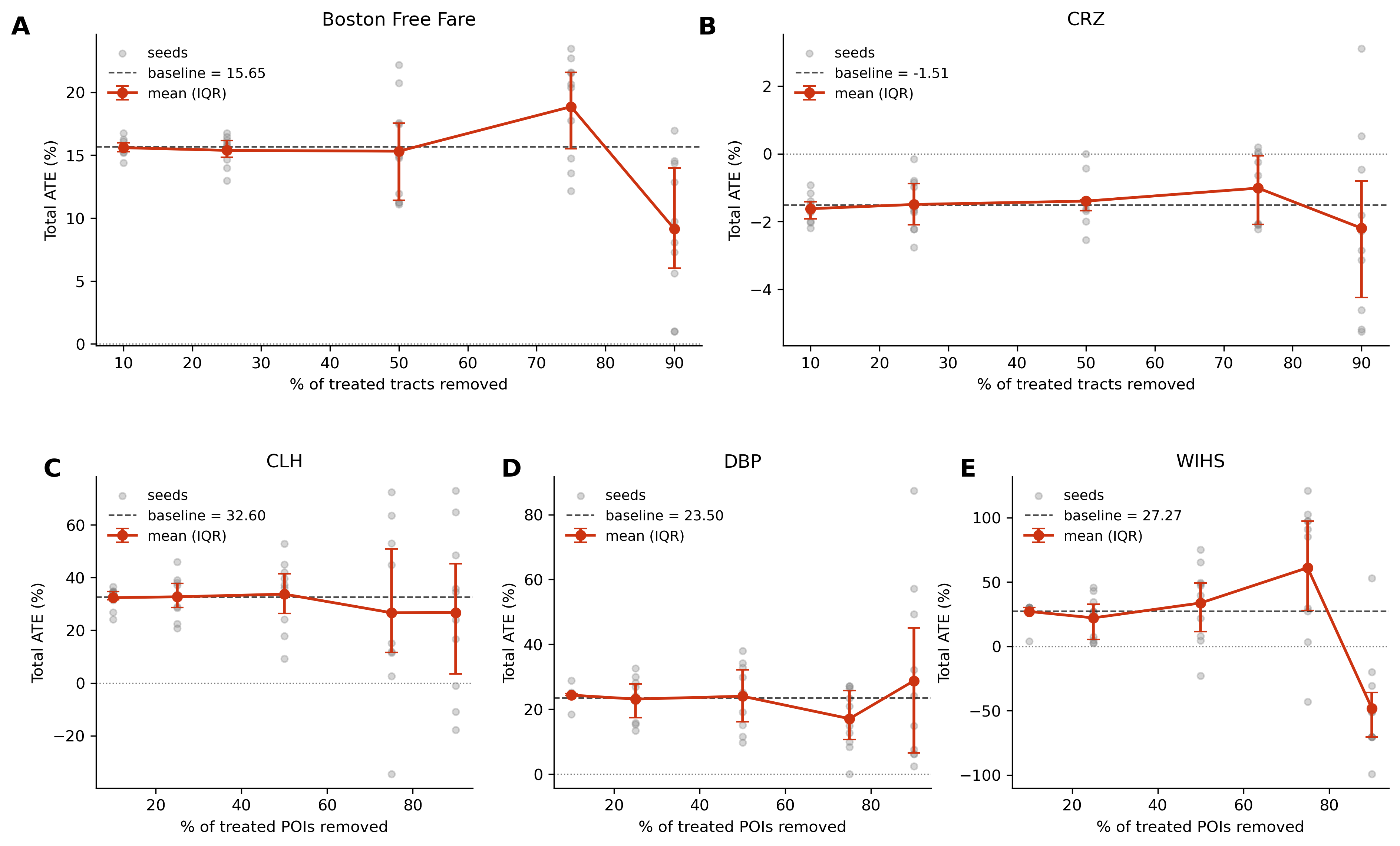}
    \caption[\textbf{NDM ATE under random treated-unit removal.}]{\textbf{NDM ATE under random treated-unit removal.} Stability of the total-visits ATE when the treated pool is randomly thinned by $\{10,25,50,75,90\}$\% with the donor pool held fixed. Grey points show the ATE for each of 10 random draws per removal percentage; red markers show the mean with whiskers spanning the IQR (Q25--Q75); the dashed line marks the full-pool baseline ATE (value given in each panel). Because the donor pool is unchanged, the placebo null is unaffected by treated-unit removal, so placebo coefficients are computed once on the full pool (100 draws) and reused to score every sub-sample. (A) Boston Free Fare. (B) CRZ. (C--E) NYC Open Streets sites (CLH, DBP, WIHS).}
   \label{supp:fig:remove_tracts_combined}
\end{figure}

Fig.~\ref{supp:fig:remove_tracts_combined} shows the ATE distribution at each removal percentage for total visits. The mean ATE stays close to the full-pool baseline up to 50\% removal. Up to that point the IQR still excludes zero wherever the full-pool estimate did, and the largest deviation is under three percentage points in every panel except WIHS, where it reaches 6.4 points. At 75\% removal the mean moves by up to 6.4 points in the other panels and by 34 points at WIHS. At 90\% removal it moves by up to 6.5 points in the other panels, while the WIHS estimate breaks down (a mean of $-48\%$), since each draw then keeps only one treated POI.

Every main-text conclusion survives across the range that keeps a usable treated pool: a significant positive uplift for Free Fare and the three Open Streets sites, and a null overall effect for the CRZ. The NDM results are therefore stable to the removal of treated units, holding up even when half the treated pool is dropped.

\small

\bibliography{bibliography}

\end{document}